\documentclass[11pt]{article}
\usepackage[letterpaper]{geometry}
\usepackage[parfill]{parskip}
\usepackage{amsmath,amsthm,amssymb,bbm}
\usepackage{mathtools}
\usepackage{cases}
\usepackage{graphicx}
\usepackage{caption}
\usepackage{subcaption}
\usepackage{tabularx}
\usepackage{microtype}
\usepackage{enumitem}
\usepackage{dsfont}
\usepackage[authoryear]{natbib}
\usepackage{algpseudocode,algorithm}

\usepackage{floatpag} 
\floatpagestyle{empty} 
\usepackage{authblk}
\usepackage{url}
\usepackage[colorlinks,citecolor=blue,urlcolor=blue,linkcolor=blue,linktocpage=true]{hyperref}
\usepackage{cleveref}
\crefformat{equation}{(#2#1#3)}
\crefrangeformat{equation}{(#3#1#4) to~(#5#2#6)}
\crefname{equation}{}{}
\Crefname{equation}{}{}

\makeatletter
\def\thm@space@setup{%
  \thm@preskip=\parskip \thm@postskip=0pt
}
\makeatother

\newtheorem{theorem}{Theorem}[section]

\newtheorem{remark}{Remark}[section]

\newtheorem*{definition*}{Definition}
\newtheorem*{remark*}{Remark}
\crefname{definition}{\textbf{definition}}{definitions}
\Crefname{definition}{Definition}{Definitions}
\crefname{assumption}{\textbf{assumption}}{assumptions}
\Crefname{assumption}{Assumption}{Assumptions}
\newcommand{\argmin}{\mathop{\mathrm{argmin}}}

\let\hat\widehat
\let\tilde\widetilde
\def\P{\mathbb{P}}
\def\E{\mathbb{E}}
\def\R{\mathbb{R}}

\def\one{\mathds{1}}

\def\SplitWidth{d}

\begin{document}

\author{Jing Lei}
\author{Max G'Sell}
\author{Alessandro Rinaldo}
\author{Ryan J. Tibshirani}
\author{Larry Wasserman}
\affil{Department of Statistics, Carnegie Mellon University}

\title{\bf\Large Distribution-Free Predictive Inference For Regression}  

\maketitle

\begin{abstract}

We develop a general framework for distribution-free predictive
inference in regression, using conformal inference. The
proposed methodology allows for the construction of
a prediction band for the response variable using any estimator of 
 the regression function.  
The resulting prediction band preserves the consistency properties
of the original estimator under standard assumptions, while
guaranteeing finite-sample marginal coverage even when these
assumptions do not hold.
We analyze and compare, both empirically and theoretically, the two
major variants of our conformal framework: full conformal inference
and split conformal inference, along with a related jackknife method.  
These methods offer different tradeoffs between statistical accuracy
(length of resulting prediction intervals) and computational
efficiency.  
As extensions, we develop a method for constructing valid in-sample
prediction intervals called {\it rank-one-out} conformal 
inference, which has essentially the same computational efficiency as
split conformal inference. We also describe an extension of our
procedures for producing 
prediction bands with locally varying length, in order to adapt to
heteroskedascity in the data.
Finally, we propose a model-free notion of variable importance, called   
{\it leave-one-covariate-out} or LOCO inference. Accompanying this  
paper is an R package {\tt conformalInference} that implements all of 
the proposals we have introduced. In the spirit of reproducibility,
all of our empirical results can also be easily (re)generated
using this package.

\end{abstract}

\newpage
\tableofcontents
\newpage
\section{Introduction}
\label{section::introduction}

Consider i.i.d.\ regression data
\begin{equation*}
Z_1,\ldots, Z_n \sim P,
\end{equation*}
where each $Z_i = (X_i,Y_i)$ is a random variable in $\mathbb{R}^d
\times \mathbb{R}$, comprised of a response variable   
$Y_i$ and a $d$-dimensional vector of features (or predictors, or
covariates) $X_i = (X_i(1),\ldots, X_i(d))$.  The feature 
dimension $d$ may be large relative to the sample size $n$ (in an 
asymptotic model, $d$ is allowed to increase with $n$). 
Let
\begin{equation*}
  \mu(x) = \E(Y \,|\, X=x), \quad x \in \mathbb{R}^d
\end{equation*}
denote the regression function.
We are interested in predicting a new response $Y_{n+1}$ from a new
feature value $X_{n+1}$, with no assumptions on $\mu$ and 
$P$.  Formally, given a nominal miscoverage level $\alpha
\in (0,1)$, we seek to constructing a prediction band 
$C\subseteq \R^d \times \R$ based on $Z_1,\ldots,Z_n$ with the
property that 
\begin{equation}
\label{eq:validity}
\P\big( Y_{n+1} \in C(X_{n+1}) \big) \geq 1-\alpha,
\end{equation}
where the probability is taken over the $n+1$ i.i.d.\ draws
\smash{$Z_1,\ldots,Z_n,Z_{n+1} \sim P$}, and for a point $x \in
\R^d$ we denote
$C(x) = \{y \in \R : (x,y) \in C \}$.
The main goal of this paper is to construct prediction bands as in
\eqref{eq:validity} that 
have finite-sample (nonasymptotic) validity, without assumptions on
$P$.  A second goal is to construct model-free
inferential statements about the importance of each
covariate in the prediction model for $Y_{n+1}$ given $X_{n+1}$.

Our leading example is high-dimensional regression,
where $d \gg n$ and a linear function is used to approximate
$\mu$ (but the linear model is not necessarily assumed to be correct).  
Common approaches in this setting include 
greedy methods like forward stepwise regression, and $\ell_1$-based 
methods like the lasso. There is an enormous
amount of work dedicated to studying various properties of these
methods, but to our knowledge, there is very little work on prediction 
sets. Our framework provides proper prediction sets for these methods,
and for essentially any high-dimensional regression method.  It also
covers classical linear regression and nonparametric regression
techniques.  The basis of our framework is 
{\em conformal prediction}, a method invented by \citet{VovkGS05}.  

\subsection{Related Work}

\paragraph{Conformal inference.}
The conformal prediction framework was originally proposed as a
sequential approach for forming prediction intervals, by
\citet{VovkGS05,vovk2009line}. The basic idea is  
simple. Keeping the regression setting introduced above and given a
new independent draw $(X_{n+1},Y_{n+1})$ from $P$, in order to 
decide if a value $y$ is to be included in $C(X_{n+1})$,
we consider testing the null hypothesis that $Y_{n+1}=y$ and construct 
a valid $p$-value based on the empirical quantiles of the augmented
sample 
$(X_1,Y_1),\ldots,(X_n,Y_n),(X_{n+1},Y_{n+1})$ with $Y_{n+1}=y$ (see
\Cref{sec::method} below for details).  The
data augmentation step makes the procedure immune to overfitting, so
that the resulting prediction band always has valid average coverage
as in \eqref{eq:validity}.
Conformal inference has also been studied as a batch (rather than
sequential) method, in various settings. For example,
\citet{burnaev2014efficiency} considered low-dimensional least squares
and ridge regression models. \citet{LeiRW13} used conformal prediction
to construct statistically near-optimal tolerance
regions. \citet{LeiW14} extended this result to low-dimensional
nonparametric regression.  Other extensions, such as classification
and clustering, are explored in \citet{Lei14,LeiRW14}.

There is very little work on prediction sets in high-dimensional
regression.  \citet{Hebiri10} described an approximation of the
conformalized lasso estimator.  This approximation leads to a big
speedup over the original conformal prediction method build on top of 
the lasso, but loses the key appeal of conformal inference in the
first place---it fails to offer finite-sample coverage.  
Recently \citet{steinberger2016} analyzed a jackknife prediction
method in the high-dimensional setting, extending results
in low-dimensional regression due to \citet{Rothman}.
However, this jackknife approach is only guaranteed to have asymptotic
validity when the base estimator (of the regression parameters) 
satisfies strong asymptotic mean squared error and stability
properties.  This is further discussed in Section
\ref{sec:jack_conformal}.  In our view, a simple, computationally
efficient, and yet powerful method that seems to have been overlooked
is {\it split conformal inference} (see
\citet{LeiRW14,papadopoulos2002inductive}, or  
Section \ref{sec:split_conformal}).  When combined with, for example, the
lasso estimator, the total cost of forming split conformal prediction 
intervals is dominated by the cost of fitting the lasso, and the method
always provides finite-sample coverage, in any setting---regardless of 
whether or not the lasso estimator is consistent.  

\paragraph{High-dimensional inference.}
A very recent and exciting research thread in the field of high-dimensional inference
is concerned with the construction of confidence 
intervals for (fixed) population-based 
targets, or (random) post-selection targets. 
In the first class, population-based approaches, the linear model is
assumed to be true and the focus is on providing confidence intervals
for the coefficients in this model
(see, e.g.,
\citet{BelloniCCH12,buhlsignif,zhangconf,vdgsignif,montahypo2}). 
In the second class, post-selection approaches, the focus is on
covering coefficients in the best linear approximation to $\mu$
given a subset of selected covariates (see, e.g., 
\citet{posi,exactlasso,exactlar,optimalinf,tian2015asymptotics,tian2015selective}). 
These inferential approaches are all interesting, 
and they serve different purposes (i.e., the purposes behind the two
classes are different). One common thread, however, is that all of
these methods rely on nontrivial assumptions---even if the linear
model need not be assumed true, conditions are typically placed (to a
varying degree) on the quality of the regression estimator under
consideration, the error distribution, the knowledge or 
estimability of error variance, the homoskedasticity of errors, etc.
In contrast, we describe two prediction-based methods for variable
importance in Section \ref{sec::varimp}, which do not rely on such 
conditions at all.

\subsection{Summary and Outline}

In this paper, we make several methodological and theoretical 
contributions to conformal inference in regression.
\begin{itemize}
\item We provide a general introduction to conformal inference
  (\Cref{sec::method}), a generic tool to construct distribution-free,
  finite-sample prediction sets.  We specifically consider the context
  of high-dimensional regression, arguably the scenario where
  conformal inference is most useful, due to the strong assumptions
  required by 
  existing inference methods.   

\item We provide new theoretical insights for conformal inference:
  accuracy guarantees for its finite-sample coverage
  (Theorems \ref{thm:conf_valid}, \ref{thm:split_valid}), and
  distribution-free asymptotic, in-sample coverage guarantees
  (Theorems \ref{thm:split_asymptotic}, \ref{thm:roo_asymptotic}).
\item  We also show that versions of conformal inference approximate 
  certain oracle methods (\Cref{sec::theory}). In doing so, we provide 
  near-optimal bounds on the length of the prediction interval under
  standard assumptions. Specifically, we show the following.
  \begin{enumerate}
    \item If the base estimator is stable under resampling and small
      perturbations, then the conformal prediction bands are close to an
      oracle band that depends on the estimator
      (Theorems \ref{lemma::split}, \ref{thm:full_practical}).
    \item If the base estimator is consistent, then the conformal
      prediction bands are close to a super oracle band which has the 
      shortest length among all valid prediction bands (Theorems 
      \ref{thm:split_super}, \ref{thm::consistent}).
    \end{enumerate}
 
\item We conduct extensive simulation studies
(\Cref{sec::experiments}) to assess the two major variants of
conformal inference: the full and split conformal methods, along with
a related jackknife method. These simulations can be reproduced using 
our accompanying R package {\tt conformalInference} 
(\url{https://github.com/ryantibs/conformal}), which provides an
implementation of all the methods studied in this paper (including the
extensions and variable importance measures described below).  

\item We develop two extensions of conformal inference
  (\Cref{sec::extensions}), allowing for more informative and flexible
  inference: prediction intervals with in-sample coverage, and
  prediction intervals with varying local length. 

\item We propose two new, model-free, prediction-based approaches for   
  inferring variable importance based on {\it leave-one-covariate-out}
  or LOCO inference (\Cref{sec::varimp}).
\end{itemize}

\section{Conformal Inference}
\label{sec::method}

The basic idea behind the theory of conformal prediction is related to
a simple result about sample quantiles.  Let 
$U_1,\ldots,U_n$ be i.i.d.\ samples of a scalar random variable (in fact, the
arguments that follow hold with the i.i.d.\ assumption replaced by the
weaker assumption of exchangeability). 
For a given miscoverage level $\alpha \in (0,1)$, and
another i.i.d.\ sample $U_{n+1}$, note that   
\begin{equation}
\label{eq:sample_quantile}
\P(U_{n+1} \leq \hat{q}_{1-\alpha}) \geq 
1-\alpha, 
\end{equation} 
where we define the sample quantile \smash{$\hat{q}_{1-\alpha}$} based 
on $U_1,\ldots,U_n$ by
\begin{equation*}
\hat{q}_{1-\alpha} = \begin{cases}
U_{(\lceil (n+1)(1-\alpha)\rceil)} & \text{if 
$\lceil (n+1)(1-\alpha)\rceil \leq n$} \\
\infty & \text{otherwise},
\end{cases}
\end{equation*}
and $U_{(1)} \leq \ldots \leq U_{(n)}$ denote the order
statistics of $U_1,\ldots,U_n$. The finite-sample coverage property in 
\eqref{eq:sample_quantile} is easy to verify: by exchangeability,
the rank of $U_{n+1}$ among $U_1,\ldots,U_n,U_{n+1}$ is uniformly distributed
over the set $\{1,\ldots,n+1\}$. 

In our regression problem, where we observe i.i.d.\ samples
$Z_i=(X_i,Y_i) \in \R^d \times \R \sim P$, $i=1,\ldots,n$, we might 
consider the following naive method for constructing a prediction
interval for $Y_{n+1}$ at the new feature value $X_{n+1}$, where
$(X_{n+1},Y_{n+1})$ is an independent draw from $P$.  Following  
 the idea described above, we can form the prediction interval defined
by 
\begin{equation}
    \label{eq:naive}
C_{\mathrm{naive}} (X_{n+1}) =  
\Big[\hat\mu(X_{n+1}) - \hat{F}_n^{-1}(1-\alpha), \;
\hat\mu(X_{n+1}) + \hat{F}_n^{-1}(1-\alpha) \Big],
\end{equation}
where \smash{$\hat\mu$} is an estimator of the underlying 
regression function and \smash{$\hat{F}_n$} the empirical distribution
of the fitted residuals \smash{$|Y_i-\hat\mu(X_i)|$}, $i=1,\ldots,n$, 
and \smash{$\hat{F}_n^{-1}(1-\alpha)$} the $(1-\alpha)$-quantile of   
\smash{$\hat{F}_n$}. This is approximately valid for large samples,
provided that the estimated regression function \smash{$\hat\mu$} is
accurate (i.e., enough for the estimated $(1-\alpha)$-quantile   
\smash{$\hat{F}_n^{-1}(1-\alpha)$} of the fitted residual distribution 
to be close the $(1-\alpha)$-quantile of the population residuals 
$|Y_i-\mu(X_i)|$, $i=1,\ldots,n$).  Guaranteeing such an accuracy for  
\smash{$\hat\mu$} generally requires appropriate 
regularity conditions, both on the underlying data distribution $P$,
and on the estimator \smash{$\hat\mu$} itself, such as a correctly
specified model and/or an appropriate choice of tuning parameter.   

\subsection{Conformal Prediction Sets}
\label{sec:full_conformal}

In general, the naive method \eqref{eq:naive} can grossly undercover since the fitted 
residual distribution can often be biased downwards. Conformal
prediction  
intervals \citep{VovkGS05,vovk2009line,LeiRW13,LeiW14} overcome the
deficiencies of the naive intervals, and, somewhat remarkably, are
guaranteed to deliver proper finite-sample coverage without any
assumptions on $P$ or \smash{$\hat\mu$} (except that
\smash{$\hat\mu$} act a symmetric function of the data points).   

Consider the following strategy: for each value $y \in \R$, we
construct an augmented regression estimator \smash{$\hat\mu_y$}, which 
is trained on the augmented data set $Z_1,\ldots,Z_n,(X_{n+1},y)$. Now
we define 
\begin{equation}
\label{eq:conf_resid} 
R_{y,i} = |Y_i -  \hat\mu_y(X_i)|, \; i=1,\ldots,n 
\quad\text{and}\quad
R_{y,n+1} = |y - \hat\mu_y(X_{n+1})|,
\end{equation}
and we rank $R_{y,n+1}$ among the remaining fitted residuals
$R_{y,1},\ldots,R_{y,n}$, computing 
\begin{equation}
\label{eq:conf_rank}
\pi(y) = \frac{1}{n+1}\sum_{i=1}^{n+1} \one\{R_{y,i}\leq R_{y,n+1}\}  
= \frac{1}{n+1} +
\frac{1}{n+1}\sum_{i=1}^n \one\{R_{y,i}\leq R_{y,n+1}\}, 
\end{equation}
the proportion of points in the augmented sample whose fitted
residual is smaller than the last one, $R_{y,n+1}$. Here
$\one\{\cdot\}$ is the indicator function. By exchangeability of the
data points and the symmetry of $\hat\mu$, when 
evaluated at $y=Y_{n+1}$, we see that the constructed statistic
$\pi(Y_{n+1})$ is uniformly distributed over the set
$\{1/(n+1),2/(n+1),\ldots,1\}$, which implies
\begin{equation}
\label{eq:pi_pvalue}
\P\Big( (n+1)\pi(Y_{n+1})\leq \lceil(1- \alpha)(n+1)\rceil\Big) 
\geq 1-\alpha. 
\end{equation}
We may interpret the above display as saying that
$1-\pi(Y_{n+1})$ provides a valid (conservative) p-value for testing the null
hypothesis that $H_0: Y_{n+1}=y$. 

By inverting such a test over all possibly values of $y \in
\mathbb{R}$, the property
\eqref{eq:pi_pvalue} immediately leads to our conformal prediction 
interval at $X_{n+1}$, namely
\begin{equation}
\label{eq:conf_int}
C_{\mathrm{conf}}(X_{n+1}) = 
\Big\{y \in \mathbb{R} : (n+1)\pi(y)\le \lceil(1- \alpha)(n+1)\rceil\Big\}. 
\end{equation}
The steps in \eqref{eq:conf_resid}, \eqref{eq:conf_rank},
\eqref{eq:conf_int} must be repeated each time we want to produce a
prediction interval (at a new feature value).  In practice, we
must also restrict our attention in \eqref{eq:conf_int} to a discrete grid
of trial values $y$.  For completeness, this is summarized in
Algorithm \ref{alg:conf}.  

\begin{algorithm}[tb]
\caption{Conformal Prediction}
\label{alg:conf}
\begin{algorithmic}
\State{\bf Input:} Data $(X_i,Y_i)$, $i=1,\ldots,n$, miscoverage
  level $\alpha \in (0,1)$, regression algorithm $\mathcal A$, points  
    $\mathcal{X}_{\mathrm{new}} = \{X_{n+1},X_{n+2},\ldots\}$ at which
    to construct prediction intervals, and values 
    $\mathcal{Y}_{\mathrm{trial}} = \{y_1,y_2,\ldots\}$ to act as
    trial values 
\State{\bf Output:} Predictions intervals, at each element of 
  $\mathcal{X}_{\mathrm{new}}$ 
\For{$x \in \mathcal{X}_{\mathrm{new}}$}
\For{$y \in \mathcal{Y}_{\mathrm{trial}}$}
\State \smash{$\hat\mu_y = \mathcal
A\big(\{(X_1,Y_1),\ldots,(X_n,Y_n),(x,y)\}\big)$}  
\State \smash{$R_{y,i} = |Y_i-\hat\mu_y(X_i)|$}, $i=1,\ldots,n$, and  
\smash{$R_{y,n+1} = |y - \hat\mu_y(x)|$}
\State $\pi(y) = (1 + \sum_{i=1}^n \one\{R_{y,i}\leq
R_{y,n+1})\}/(n+1)$ 
\EndFor
\State \smash{$C_{\mathrm{conf}}(x) = \{ y \in  
\mathcal{Y}_{\mathrm{trial}} : (n+1)\pi(y) \leq \lceil ( 1-\alpha)(n+1) \rceil \}$}
\EndFor
\State Return \smash{$C_{\mathrm{conf}}(x)$}, for   
each $x \in \mathcal{X}_{\mathrm{new}}$
\end{algorithmic}
\end{algorithm}

By construction, the conformal prediction interval in
\eqref{eq:conf_int} has valid finite-sample coverage; this interval 
is also accurate, meaning that it does not substantially over-cover.
These are summarized in the following theorem, whose proof is in
\Cref{sec:proof_1}. 

\begin{theorem}
\label{thm:conf_valid}
If $(X_i,Y_i)$, $i=1,\ldots,n$ are i.i.d., then for an new i.i.d.\
pair $(X_{n+1},Y_{n+1})$, 
\begin{equation*}
\P\big(Y_{n+1} \in C_{\mathrm{conf}}(X_{n+1})\big) 
\geq 1-\alpha,
\end{equation*}
for the conformal prediction band
$C_{\mathrm{conf}}$ constructed in 
\eqref{eq:conf_int} (i.e., Algorithm \ref{alg:conf}). If we assume 
additionally that for all $y \in \mathbb{R}$, the fitted absolute residuals
\smash{$R_{y,i}=|Y_i-\hat\mu_y(X_i)|$}, 
$i=1,\ldots,n$ have a continuous joint distribution, then it also holds
that  
\begin{equation*}
\P\big(Y_{n+1} \in C_{\mathrm{conf}}(X_{n+1})\big)
\leq 1-\alpha+\frac{1}{n+1}. 
\end{equation*}
\end{theorem}

\begin{remark}
The first part of the theorem, on the finite-sample validity of 
conformal intervals in regression, is a standard property of all 
conformal inference procedures and is due to Vovk.
The second part---on the anti-conservativeness of
conformal intervals---is new.  For the second part only, we require
that the residuals have a continuous distribution, which
is quite a weak assumption, and is used to avoid ties when ranking the
(absolute) residuals.  By using a random tie-breaking rule, this
assumption could be avoided entirely.
In practice, the coverage of conformal intervals is
highly concentrated around $1-\alpha$, as confirmed by the experiments 
in \Cref{sec::experiments}.
Other than the continuity assumption, no assumptions are needed in
\Cref{thm:conf_valid} about the  
the regression estimator \smash{$\hat\mu$} or the data generating
distributions $P$. This is a somewhat remarkable 
and unique property of conformal inference, and is not true for the 
jackknife method, as discussed in 
\Cref{sec:jack_conformal} (or, say, for the methods used to produce
confidence intervals  for the coefficients in high-dimensional
linear model).
\end{remark}

\begin{remark}
Generally speaking, as we improve our estimator \smash{$\hat\mu$} 
of the underlying regression function $\mu$, the resulting conformal
prediction interval decreases in length.  Intuitively, this happens
because a more accurate \smash{$\hat\mu$} leads to
smaller residuals, and conformal
intervals are essentially defined by the quantiles of the (augmented)
residual distribution. \Cref{sec::experiments} gives empirical
examples that support this intuition.
\end{remark}

\begin{remark}
The probability statements in \Cref{thm:conf_valid} are taken
over the i.i.d.\ samples $(X_i,Y_i)$, $i=1,\ldots,n,n+1$,
and thus they assert average (or marginal) coverage guarantees.  This
should not be confused with 
\smash{$\P(Y_{n+1} \in C(x) \,|\, X_{n+1}=x) \geq 1-\alpha$} for all
$x \in \R^d$, i.e., conditional coverage, 
which is a much stronger property and cannot be achieved by
finite-length 
prediction intervals without regularity and consistency assumptions on
the model and the estimator \citep{LeiW14}. Conditional coverage does
hold asymptotically under certain conditions; see \Cref{thm::consistent} in
\Cref{sec::theory}. 
\end{remark}



\begin{remark}
\label{rem:conformity_score}
\Cref{thm:conf_valid} still holds if we replace each
$R_{y,i}$ by 
\begin{equation}
\label{eq:conf_score_gen}
f\big((X_1,Y_1),\ldots,(X_{i-1},Y_{i-1}),(X_{i+1},Y_{i+1}),\ldots,(X_{n+1},y);
\;(X_{i},Y_{i})\big), 
\end{equation}
where $f$ is any function that is symmetric in its first $n$
arguments. Such a function $f$ is called the 
{\em conformity score}, in the context of conformal 
inference. For example, the value in \eqref{eq:conf_score_gen} can be
an estimated joint density function evaluated at $(X_i,Y_i)$,
or conditional density function at $(X_i,Y_i)$ (the latter is
equivalent to the absolute residual $R_{y,i}$ when $Y-\E(Y|X)$ is 
independent of $X$, and has a symmetric distribution with decreasing
density on $[0,\infty)$.) We will discuss a special locally-weighted
conformity score in \Cref{sec:local_weight}. 
\end{remark}

\begin{remark} 
We generally use the term ``distribution-free'' to refer to the
finite-sample coverage property, assuming only i.i.d.\ data. 
Although conformal prediction provides valid coverage for all
distributions and all symmetric estimators under only the i.i.d.\ 
assumption, the length of the conformal interval depends on 
the quality of the initial estimator, and in \Cref{sec::theory} we
provide theoretical insights on this relationship.  
\end{remark}

\subsection{Split Conformal Prediction Sets}
\label{sec:split_conformal}

The original conformal prediction method studied in the last
subsection is computationally intensive. For any $X_{n+1}$ and $y$, in 
order to tell if $y$ is to be included in
$C_{\mathrm{conf}}(X_{n+1})$, we retrain the model on the
augmented data set (which includes 
the new point $(X_{n+1},y)$), and recompute and reorder the absolute 
residuals. In some applications, where $X_{n+1}$ is not 
necessarily observed, prediction intervals are build by evaluating
\smash{$\one \{y\in C_{\mathrm{conf}}(x)\}$} over all pairs of $(x,y)$
on a fine grid, as in \Cref{alg:conf}. In the special cases 
of kernel density estimation and kernel regression, simple and
accurate approximations to the full conformal prediction 
sets are described in \citet{LeiRW13,LeiW14}. In low-dimensional linear
regression, the Sherman-Morrison updating scheme can be used to reduce
the complexity of the full conformal method, by saving on the cost of solving
a full linear system each time the query point
$(x,y)$ is changed.  But in high-dimensional regression, where we
might use relatively sophisticated (nonlinear) estimators such as the
lasso, performing efficient full conformal inference is still an open
problem.  

Fortunately, there is an alternative approach, which we call 
{\em split conformal prediction}, that is completely general, and 
whose computational cost is a small fraction of the full
conformal method. The split conformal 
method separates the fitting and ranking steps using sample
splitting, and its computational cost is simply that of the fitting
step. Similar ideas have appeared in the online prediction literature
known under the name {\em inductive conformal inference}  
\citep{papadopoulos2002inductive,VovkGS05}. 
The split conformal algorithm summarized in
\Cref{alg:split} is adapted from \citet{LeiRW14}.  Its
key coverage properties are given in \Cref{thm:split_valid}, proved
in \Cref{sec:proof_1}. 
(Here, and henceforth when discussing split conformal inference, we 
assume that the sample size $n$ is even, for simplicity, as only very
minor changes are needed when $n$ is odd.)

\begin{algorithm}[tb]
\caption{Split Conformal Prediction}
\label{alg:split}
\begin{algorithmic}
\State{\bf Input:} Data $(X_i,Y_i)$, $i=1,\ldots,n$, miscoverage level
$\alpha \in (0,1)$, regression algorithm $\mathcal A$
\State{\bf Output:} Prediction band, over $x \in \R^d$
\State Randomly split $\{1,\ldots,n\}$ into two equal-sized
subsets $\mathcal I_1$, $\mathcal I_2$
\State \smash{$\hat\mu = \mathcal A\big(\{(X_i,Y_i): i \in \mathcal
  I_1\}\big)$} 
\State \smash{$R_i = |Y_i-\hat\mu(X_i)|$, $i \in \mathcal I_2$}
\State $\SplitWidth =$ the $k$th smallest value in $\{R_i : i \in 
\mathcal I_2\}$, where $k=\lceil(n/2 +1)(1-\alpha)\rceil$
\State Return \smash{$C_{\mathrm{split}}(x) =
  [\hat\mu(x)-\SplitWidth, \hat\mu(x)+ \SplitWidth]$}, for all $x \in \R^d$ 
\end{algorithmic}
\end{algorithm}

\begin{theorem}
\label{thm:split_valid}
If $(X_i,Y_i)$, $i=1,\ldots,n$ are i.i.d., then for an new i.i.d.\
draw $(X_{n+1},Y_{n+1})$, 
\begin{equation*}
\P\big(Y_{n+1} \in C_{\mathrm{split}}(X_{n+1})\big) 
\geq 1-\alpha,
\end{equation*}
for the split conformal prediction band
$C_{\mathrm{split}}$ constructed in 
\Cref{alg:split}. Moreover, if we assume 
additionally that the residuals $R_i$, $i \in \mathcal I_2$ have a
continuous joint distribution, then
\begin{equation*}
\P\big(Y_{n+1} \in C_{\mathrm{split}}(X_{n+1})\big) 
\leq 1-\alpha+\frac{2}{n+2}.
\end{equation*}
\end{theorem}

In addition to being extremely efficient compared to the original
conformal method, split conformal inference can also hold an advantage 
in terms of memory requirements.  For example, if the regression
procedure $\mathcal A$ (in the notation of 
\Cref{alg:split}) involves variable selection, like the
lasso or forward stepwise regression, then we only need to store the  
selected variables when we evaluate the fit at new points $X_i$, $i \in 
\mathcal I_2$, and compute residuals, for the ranking step.  This
can be a big savings in memory when the original variable set is very 
large, and the selected set is much smaller.

Split conformal prediction intervals also provide an approximate
in-sample coverage guarantee, making them easier to illustrate and 
interpret using the given sample $(X_i,Y_i)$, $i=1,\ldots,n$, without
need to obtain future draws.  This is described next.  


\begin{theorem}
\label{thm:split_asymptotic}
Under the conditions of \Cref{thm:split_valid}, there is an
absolute constant $c>0$ such that, for any $\epsilon>0$,
\begin{equation*}
\P\Bigg(\bigg|\frac{2}{n}\sum_{i \in\mathcal I_2} \one\{Y_i\in
C_{\mathrm{split}}(X_i)\} - (1-\alpha) \bigg| \geq \epsilon\Bigg) \leq 
2\exp\Big(cn^2(\epsilon-4/n)_+^2\Big).
\end{equation*}
\end{theorem}

\begin{remark}
\Cref{thm:split_asymptotic} implies ``half sample''
in-sample coverage.  It is straightforward to extend 
this result to the whole sample, by constructing another
split conformal prediction band, but with the roles of $\mathcal I_1,  
\mathcal I_2$ reversed. This idea is further explored and extended
in \Cref{sec:in_sample}, where we derive 
\Cref{thm:split_asymptotic} as a corollary of a more  
general result. Also, 
for a related result, see Corollary 1 of \cite{VVV}.
\end{remark}

\begin{remark}
Split conformal inference can also be implemented using an unbalanced 
split, 
with $|\mathcal I_1|=\rho n$ and $|\mathcal I_2|=(1-\rho)n$ for some
$\rho \in(0,1)$ (modulo rounding issues).  In some situations, e.g., 
when the regression procedure is complex, it may be beneficial to
choose $\rho>0.5$ so that the trained estimator $\hat\mu$ is more
accurate.  In this paper, we focus on $\rho=0.5$ for simplicity, and
do not pursue issues surrounding the choice of $\rho$.
\end{remark}

\subsection{Multiple Splits}\label{sec:multiple_split}

Splitting improves dramatically on the speed of conformal inference,
but introduces extra randomness into the procedure. One way to reduce
this extra randomness is to combine inferences from several splits.
Suppose that we split the training data $N$ times, yielding split conformal
prediction intervals $C_{{\rm split},1},\ldots, C_{{\rm split},N}$ where each interval is
constructed at level $1-\alpha/N$. Then, we define
\begin{equation}
\label{eq:multi_split}
C_{\mathrm{split}}^{(N)}(x) = \bigcap_{j=1}^N C_{{\rm split},j}(x),
\quad \text{over $x \in \R^d$}.  
\end{equation}
It follows, using a simple Bonferroni-type argument, that the
prediction band \smash{$C_{\mathrm{split}}^{(N)}$}
has marginal coverage level at least $1-\alpha$. 

Multi-splitting as described above
decreases the variability from splitting.  But this may come at a 
price: it is possible that the length of
\smash{$C_{\mathrm{split}}^{(N)}$} grows 
with $N$, though this is not immediately  
obvious. Replacing $\alpha$ by $\alpha/N$ certainly makes the
individual split conformal intervals larger. However, taking an
intersection reduces the size of the final interval. Thus there is a 
``Bonferroni-intersection tradeoff.'' 

The next result shows that, under rather 
general conditions as detailed in \Cref{sec::theory}, the Bonferroni
effect dominates and we hence get larger intervals as $N$
increases. Therefore, we suggest using a single split. The proof  
is given in \Cref{sec:proof_3}. 

\begin{theorem}\label{thm:mult_split}
Under Assumptions A0, A1, and A2 with $\rho_n=o(n^{-1})$ (these are 
described precisely in \Cref{sec::theory}), if $|Y-\tilde\mu(X)|$ has 
continuous distribution, then with probability tending to $1$ as $n 
\to \infty$, \smash{$C_{\rm split}^{(N)}(X)$} is wider than
\smash{$C_{\rm split}(X)$}.  
\end{theorem}

\begin{remark}
Multiple splits have also been considered by other authors, e.g.,
\citet{stabselect}. However, the situation there is rather different,
where the linear model is assumed correct and inference is performed 
on the coefficients in this linear model. 
\end{remark}

\subsection{Jackknife Prediction Intervals}
\label{sec:jack_conformal}

Lying between the computational complexities of the full and
split conformal methods is {\em jackknife prediction}.
This method uses the quantiles of leave-one-out residuals to
define prediction intervals, and is summarized in
\Cref{alg:jack}. 

An advantage of the jackknife method over the split conformal method
is that it utilizes more of the training data when constructing the
absolute residuals, and subsequently, the quantiles.  
This means that it can often produce intervals of
shorter length.  A clear disadvantage, however, is that its
prediction intervals are not guaranteed to have valid coverage in
finite samples.  In fact, even asymptotically, its coverage properties
do not hold without requiring nontrivial conditions on the base
estimator.  We note that, by symmetry, the jackknife method has the
finite-sample in-sample coverage property
\begin{equation*}
\P\big(Y_i\in C_{\mathrm{jack}}(X_i)\big) \geq 
1-\alpha, \quad \text{for all $i=1,\ldots,n$}.
\end{equation*}
But in terms of out-of-sample coverage (true predictive inference),
its properties are much more fragile.
\citet{Rothman} show that in a low-dimensional linear regression
setting, the jackknife method produces asymptotic valid intervals
under regularity conditions strong enough that they also imply
consistency of the linear regression estimator.  More recently, 
\citet{steinberger2016} establish asymptotic validity of the
jackknife intervals in a high-dimensional regression
setting; they do not require consistency of the base estimator
\smash{$\hat\mu$} per say, but they do require a uniform asymptotic
mean squared error bound (and an asymptotic stability condition) on
\smash{$\hat\mu$}. 
The conformal method requires no such conditions.
Moreover, the analyses in
\citet{Rothman,steinberger2016} assume a standard linear model setup,
where the regression function is itself a linear function of the
features, the features are independent of the errors, and the errors
are homoskedastic; none of these conditions are needed in order for
the split conformal method (and full conformal method) to have finite
simple validity.  

\begin{algorithm}[tb]
\caption{Jackknife Prediction Band}
\label{alg:jack}
\begin{algorithmic}
\State{\bf Input:} Data $(X_i,Y_i)$, $i=1,\ldots,n$, miscoverage level 
$\alpha \in (0,1)$, regression algorithm $\mathcal A$
\State{\bf Output:} Prediction band, over $x \in \R^d$
\For{$i \in \{1,\ldots,n\}$}
\State \smash{$\hat\mu^{(-i)} = \mathcal A\big(\{ (X_\ell,Y_\ell): 
  \ell \not=i \}\big)$}
\State \smash{$R_i = |Y_i-\hat\mu^{(-i)}(X_i)|$}
\EndFor
\State $d =$ the $k$th smallest value in $\{R_i : i \in
\{1,\ldots,n\}\}$, where $k=\lceil n(1-\alpha) \rceil$ 
\State Return \smash{$C_{\mathrm{jack}}(x) =
  [\hat\mu(x)-d, \hat\mu(x)+ d]$}, for all $x \in \R^d$ 
\end{algorithmic}
\end{algorithm}


\section{Statistical Accuracy}
\label{sec::theory}

Conformal inference offers reliable coverage under no assumptions
other than i.i.d.\ data. 
In this section, we investigate the statistical accuracy of conformal
prediction intervals by bounding the length of the resulting intervals
$C(X)$. Unlike coverage guarantee, such statistical accuracy must be
established under appropriate 
regularity conditions on both the model and the fitting method. 
Our analysis starts from a very mild set of conditions, and moves
toward the standard assumptions typically made in high-dimensional
regression, where we show that conformal methods achieve near-optimal
statistical efficiency. 

We first collect some common assumptions and notation that will be
used throughout this section. Further assumptions will be stated when
they are needed. 

\begin{enumerate}
  \item [] {\bf Assumption A0} (i.i.d.\ data){\bf.} 
    We observe i.i.d.\ data $(X_i,Y_i)$, $i=1,\ldots,n$ 
    from a common distribution $P$ on $\R^d\times \R$,  
    with mean function $\mu(x)=\E(Y \,|\, X=x)$, 
    $x \in \R^d$. 
\end{enumerate}  
Assumption A0 is our most basic assumption used throughout the 
paper.

\begin{enumerate}
  \item [] {\bf Assumption A1} (Independent and symmetric noise){\bf.} 
    For $(X,Y) \sim P$, the noise variable $\epsilon=Y-\mu(X)$ is
    independent of $X$, and the 
    density function of $\epsilon$ is symmetric about $0$ and
    nonincreasing on $[0,\infty)$. 
\end{enumerate}  

  Assumption A1 is weaker than the assumptions usually made
  in the regression literature. In particular, we do not even require
  $\epsilon$ to have a finite first moment.  The symmetry and
  monotonicity conditions are for convenience, and can be
  dropped by considering appropriate quantiles or density level sets
  of $\epsilon$.  The continuity of the distribution of $\epsilon$
  also ensures that with probability 1 the fitted residuals will all
  be distinct, making inversion of empirical distribution function
  easily tractable. 
  We should note that, our other assumptions, such as the stability or
  consistency of the base estimator (given below), may implicitly impose
  some further moment conditions on $\epsilon$; thus when these
  further assumptions are in place, our above assumption on $\epsilon$
  may be comparable to the standard ones. 
  
  %

\paragraph{Two oracle bands.} To quantify the accuracy of the prediction bands
constructed with the full and split conformal methods,
we will compare their lengths to the length of the idealized prediction bands
obtained by two oracles: the ``super oracle" and a regular oracle. The
super oracle has complete knowledge of 
the regression function $\mu(x)$ and the error distribution, while a regular
oracle has knowledge only of the residual distribution, i.e., of the distribution
of \smash{$Y - \hat\mu_n(X)$}, where $(X,Y) \sim P$ is independent of
the given data $(X_i,Y_i)$, $i=1,\ldots,n$ used to compute the
regression estimator \smash{$\hat\mu_n$} (and our notation for the
estimator and related quantities in this section emphasizes the sample
size $n$).

Assumptions A0 and A1 imply that the super oracle prediction band is 
\begin{equation*}
  C_s^*(x) = [\mu(x) - q_{\alpha}, \mu(x) + q_{\alpha}], \;
  \text{where $q_\alpha$ is the $\alpha$ upper quantile of 
    $\mathcal{L}(|\epsilon|)$.}   
\end{equation*}
The band $C_s^*(x)$ is optimal in the following sense: (i) it is has
valid conditional coverage: $\P(Y\in C(x)\,|\, X=x) \geq 1-\alpha$,
(ii) it has the shortest length among all bands with conditional
coverage, and (iii) it has the shortest average length among all bands 
with marginal coverage \citep{LeiW14}.  

With a base fitting algorithm $\mathcal A_n$ and a sample of size $n$,
we can mimic the super oracle by substituting $\mu$ with
\smash{$\hat\mu_n$}. In order to have valid prediction, the band needs
to accommodate randomness of \smash{$\hat\mu_n$} and the new
independent sample $(X,Y)$. Thus it is natural to consider 
the oracle
$$
C_o^*(x) = [\hat\mu_n(x)- q_{n,\alpha}, 
\hat\mu_n(x) + q_{n,\alpha}], \; 
\text{where $q_{n,\alpha}$ is
the $\alpha$ upper quantile of $\mathcal{L}(|Y-\hat\mu_n(X)|)$.} 
$$
We note that the definition of $q_{n,\alpha}$ is unconditional, so the
randomness is regarding the $(n+1)$ pairs
$(X_1,Y_1),\ldots,(X_n,Y_n),(X,Y)$.  The band $C_o^*(x)$ is still
impractical because the distribution of 
\smash{ $|Y-\hat\mu_n(X)|$} is unknown but its quantiles can be
estimated. Unlike the super oracle band, in general the oracle band 
only offers marginal coverage:  $\mathbb{P}(Y\in C_o^*(X))\geq
1-\alpha$, over the randomness of the $(n+1)$ pairs. 

Our main theoretical results in this section can be summarized as
follows. 
\begin{enumerate}
  \item If the base estimator is consistent, then the two oracle bands
    have similar lengths (\Cref{sec:compare_oracle}). 
  \item If the base estimator is stable under resampling and small
    perturbations, then the conformal prediction bands are close to
    the oracle band (\Cref{sec:oracle_approximation}).  
  \item If the base estimator is consistent, then the conformal
    prediction bands are close to the super oracle
    (\Cref{sec:kool-aid}). 
\end{enumerate}
The proofs for these results are deferred to \Cref{sec:proof_3}.  

\subsection{Comparing the Oracles}\label{sec:compare_oracle}

Intuitively, if \smash{$\hat\mu_n$} is close to $\mu$, then the two
oracle bands should be close. Denote by
$$
\Delta_n(x) = \hat\mu_n(x)-\mu(x)
$$
the estimation error.  We now have the following result.

\begin{theorem}[Comparing the oracle bands]
\label{thm:compare_oracle}
Under Assumptions A0, A1, let $F,f$ be the distribution and density 
functions of $|\epsilon|$. 
Assume further that $f$ has continuous derivative that is uniformly
bounded by $M>0$. Let $F_n,f_n$ be the distribution
and density functions of \smash{$|Y-\hat\mu_n(X)|$}. Then we have  
\begin{equation}\label{eq:Linfty}
\sup_{t > 0} |F_n(t) - F(t)|  \leq (M/2) \mathbb E \Delta_n^2(X), 
\end{equation}
where the expectation is taken over the randomness of
\smash{$\hat\mu_n$} and $X$. 

Moreover, if
$f$ is lower bounded by $r>0$ on \smash{$(q_\alpha-
  \eta,q_\alpha+\eta)$} for some 
\smash{$\eta>(M/2r) \mathbb E \Delta_n^2(X)$}, then    
\begin{equation}\label{eq::next}
|q_{n,\alpha}-q_\alpha| \leq (M/2r) \mathbb{E}\Delta_n^2(X).    
\end{equation}
\end{theorem}

In the definition of the oracle bands, the width (i.e., the length, we 
will use these two terms interchangeably) is $2q_\alpha$ for
the super oracle and $2 q_{n,\alpha}$ for the oracle. 
\Cref{thm:compare_oracle} indicates that the oracle bands have similar
width, with a difference proportional to
$\E\Delta_n^2(X)$.  It is
worth mentioning that we do not even require the estimate
\smash{$\hat\mu_n$} to be consistent.  Instead,
\Cref{thm:compare_oracle} applies whenever $\E\Delta_n^2(X)$ is
smaller than some constant, as specified by the triplet $(M,r,\eta)$
in the theorem. Moreover, it is also worth noting that the estimation 
error $\Delta_n(X)$ has only a second-order impact
on the oracle prediction band.  This is due to the assumption that
$\epsilon$ has symmetric density.

\subsection{Oracle Approximation Under Stability Assumptions}
\label{sec:oracle_approximation}

Now we provide sufficient conditions under which the split conformal
and full conformal intervals approximate the regular oracle.

\paragraph{Case I: Split conformal.} For the split conformal analysis,
our added assumption is on sampling stability.


\begin{enumerate}
  \item [] {\bf Assumption A2} (Sampling stability){\bf.}  
    For large enough $n$, 
    $$
    \mathbb P(\|\hat\mu_n-\tilde\mu\|_\infty \geq \eta_n) \leq \rho_n, 
    $$
    for some sequences satisfying $\eta_n=o(1)$, $\rho_n=o(1)$ as
    $n \rightarrow \infty$, and some function $\tilde\mu$. 
\end{enumerate}

We do not need to assume that \smash{$\tilde\mu$} is close to the true
regression function $\mu$.  We only need the estimator
\smash{$\hat\mu_n$} to concentrate around \smash{$\tilde\mu$}.  
This is just a stability assumption rather than consistency
assumption. For example, this is satisfied in nonparametric regression
under over-smoothing.  When \smash{$\tilde\mu=\mu$}, this becomes a 
sup-norm consistency assumption, and is satisfied, for example, by
lasso-type estimators under standard assumptions, fixed-dimension
ordinary least squares with bounded predictors, and
standard nonparametric regression estimators on a compact domain. 
Usually $\eta_n$ has the form of $c(\log n/n)^{-r}$, and $\rho_n$ is
of order $n^{-c}$, for some fixed $c>0$ (the choice of the constant
$c$ is arbitrary and will only impact the constant term in front of
$\eta_n$).

When the sampling stability fails to hold, conditioning on 
\smash{$\hat\mu_n$}, the residual \smash{$Y-\hat\mu_n(X)$} may have a   
substantially different distribution than $F_n$, and the split
conformal interval can be substantially different from the oracle
interval.    

\begin{remark}
\label{rem:sup-norm}
  The sup-norm bound required in Assumption A2 can be weakened to an
  \smash{$\ell_{p,X}$} norm bound with $p>0$ where
  \smash{$\ell_{p,X}(g)=\left(\mathbb E_X |g(X)|^p \right)^{1/p}$} for any
  function $g$.  The idea is that when \smash{$\ell_{p,X}$} norm bound
  holds, by Markov's inequality the $\ell_\infty$ norm bound holds
  (with another vanishing sequence $\eta_n$) except on a small set
  whose probability is vanishing.  Such a small set will have
  negligible impact on the quantiles.  An example of this argument is
  given in the proof of \Cref{thm:split_super}.  
\end{remark}


\begin{theorem}[Split conformal approximation of oracle]
  \label{lemma::split} 
Fix $\alpha \in (0,1)$, and let \smash{$C_{n,{\rm
split}}$} and \smash{$\nu_{n,{\rm split}}$} denote the split conformal 
interval and its width. Under Assumptions A0, A1, A2, 
assume further that \smash{$\tilde{f}$}, the density of
\smash{$|Y-\tilde\mu(X)|$}, is lower bounded away from zero in an open   
neighborhood of its $\alpha$ upper quantile. Then 
$$
\nu_{n,{\rm split}} - 2q_{n,\alpha} = O_\P(\rho_n+\eta_n+n^{-1/2}).
$$
\end{theorem}

\paragraph{Case II: Full conformal.}
Like the split conformal analysis, our analysis for the
full conformal band to approximate the oracle also will require 
sampling stability as in Assumption A2.  But it will also require a
perturb-one sensitivity condition. 

Recall that for any candidate value $y$, we will fit the regression
function with augmented data, where the $(n+1)$st data point is
$(X,y)$.  We denote this fitted regression function by
\smash{$\hat\mu_{n,(X,y)}$}.  Due to the arbitrariness of $y$, we must
limit the range of $y$ under consideration.  Here we restrict our
attention to $y \in \mathcal Y$; we can think of a typical case for  
$\mathcal Y$ as a compact interval of fixed length. 

\begin{enumerate}
  \item [] \textbf{Assumption A3} (Perturb-one sensitivity){\bf.}
 For large enough $n$, 
 \begin{equation*}
   \mathbb P\left(\sup_{y\in\mathcal
       Y}\|\hat\mu_n-\hat\mu_{n,(X,y)}\|_\infty\ge\eta_n\right)\le
   \rho_n,
 \end{equation*}
 for some sequences satisfying $\eta_n=o(1)$, $\rho_n=o(1)$ as
 $n \rightarrow \infty$.
  \end{enumerate}

  The perturb-one sensitivity condition requires that the fitted
  regression function does not change much if we only perturb the $y$
  value of the last data entry.  It is satisfied, for example, by
  kernel and local polynomial regression with a large enough 
  bandwidth, least squares regression with a well-conditioned design,  
  ridge regression, and even the lasso under certain conditions
  \citep{ThakurtaS13}. 
  
  For a similar reason as in \Cref{rem:sup-norm}, we can weaken the
  $\ell_\infty$ norm requirement to an \smash{$\ell_{p,X}$} norm bound
  for any $p>0$.   

\newcommand{\muhat}{\hat{\mu}}
\newcommand{\muXY}{\tilde{\mu}_{X,Y}}
\newcommand{\muXy}{\tilde{\mu}_{X,y}}
\newcommand{\muni}{\tilde{\mu}_{X,Y}^{(-i)}}
\newcommand{\musi}{\hat{\mu}^{(*i)}}


\begin{theorem}[Full conformal approximation of oracle]\label{thm:full_practical}
Under the same assumptions as in \Cref{lemma::split}, assume in addition that
$Y$ is supported on $\mathcal Y$ such that
Assumption A3 holds. Fix $\alpha \in (0,1)$, and let \smash{$C_{n,{\rm  
    conf}}(X)$} and \smash{$\nu_{n,{\rm  conf}}(X)$} be the conformal
prediction interval and its width at $X$. Then
$$
\nu_{n,{\rm conf}}(X) - 2 q_{n,\alpha} = O_\P(\eta_n + \rho_n +
n^{-1/2}). 
$$
\end{theorem}

\subsection{Super Oracle Approximation Under Consistency Assumptions} 
\label{sec:kool-aid} 

Combining the results in
\Cref{sec:compare_oracle,sec:oracle_approximation}, we immediately get 
\smash{$\nu_{n,{\rm split}}-2q_\alpha=o_\P(1)$} and 
\smash{$\nu_{n,{\rm conf}}-2q_\alpha=o_\P(1)$} when \smash{$\mathbb E
  \Delta_n^2(X)=o(1)$}.  In fact, when 
the estimator \smash{$\hat\mu_n$} is consistent, we can further
establish conditional coverage results for conformal prediction bands.
That is, they have not only near-optimal length, but also near-optimal
location.  

The only additional assumption we need here is consistency of
\smash{$\hat\mu_n$}.  A natural condition would be \smash{$\mathbb E
  \Delta_n^2(X)=o(1)$}.  Our analysis  uses an even weaker
assumption. 

\begin{enumerate}
  \item [] \textbf{Assumption A4} (Consistency of base
    estimator){\bf.} 
    For $n$ large enough, 
$$
\mathbb P\left(\mathbb E_X\Big[(\hat\mu_n(X)-\mu(X))^2 \,|\,
    \hat\mu_n\Big]\ge \eta_n\right)\leq \rho_n,
$$
for some sequences satisfying $\eta_n=o(1)$, $\rho_n=o(1)$ as $n \to
\infty$. 
\end{enumerate}
It is easy to verify that Assumption A4 is implied by the condition
$\mathbb E\Delta^2_n(X)=o(1)$, using Markov's inequality.  Many
consistent estimators in the literature have this property, such as
the lasso under a sparse eigenvalue condition for the design (along 
with appropriate tail bounds on the distribution of $X$), and 
kernel and local polynomial regression on a compact domain.  

We will show that conformal bands are close to the super oracle, and
hence have approximately correct asymptotic conditional coverage,
which we formally define as follows. 

\begin{definition*}[Asymptotic conditional coverage]
  We say that a sequence $C_n$ of (possibly) random prediction bands  
  has \emph{asymptotic conditional coverage} at the level
  $(1-\alpha)$ if there exists a sequence of (possibly) random sets
  $\Lambda_n\subseteq \mathbb R^d$ such that 
  $\mathbb P(X\in \Lambda_n \,|\, \Lambda_n)=1-o_\P(1)$ and 
  $$
  \inf_{x\in \Lambda_n}\Big|\mathbb P(Y\in C_n(x)\,|\,
  X=x)-(1-\alpha)\Big|=o_\P(1). 
  $$
\end{definition*}

Now we state our result for split conformal.

\begin{theorem}[Split conformal approximation of super oracle]
\label{thm:split_super} 
  Under Assumptions A0, A1, A4, assuming in addition that $|Y-\mu(X)|$
  has density bounded away from zero in an open neighborhood of its
  $\alpha$ upper quantile, the split conformal interval satisfies
  \begin{equation*}
L(C_{n,{\rm split}}(X) \,\triangle\, C^*_{s}(X)) = o_\P(1),
  \end{equation*}
  where $L(A)$ denotes the Lebesgue measure of a set $A$, and $A
  \,\triangle\, B$ the symmetric difference between sets $A,B$.  
  Thus, \smash{$C_{n,{\rm split}}$} has asymptotic conditional
  coverage at the level $1-\alpha$.  
\end{theorem}

\begin{remark}\label{rem:non-consistent-super}
  The proof of \Cref{thm:split_super} can be modified to account fro
  the case when $\eta_n$ does not vanish; in this case we do not have
  consistency but the error will contain a term involving $\eta_n$.  
\end{remark}

The super oracle approximation for the full conformal prediction band
is similar, provided that the perturb-one sensitivity condition holds.

\begin{theorem}[Full conformal approximation of super oracle]
\label{thm::consistent}
Under the same conditions as in \Cref{thm:split_super}, and in 
addition Assumption A3, we have
$$
L(C_{n,{\rm conf}}(X)\,\triangle\, C_s^*) = o_\P(1),
$$
and thus \smash{$C_{n,{\rm conf}}$} has asymptotic conditional
coverage at the level $1-\alpha$. 
\end{theorem}

\subsection{A High-dimensional Sparse Regression Example} 
We consider a high-dimensional linear regression setting, to
illustrate the general theory on the width of conformal prediction
bands under stability and consistency of the base estimator.  The
focus will be finding conditions that imply (an
appropriate subset of) Assumptions 
A1 through A4.  The width of the conformal prediction band for 
low-dimensional nonparametric regression has already been studied in 
\citet{LeiW14}.    

We assume that the data are i.i.d.\ replicates from the model $Y=X^T 
\beta+\epsilon$, with $\epsilon$ being independent of $X$ with mean
$0$ and variance $\sigma^2$.  For convenience we will assume that  
the supports of $X$ and $\epsilon$ are $[-1,1]^p$ and $[-R,R]$,
respectively, for a constant $R>0$. Such a boundedness
condition is used for simplicity, and is only required for the
strong versions of the sampling stability condition (Assumption A2)
and perturb-one sensitivity condition (Assumption A4), which are
stated under the sup-norm. Boundedness can be relaxed by using
appropriate tail conditions on $X$ and $\epsilon$, together with the
weakened $\ell_p$ norm versions of Assumptions A2 and A4.  

Here $\beta\in \mathbb R^d$ is assumed to be a sparse vector with
$s\ll \min\{n,d\}$ nonzero entries.  We are mainly interested in the
high-dimensional setting where both $n$ and $d$ are large, but $\log
d/n$ is small. When we say ``with high probability'', we mean with
probability tending to 1 as $\min\{n,d\} \to \infty$ and $\log
d/n \rightarrow 0$.   

Let $\Sigma$ be the covariance matrix of $X$.  For
$J,J'\subseteq\{1,\ldots,d\}$, let $\Sigma_{JJ'}$ denote the submatrix
of $\Sigma$ with corresponding rows in $J$ and columns in $J'$, and 
$\beta_J$ denotes the subvector of $\theta$ corresponding to
components in $J$. 

The base estimator we consider here is the lasso \citep{lasso},
defined by 
\begin{align*}
\hat\beta_{n,{\rm lasso}} = \argmin_{\beta \in \R^d} \; 
\frac{1}{2n} \sum_{i=1}^n(Y_i-X_i^T\beta)^2 + \lambda \|\beta\|_1,  
\end{align*}
where $\lambda \geq 0$ is a tuning parameter. 

\paragraph{Case I: Split conformal.} For the split conformal method,
sup-norm prediction consistency has been widely studied in the
literature.  Here we follow the arguments in \citet{BickelRT09} (see
also \citet{BuneaTW07}) and make the following assumptions: 
\begin{itemize}
  \item The support $J$ of $\beta$ has cardinality $s<\min\{n,d\}$,
    and 
  \item The covariance matrix $\Sigma$ of $X$ satisfies the restricted
    eigenvalue condition, for $\kappa>0$:
    $$
    \min_{v\,:\,\|v_J\|_2=1, \, \|v_{J^c}\|_1\leq 3\|v_J\|_1} v^T
    \Sigma v \geq \kappa^2.
    $$
  \end{itemize}
  Applying Theorem 7.2 of \citet{BickelRT09}\footnote{The exact
    conditions there are slightly different. For example, the noise is
    assumed to be Gaussian and the columns of the design matrix are
    normalized. But the proof essentially goes through in our present
    setting, under small modifications.}, for any constant $c>0$, if  
  \smash{$\lambda=C\sigma\sqrt{\log d/n}$} for some constant $C>0$ large
  enough depending on $c$, we have, with probability at least
  $1-d^{-c}$ and another constant $C'>0$,
  $$\|\hat \beta_{n,{\rm lasso}}-\beta\|_1\le C' \kappa^2 R
  s\sqrt{\log d/n}.$$ 
As a consequence, Assumptions A2 and A4 hold with  
\smash{$\tilde\mu(x)=x^T\beta$}, 
\smash{$\eta_n=C' \kappa^2 R s\sqrt{\log d/n}$}, and 
\smash{$\rho_n=d^{-c}$}. 

\paragraph{Case II: Full conformal.} For the full conformal method,
we also need to establish the much stronger perturb-one sensitivity
bound (Assumption A3). 
Let $\hat\beta_{n,{\rm lasso}}(X,y)$ denote the lasso solution
obtained using the augmented data $(X_1,Y_1), \ldots, (X_n,Y_n),
(X,y)$. To this end, we invoke the model selection stability result in
\citet{ThakurtaS13}, and specifically, we assume the following (which
we note is enough to ensure support recovery by the lasso estimator):  
\begin{itemize}
  \item There is a constant $\Phi\in(0,1/2)$ such that the
    absolute values of all nonzero entries of $\beta$ are in 
    $[\Phi,1-\Phi]$. (The lower bound is necessary for support 
    recovery and the upper bound can be relaxed to any constant  
    by scaling.)
  \item There is a constant $\delta\in (0,1/4)$ such that
    \smash{$\|\Sigma_{J^c J}\Sigma_{JJ}^{-1}{\rm
        sign}(\beta_J)\|_{\infty}\le 1/4 - \delta$}, where we denote
    by ${\rm sign}(\beta_J)$ the vector of signs of  
    each coordinate of $\beta_J$. (This is the strong
    irrepresentability condition, again necessary for support
    recovery.) 
  \item The active block of the covariance matrix $\Sigma_{JJ}$ has
    minimum eigenvalue $\Psi>0$. 
\end{itemize}
To further facilitate the presentation, we assume 
$s,\sigma,R,\Psi,\Phi$ are constants not changing with $n,d$. 

Under our boundedness assumptions on $X$ and $\epsilon$, note we can  
choose $\mathcal Y=[-s-R,s+R]$. 
Using a standard union bound argument, we can verify that, with high 
probability, the data $(X_i,Y_i)$, $i=1,\ldots,n$ satisfy the
conditions required in Theorem 8 of \cite{ThakurtaS13}.
Thus for $n,d$ large enough, with high probability, the supports of
\smash{$\hat\beta_{n,{\rm lasso}}$} and \smash{$\hat\beta_{n,{\rm
      lasso}}(X,y)$} both equal 
$J$.   Denote by \smash{$\hat\beta_{J,\rm LS}$} the (oracle) least  
squares estimator on the subset $J$ of predictor variables, and by   
\smash{$\hat\beta_{J,\rm ols}(X,y)$} this least squares estimator but
using the augmented data.  
Standard arguments show that 
\smash{$\|\hat\beta_{J,\rm ols}-\beta_J\|_\infty=o_\P(\sqrt{\log
    d/n})$}, and 
\smash{$\|\hat\beta_{J,\rm ols}-\hat\beta_{J,\rm
    ols}(X,y)\|_\infty=O_\P(s/n)$}.  Then both 
\smash{$\hat\beta_{J,\rm ols}$}
and \smash{$\hat\beta_{J,\rm ols}(X,y)$} are close to $\beta_J$, with 
$\ell_\infty$ distance \smash{$o_\P(\sqrt{\log d/n})$}.  Combining  
this with the lower bound condition on the magnitude of the entries of
$\beta_J$, and the KKT conditions for the lasso problem, we have 
\smash{$\|\hat\beta_{J,\rm lasso}(X,y)-\hat\beta_{J,\rm ols}(X,y)\|_\infty\le  
O_\P(n^{1/2})+O_\P(\lambda)=O_\P(\sqrt{\log d/n})$}.  Therefore,
Assumptions A3 and A4 hold for any $\eta_n$ such that
\smash{$\eta_n\sqrt{n/\log d}\rightarrow \infty$}, and 
$\rho_n=o(1)$. 
  
\section{Empirical Study}
\label{sec::experiments}

Now we examine empirical properties of the conformal
prediction intervals under three simulated data settings. 
Our empirical findings can be summarized as follows.
\begin{enumerate}
  \item Conformal prediction bands have nearly exact (marginal)
    coverage, even when the model is completely misspecified.
  \item In high-dimensional problems, conformal inference 
    often yields much smaller bands than conventional methods. 
  \item The accuracy (length) of the conformal prediction band is
    closely related to the quality of initial estimator, which in turn
    depends on the model and the tuning parameter. 
\end{enumerate}

In each
setting, the samples $(X_i,Y_i)$, $i=1,\ldots,n$ are generated in an
i.i.d. fashion, by
first specifying $\mu(x)=\E(Y_i \,|\, X_i=x)$, then specifying a  
distribution for $X_i=(X_i(1),\ldots,X_i(d))$, and lastly specifying a 
distribution for $\epsilon_i=Y_i-\mu(X_i)$ (from which we can form 
$Y_i=\mu(X_i)+\epsilon_i$).  These specifications are described
below.  We write $N(\mu,\sigma^2)$ for the normal
distribution with mean $\mu$ and variance $\sigma^2$,
$SN(\mu,\sigma^2,\alpha)$ for the skewed normal 
with skewness parameter $\alpha$, $t(k)$ for the
$t$-distribution with $k$ degrees of freedom, and 
$\mathrm{Bern}(p)$ for the Bernoulli distribution with success
probability $p$. 

Throughout, we will consider the following three experimental setups. 

\begin{description}[labelindent=0em, leftmargin=1em]
  \item[Setting A] {\em (linear, classical)}: the mean $\mu(x)$ is
    linear in $x$;  
    the features $X_i(1),\ldots,X_i(d)$ are i.i.d.\ $N(0,1)$; and the 
    error $\epsilon_i$ is $N(0,1)$, independent of the features. 
  \item[Setting B] {\em (nonlinear, heavy-tailed)}: like Setting A,
    but where $\mu(x)$ is nonlinear in $x$, an additive function of
    B-splines of $x(1),\ldots,x(d)$; and the error $\epsilon_i$ is
    $t(2)$ (thus, without a finite variance), independent of the
    features.    
  \item[Setting C] {\em (linear, heteroskedastic, heavy-tailed,
      correlated features)}: the mean $\mu(x)$ is linear in $x$; the
    features $X_i(1),\ldots,X_i(d)$ are first independently drawn from 
    a mixture distribution, with equal probability on the components
    $N(0,1)$, $SN(0,1,5)$, $\mathrm{Bern}(0.5)$, and then given
    autocorrelation by redefining in a sequential fashion each
    $X_i(j)$ to be a convex combination of its current value and 
    $X_i(j-1),\ldots,X_i((j-3)\wedge 1)$, for $j=1,\ldots,d$; 
    the error $\epsilon_i$ is $t(2)$, with standard deviation
    $1+2|\mu(X_i)|^3/\E(|\mu(X)|^3)$
     (hence, clearly not independent 
    of the features).
\end{description}

Setting A is a simple setup where classical methods are
expected to perform well. Setting B explores the performance when the
mean is nonlinear and the errors are heavy-tailed.  Setting C provides
a particularly difficult linear setting for estimation, with 
heavy-tailed, heteroskedastic errors and highly correlated features.
All simulation results in the following subsections are averages over
50 repetitions.  Additionally, all intervals are computed at the 90\%
nominal coverage level.  The results that follow can be directly
reproduced using the code provided at 
\url{https://github.com/ryantibs/conformal}.   

\subsection{Comparisons to Parametric Intervals from Linear
  Regression} 
\label{sec:parametric_sims}

Here we compare the conformal prediction intervals based on the
ordinary linear regression estimator to the classical parametric
prediction intervals for linear models. 
The classical intervals are valid when the true mean is linear and the  
errors are both normal and homoskedastic, or are asymptotically valid
if the errors have finite variance.  
Recall that  the full and
split conformal intervals are valid under
essentially no assumptions, whereas the jackknife method requires
at least a uniform mean squared error bound on the linear regression
estimator in order to achieve asymptotic validity  
\citep{Rothman,steinberger2016}. We empirically compare the
classical and conformal intervals across Settings A-C, in both 
low-dimensional ($n=100$, $d=10$) and high-dimensional ($n=500$,
$d=490$) problems.

In Settings A and C (where the mean is linear), the mean function was
defined by choosing $s=10$ regression coefficients to be nonzero,
assigning them values $\pm 1$ with equal probability, and mulitplying
them against the standardized predictors. In Setting B (where the mean
is nonlinear), it is defined by multiplying these coefficients  
against B-splines transforms of the standardized predictors. Note that
$d<n$ in the present high-dimensional case, so that the linear
regression estimator and the corresponding intervals are well-defined.     

In the low-dimensional problem, with a linear mean function and
normal, homoskedastic errors (Setting A, \Cref{tab:lm.lo}), all
four methods give reasonable coverage.  The parametric intervals are
shorter than the conformal intervals, as the parametric assumptions
are satisfied and $d$ is small enough for the model to be
estimated well.  The full conformal interval is shorter than the split
conformal interval, but comes at a higher computational cost.   

In the other two low-dimensional problems (Settings B and C,
\Cref{tab:lm.lo}), the assumptions supporting the classical
prediction 
intervals break down. This drives the parametric intervals 
to over-cover, thus yielding much wider intervals than those from 
the conformal methods. Somewhat surprisingly (as the linear regression
estimator in Settings B and C is far from accurate), the jackknife
intervals maintain reasonable coverage at a reasonable length.  The
full conformal intervals continue to be 
somewhat shorter than the split conformal intervals, again at a
computational cost.  Note that the conformal intervals are also using
a linear regression estimate here, yet their coverage is still right
around the nominal 90\% level; the coverage provided by the conformal
approach is robust to the model misspecification.

In the high-dimensional problems (\Cref{tab:lm.hi}), the full
conformal interval outperforms the parametric interval in terms of
both length and 
coverage across all settings, even in Setting A where the true model
is linear.  This is due to poor accuracy of linear regression
estimation when $d$ is large.  The jackknife interval also struggles,
again because the linear regression estimate itself is so poor. 
The split conformal method must be omitted here, since linear
regression is not well-defined once the sample is split ($n/2=250$,   
$d=490$).

Because of the problems that high dimensions pose for linear
regression, we also explore the use of ridge regression 
(\Cref{tab.ridge.lm.hi}).  The parametric intervals here are
derived in a similar fashion to those for ordinary linear regression 
\citep{burnaev2014efficiency}.
For all methods we used ridge regression tuning parameter $\lambda=10$,
which gives nearly optimal prediction bands in the ideal setting (Setting A).
For the split
conformal method, such a choice of $\lambda$ gives similar results
to the cross-validated choice of $\lambda$.  The results show that
the ridge penalty improves the performance of all methods, but that
the conformal methods continue to outperform the parametric one.
Moreover, the split conformal method exhibits a clear
computational advantage compared to the full conformal method, with
similar performance. With such a dramatically reduced computation
cost, we can easily combine split conformal with computationally heavy
estimators that involve cross-validation or bootstrap. The (rough)
link between prediction error and interval length will be further
examined in the next subsection. 

\begin{table}[htb]
\centering
Setting A \\ \smallskip
\begin{tabular}{|l|l|l|l|l|}
\hline
& Conformal & Jackknife & Split & Parametric \\
\hline
Coverage & 0.904 (0.005) & 0.892 (0.005) & 0.905 (0.008) & 0.9 (0.006) \\
\hline
Length & 3.529 (0.044) & 3.399 (0.04) & 3.836 (0.082) & 3.477 (0.036) \\
\hline
Time & 1.106 (0.004) & 0.001 (0) & 0 (0) & 0.001 (0) \\
\hline
\end{tabular}

\bigskip
Setting B \\ \smallskip
\begin{tabular}{|l|l|l|l|l|}
\hline
& Conformal & Jackknife & Split & Parametric \\
\hline
Coverage & 0.915 (0.005) & 0.901 (0.006) & 0.898 (0.006) & 0.933 (0.007) \\
\hline
Length & 6.594 (0.254) & 6.266 (0.254) & 7.384 (0.532) & 8.714 (0.768) \\
\hline
Time & 1.097 (0.002) & 0.001 (0) & 0.001 (0) & 0.001 (0) \\
\hline
\end{tabular}

\bigskip
Setting C \\ \smallskip
\begin{tabular}{|l|l|l|l|l|}
\hline
& Conformal & Jackknife & Split & Parametric \\
\hline
Coverage & 0.904 (0.004) & 0.892 (0.005) & 0.896 (0.008) & 0.943 (0.005) \\
\hline
Length & 20.606 (1.161) & 19.231 (1.082) & 24.882 (2.224) & 33.9 (4.326) \\
\hline
Time & 1.105 (0.002) & 0.001 (0) & 0.001 (0) & 0 (0) \\
\hline
\end{tabular}

\caption{\it Comparison of prediction intervals in low-dimensional
  problems with $n=100$, $d=10$. All quantities have been
  averaged over 50 repetitions, and the standard errors are in
  parantheses. The same is true in Tables \ref{tab:lm.hi} and
  \ref{tab.ridge.lm.hi}.}  
\label{tab:lm.lo}
\end{table}

\begin{table}[htbp]
\centering
Setting A \\ \smallskip
\begin{tabular}{|l|l|l|l|}
\hline
& Conformal & Jackknife & Parametric \\
\hline
Coverage & 0.903 (0.013) & 0.883 (0.018) & 0.867 (0.018) \\
\hline
Length & 8.053 (0.144) & 26.144 (0.95) & 24.223 (0.874) \\
\hline
Time & 167.189 (0.316) & 1.091 (0) & 0.416 (0) \\
\hline
\end{tabular}

\bigskip
Setting B \\ \smallskip
\begin{tabular}{|l|l|l|l|}
\hline
& Conformal & Jackknife & Parametric \\
\hline
Coverage & 0.882 (0.015) & 0.881 (0.016) & 0.858 (0.019) \\
\hline
Length & 53.544 (12.65) & 75.983 (15.926) & 69.309 (14.757) \\
\hline
Time & 167.52 (0.019) & 1.092 (0.001) & 0.415 (0) \\
\hline
\end{tabular}

\bigskip
Setting C \\ \smallskip
\begin{tabular}{|l|l|l|l|}
\hline
& Conformal & Jackknife & Parametric \\
\hline
Coverage & 0.896 (0.013) & 0.869 (0.017) & 0.852 (0.019) \\
\hline
Length & 227.519 (12.588) & 277.658 (16.508) & 259.352 (15.391) \\
\hline
Time & 168.531 (0.03) & 1.092 (0.002) & 0.415 (0) \\
\hline
\end{tabular}

\caption{\it Comparison of prediction intervals in high-dimensional 
  problems with $n=500$, $d=490$.} 
\label{tab:lm.hi}

\bigskip
\bigskip
Setting A \\ \smallskip
\begin{tabular}{|l|l|l|l|l|}
\hline
& Conformal & Jackknife & Split & Parametric \\
\hline
Coverage & 0.903 (0.004) & 0.9 (0.005) & 0.907 (0.005) & 1 (0) \\
\hline
Length & 3.348 (0.019) & 3.325 (0.019) & 3.38 (0.031) & 23.837 (0.107) \\
\hline
Test error & 1.009 (0.018) & 1.009 (0.018) & 1.009 (0.021) & 1.009 (0.018) \\
\hline
Time & 167.189 (0.316) & 1.091 (0) & 0.155 (0.001) & 0.416 (0) \\
\hline
\end{tabular}

\bigskip
Setting B \\ \smallskip
\begin{tabular}{|l|l|l|l|l|}
\hline
& Conformal & Jackknife & Split & Parametric \\
\hline
Coverage & 0.905 (0.006) & 0.903 (0.004) & 0.895 (0.006) & 0.999 (0) \\
\hline
Length & 5.952 (0.12) & 5.779 (0.094) & 5.893 (0.114) & 69.335 (12.224) \\
\hline
Test error & 6.352 (0.783) & 6.352 (0.783) & 11.124 (3.872) & 6.352 (0.783) \\
\hline
Time & 167.52 (0.019) & 1.092 (0.001) & 0.153 (0) & 0.415 (0) \\
\hline
\end{tabular}

\bigskip
Setting C \\ \smallskip
\begin{tabular}{|l|l|l|l|l|}
\hline
& Conformal & Jackknife & Split & Parametric \\
\hline
Coverage & 0.906 (0.004) & 0.9 (0.004) & 0.902 (0.005) & 0.998 (0.001) \\
\hline
Length & 15.549 (0.193) & 14.742 (0.199) & 15.026 (0.323) & 249.932 (9.806) \\
\hline
Test error & 158.3 (48.889) & 158.3 (48.889) & 114.054 (19.984) & 158.3 (48.889) \\
\hline
Time & 168.531 (0.03) & 1.092 (0.002) & 0.154 (0) & 0.415 (0) \\
\hline
\end{tabular}

\caption{\it Comparison of prediction intervals in high-dimensional 
  problems with $n=500$, $d=490$, using ridge regularization.}
\label{tab.ridge.lm.hi}
\end{table}

\subsection{Comparisons of Conformal Intervals Across Base  
  Estimators}
\label{sec:different_base_sims}

We explore the behavior of the conformal intervals across
a variety of base estimators.
We simulate data from Settings A-C, in both low ($n=200$, $d=20$) and
high ($n=200$, $d=2000$) dimensions, and in each case we
apply forward stepwise regression \citep{stepwise}, the lasso
\citep{lasso}, the elastic net \citep{enet}, sparse additive models
\citep[SPAM,][]{Ravikumar09}, 
and random forests \citep{Breiman01}.

\begin{figure}[p]
\centering
\includegraphics[width=0.32\textwidth]{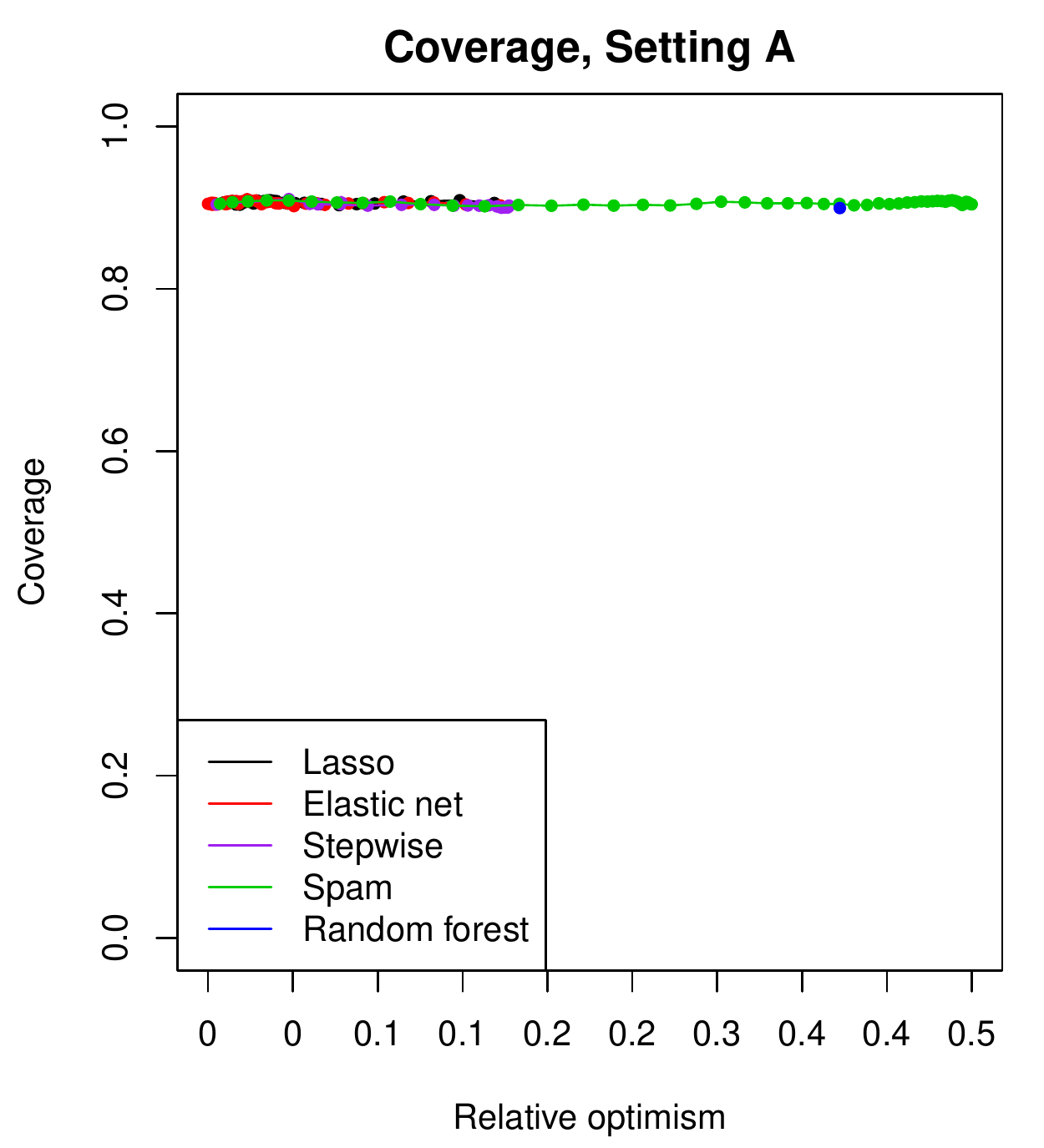} 
\includegraphics[width=0.32\textwidth]{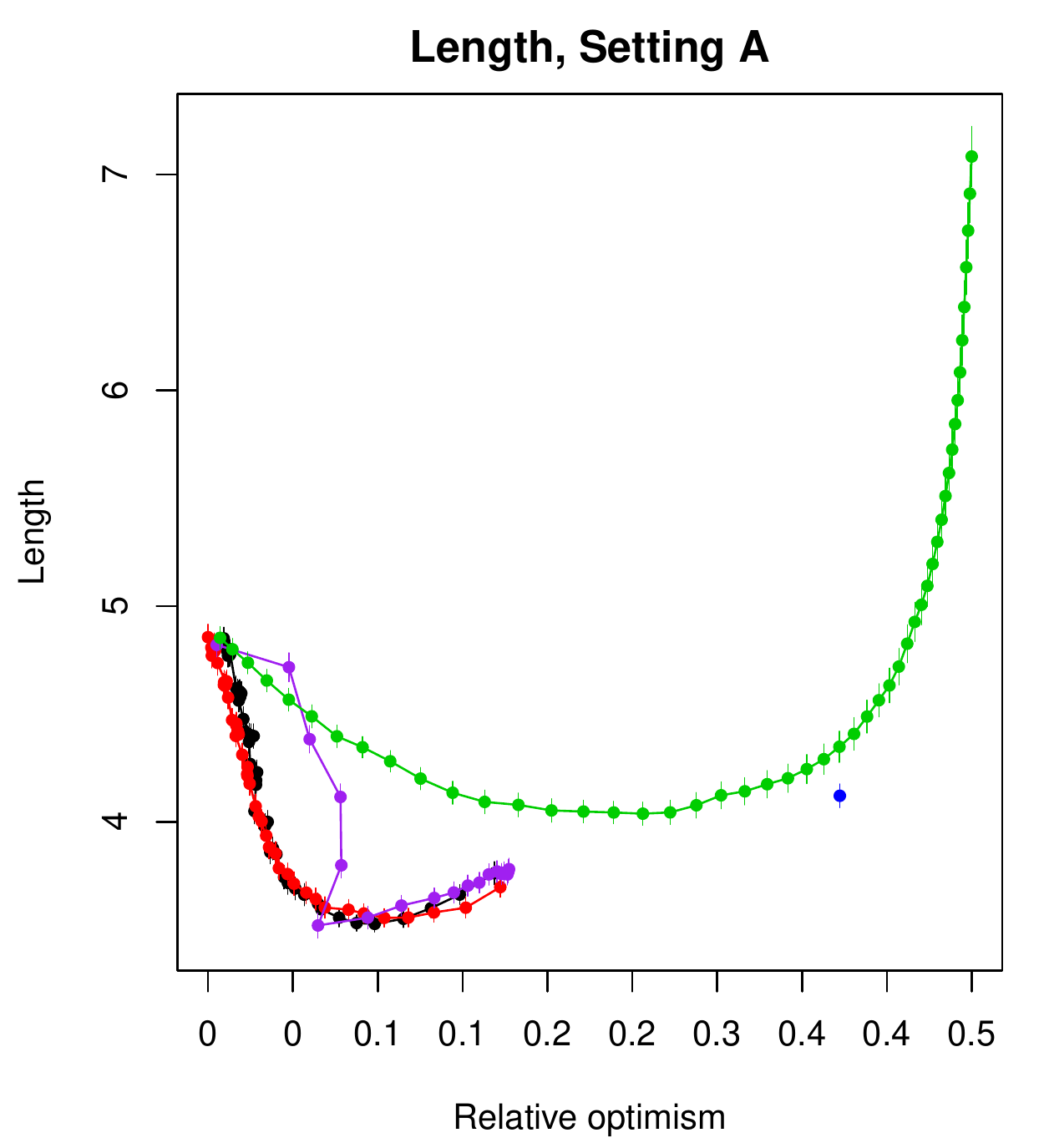} 
\includegraphics[width=0.32\textwidth]{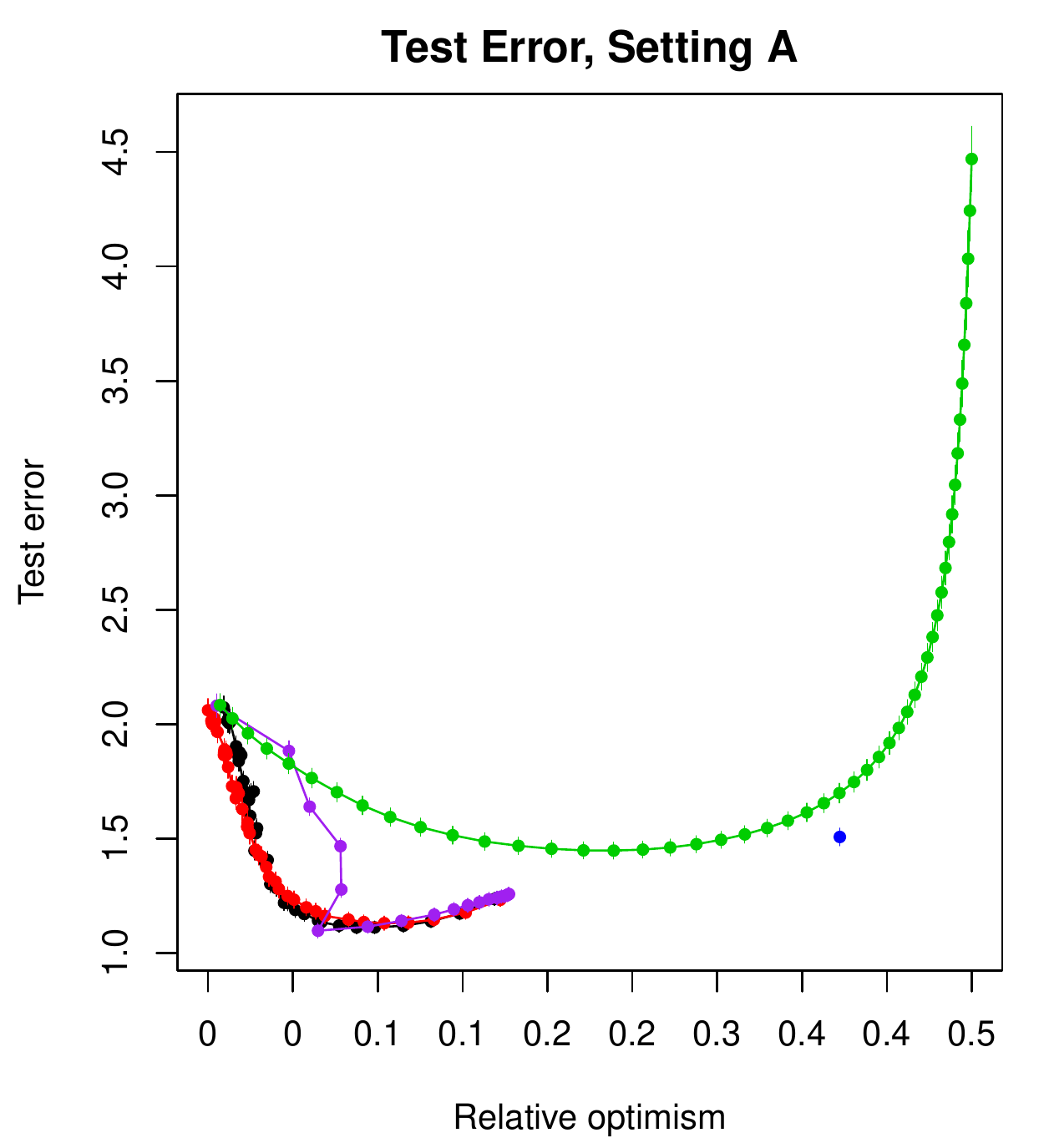} \\
\includegraphics[width=0.32\textwidth]{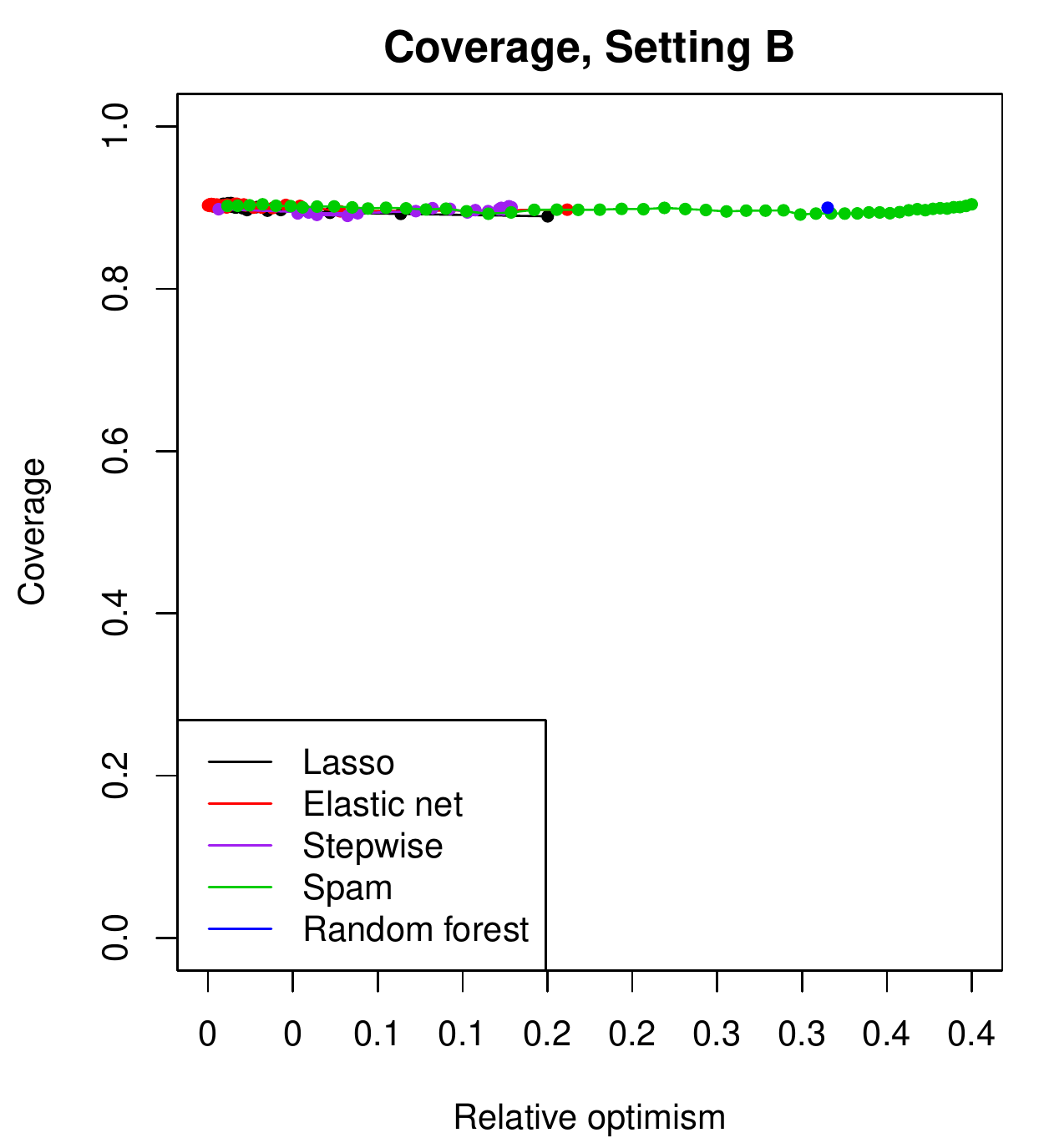} 
\includegraphics[width=0.32\textwidth]{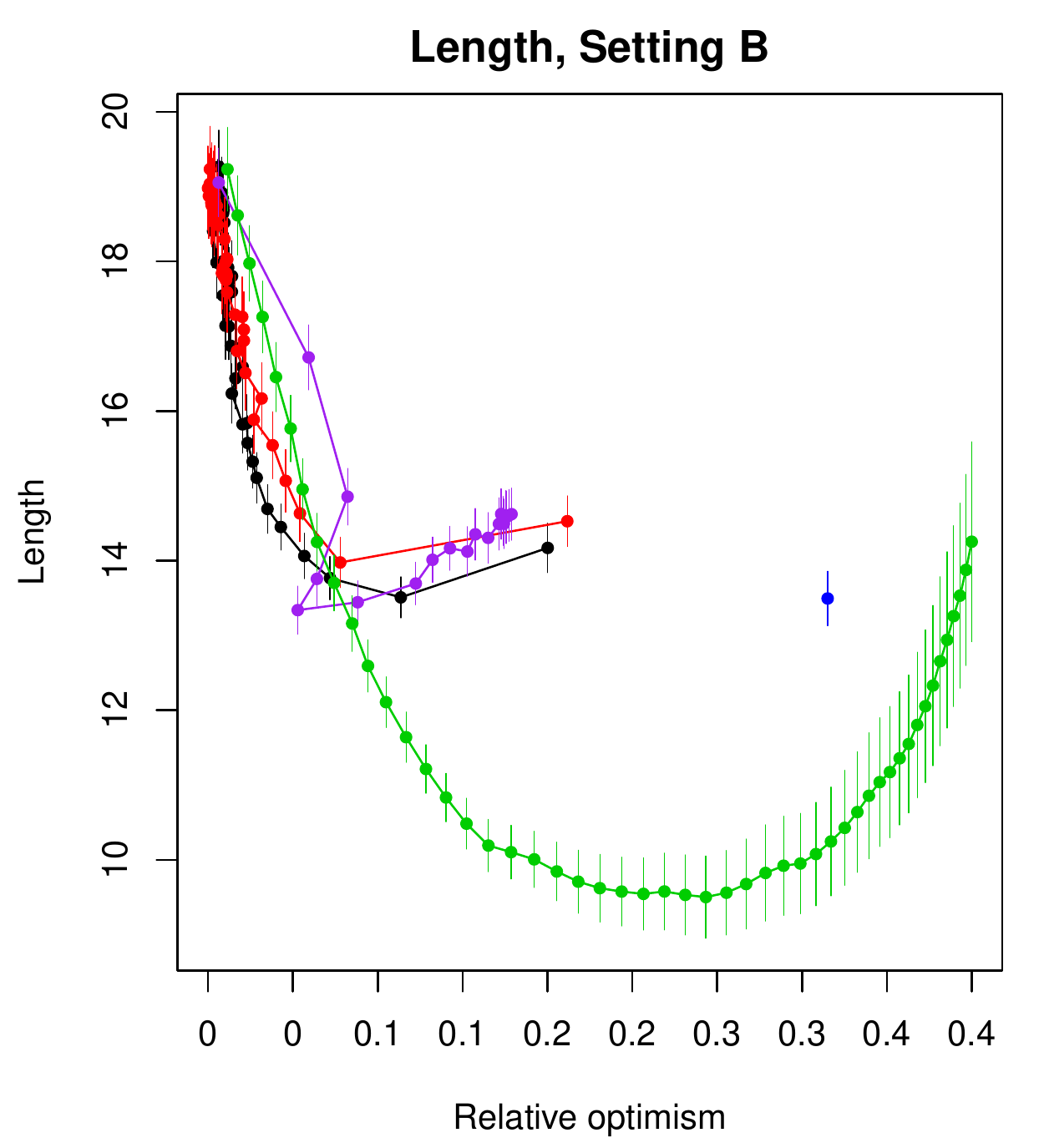} 
\includegraphics[width=0.32\textwidth]{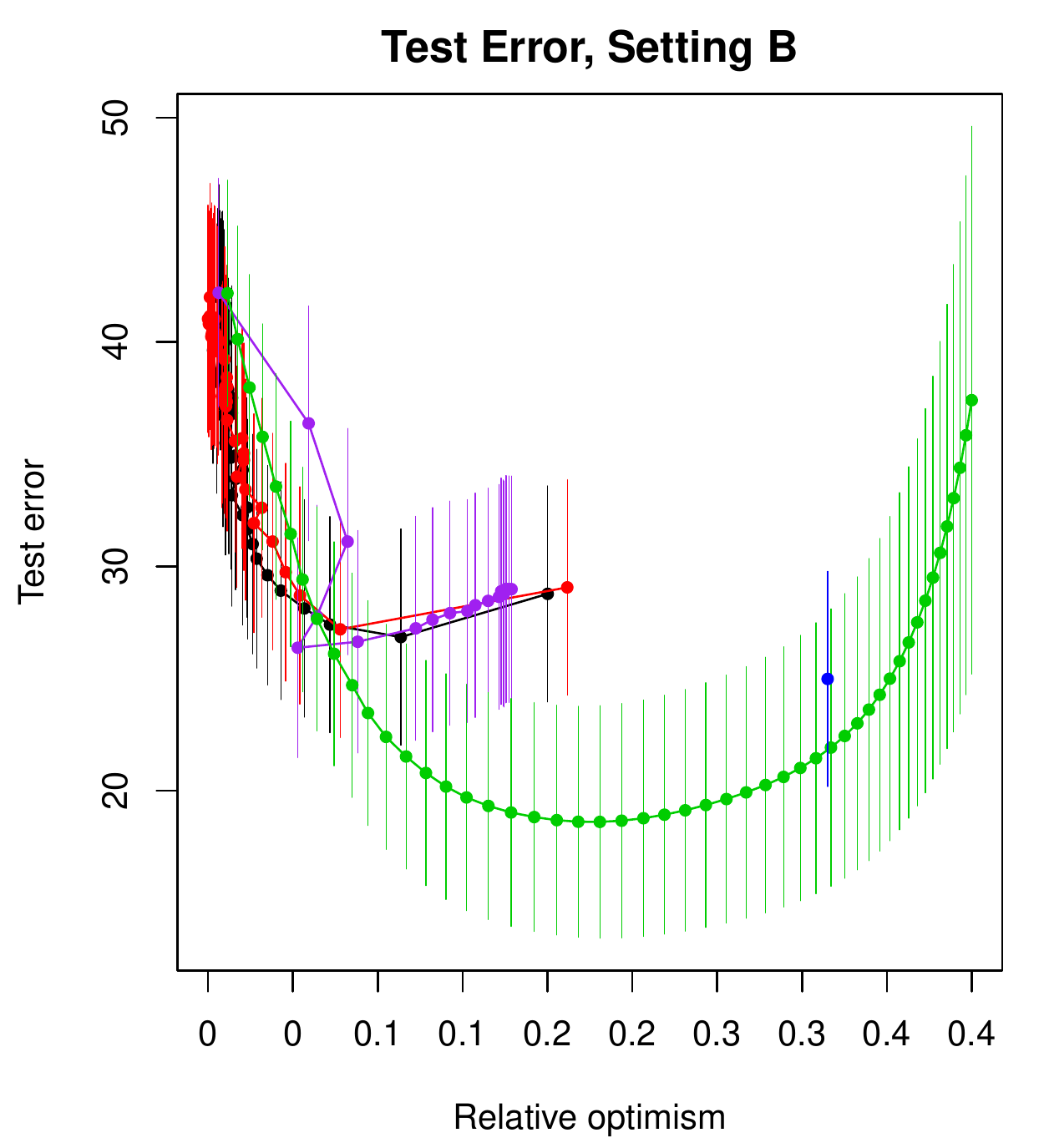} \\
\includegraphics[width=0.32\textwidth]{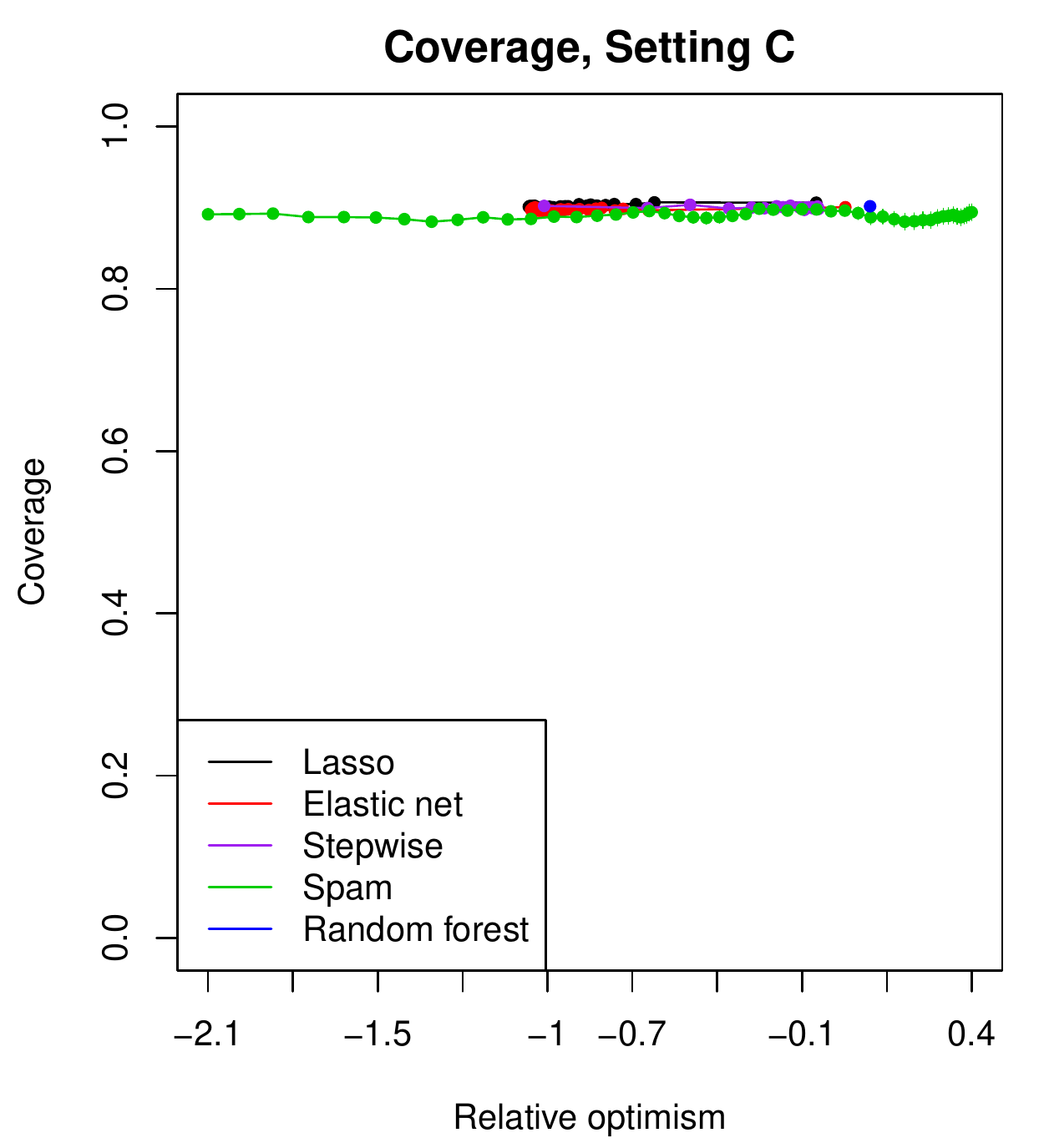} 
\includegraphics[width=0.32\textwidth]{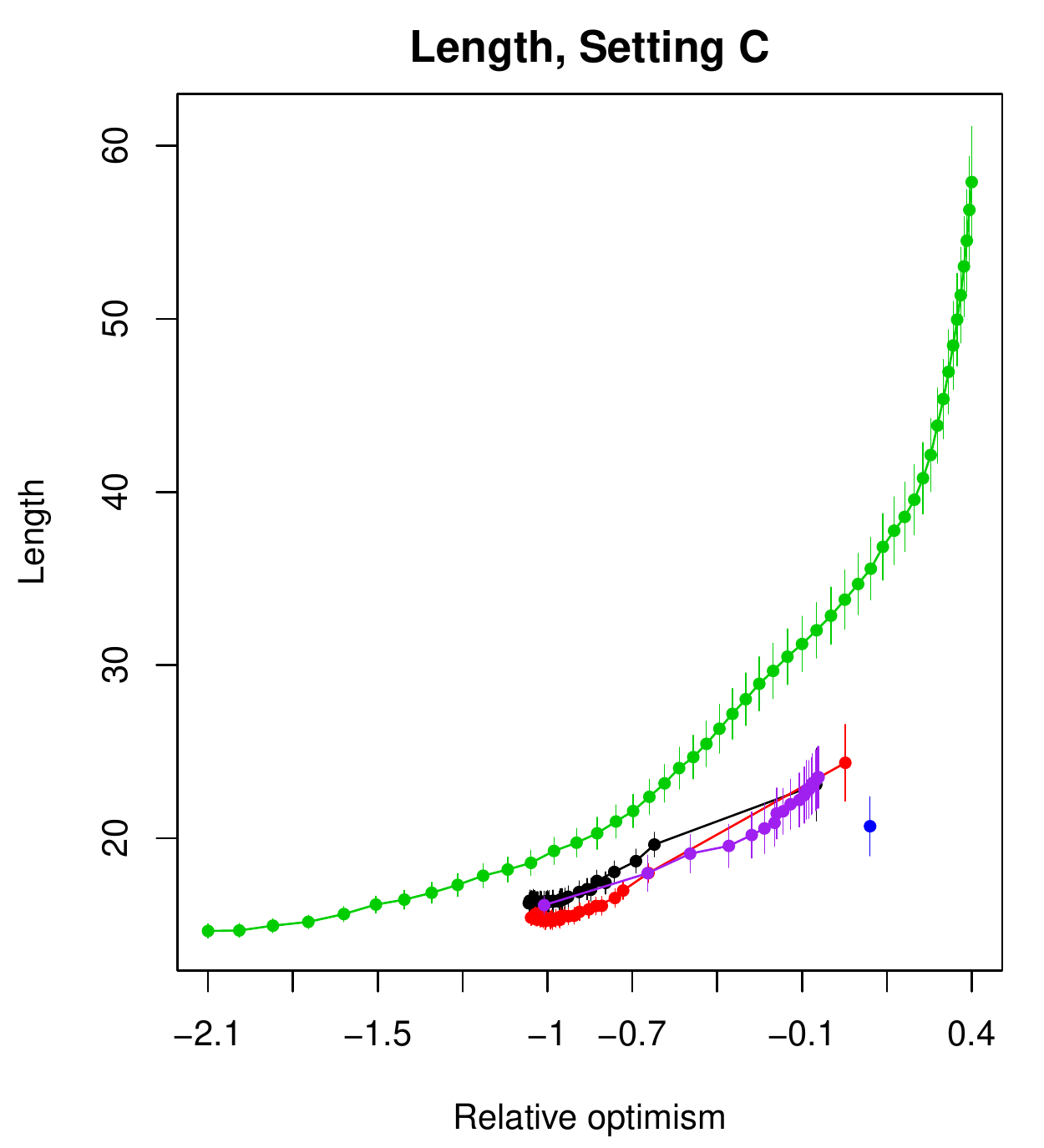} 
\includegraphics[width=0.32\textwidth]{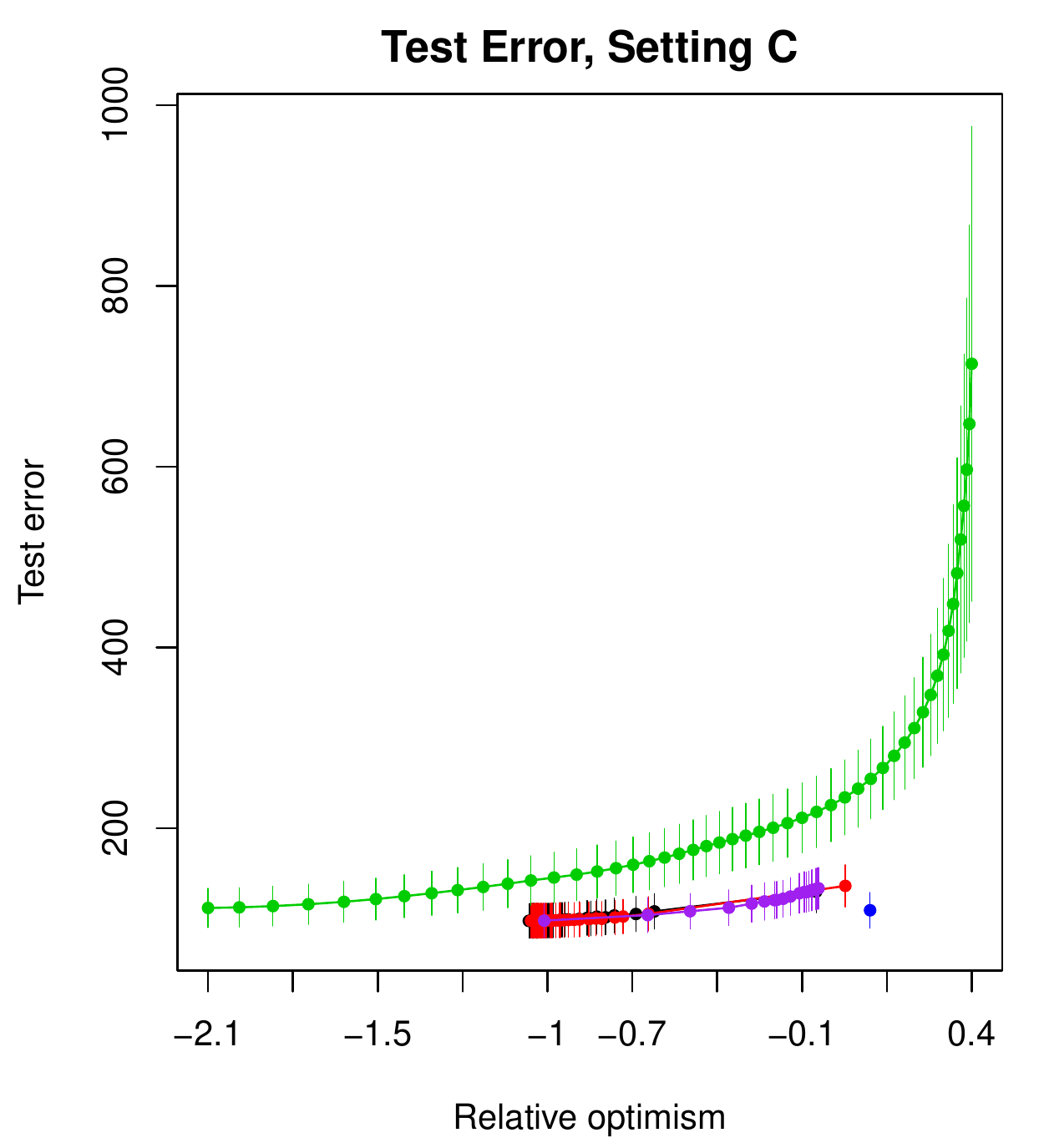} 
\caption{\it Comparison of conformal prediction intervals in
  low-dimensional problems with $n=200$, $d=20$, across a variety of
  base estimators.}
\label{fig:many.lo}
\end{figure}

\begin{figure}[p]
\centering
\includegraphics[width=0.32\textwidth]{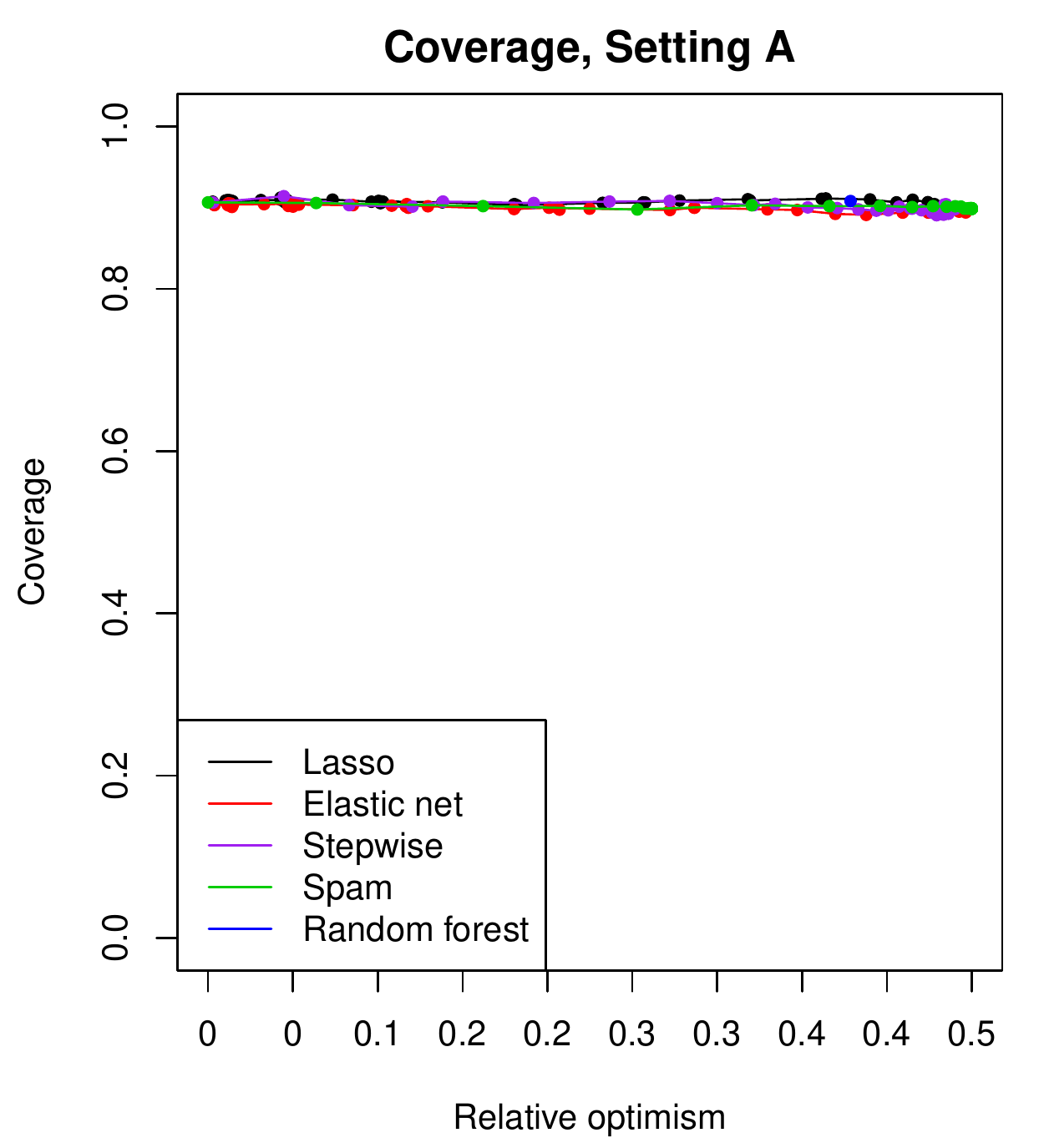} 
\includegraphics[width=0.32\textwidth]{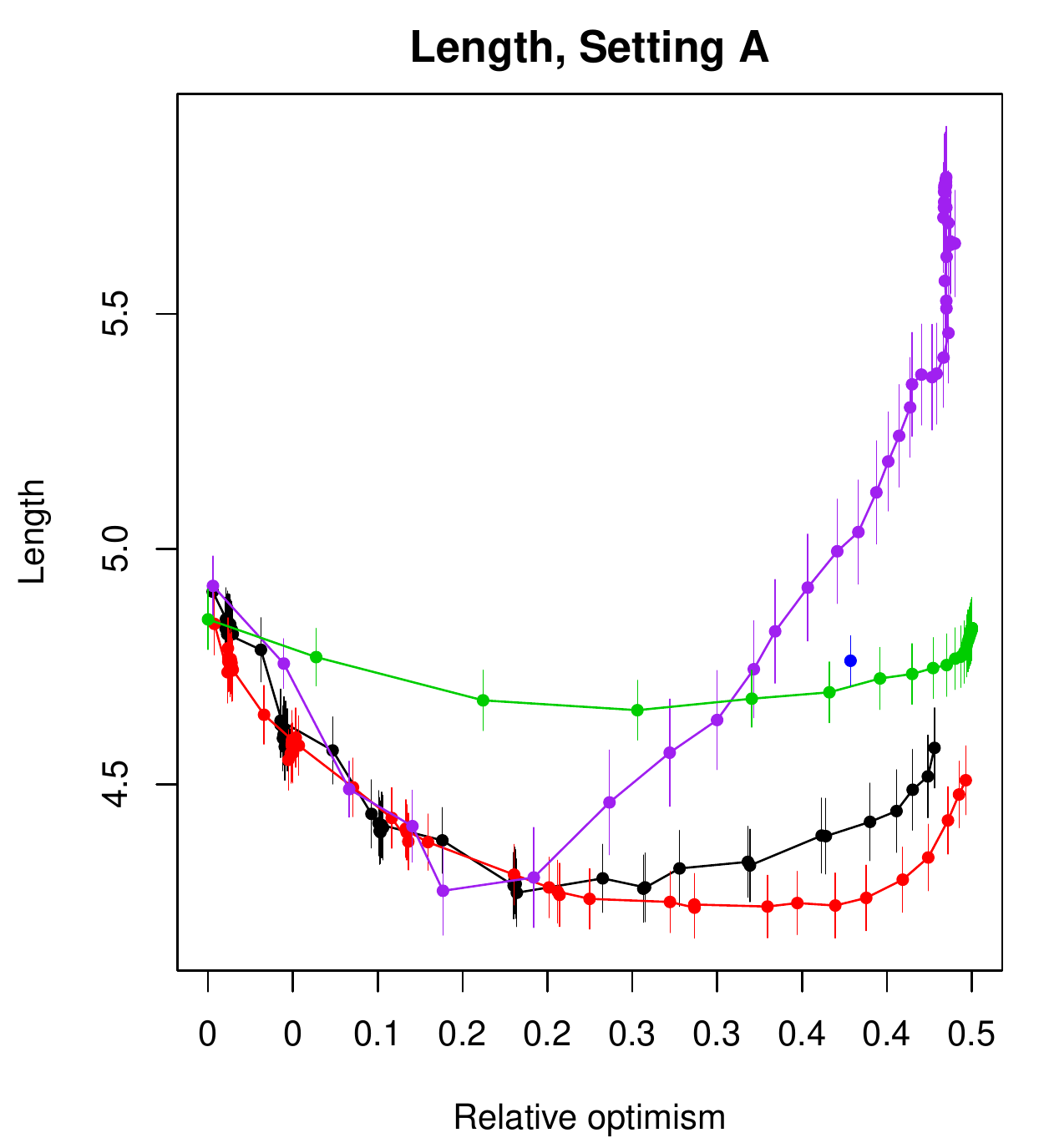} 
\includegraphics[width=0.32\textwidth]{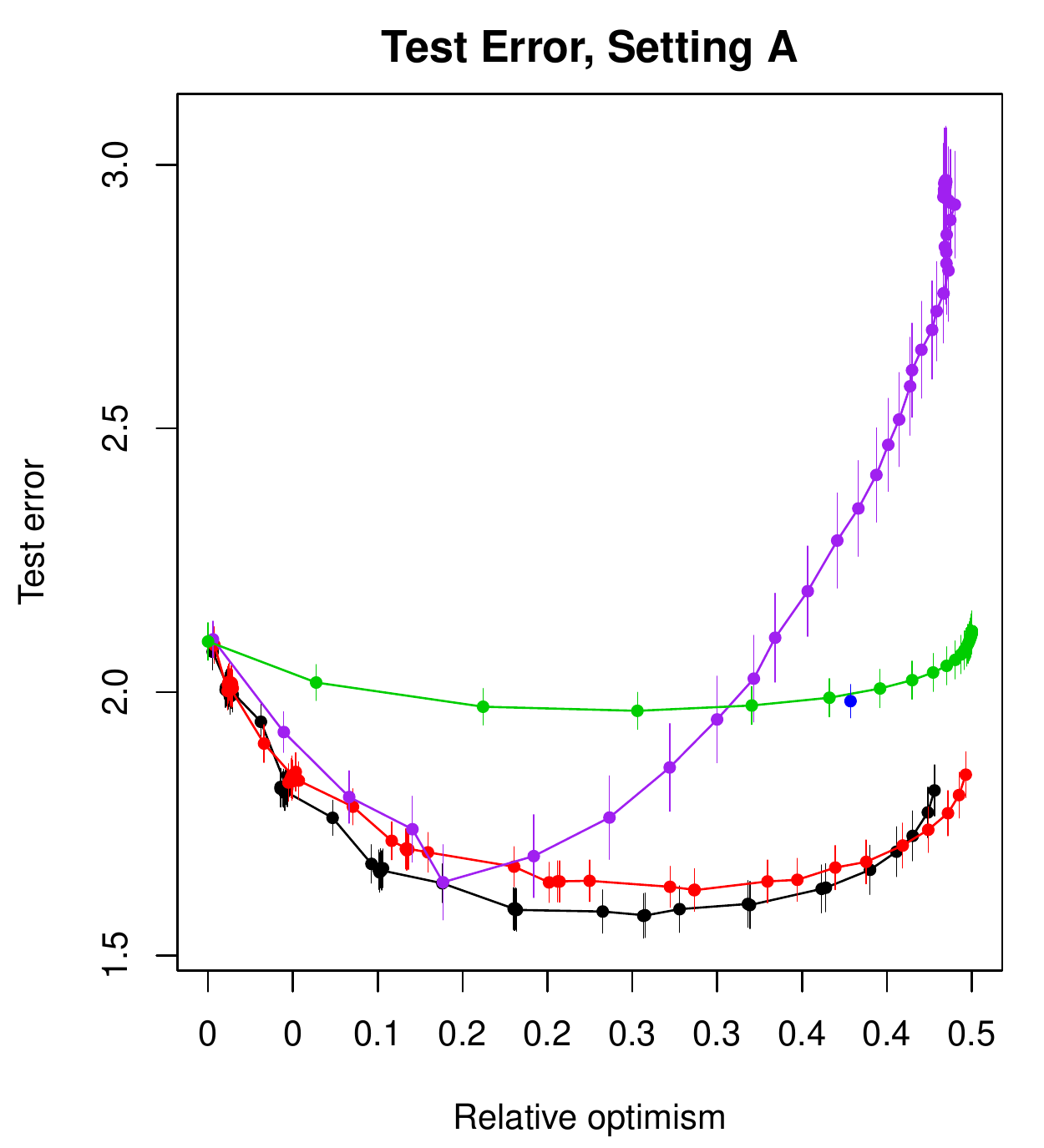} \\
\includegraphics[width=0.32\textwidth]{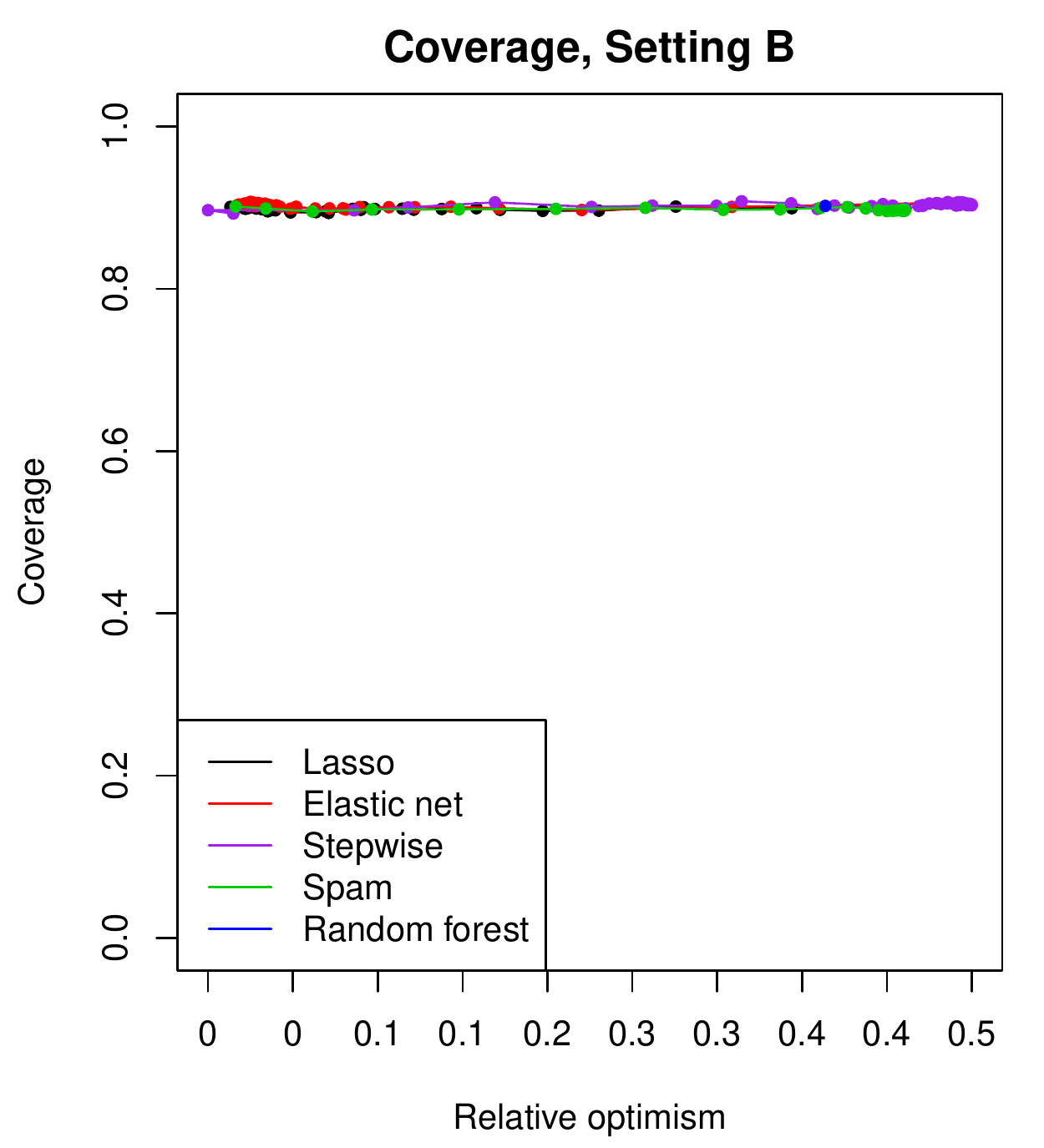} 
\includegraphics[width=0.32\textwidth]{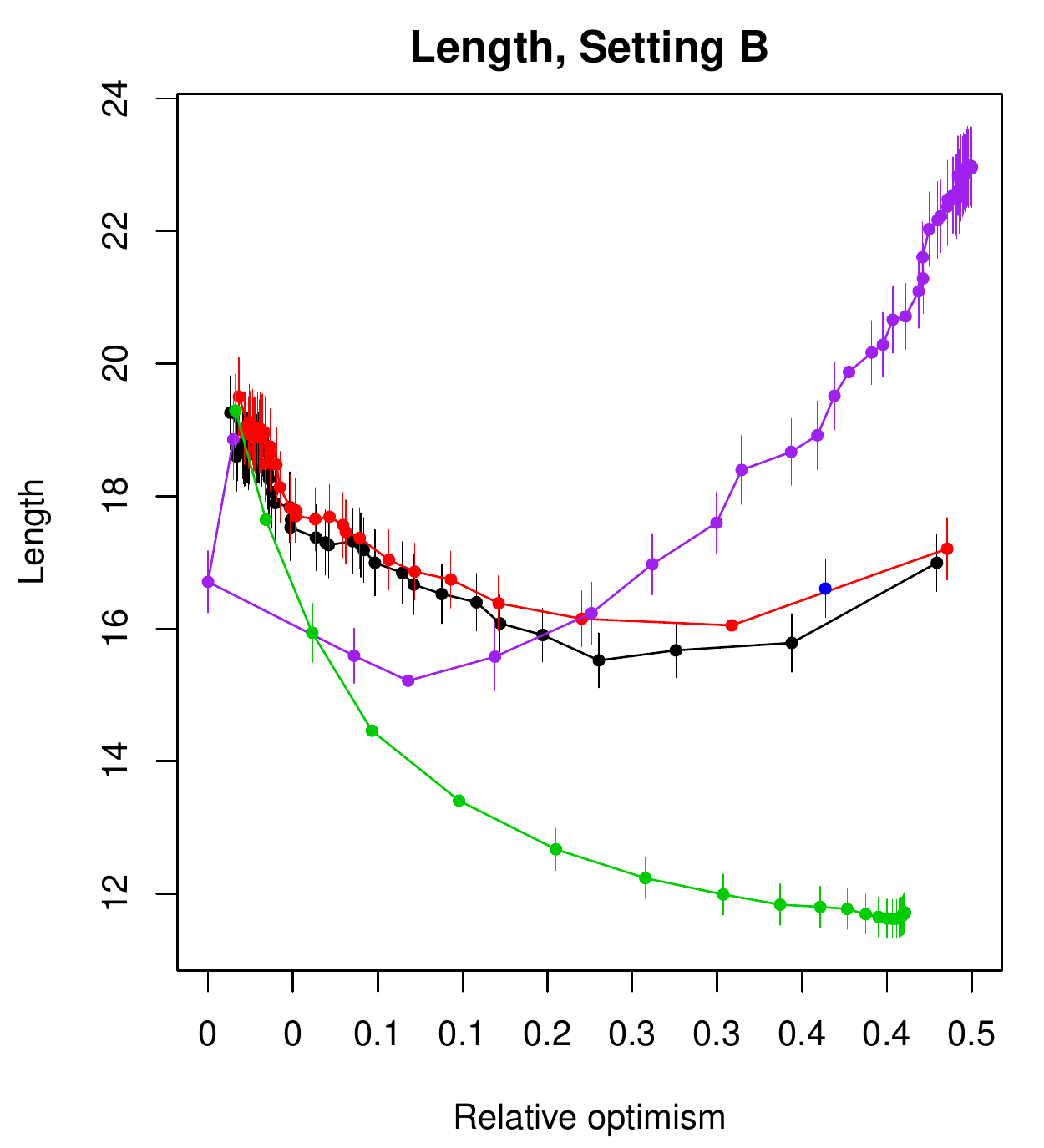} 
\includegraphics[width=0.32\textwidth]{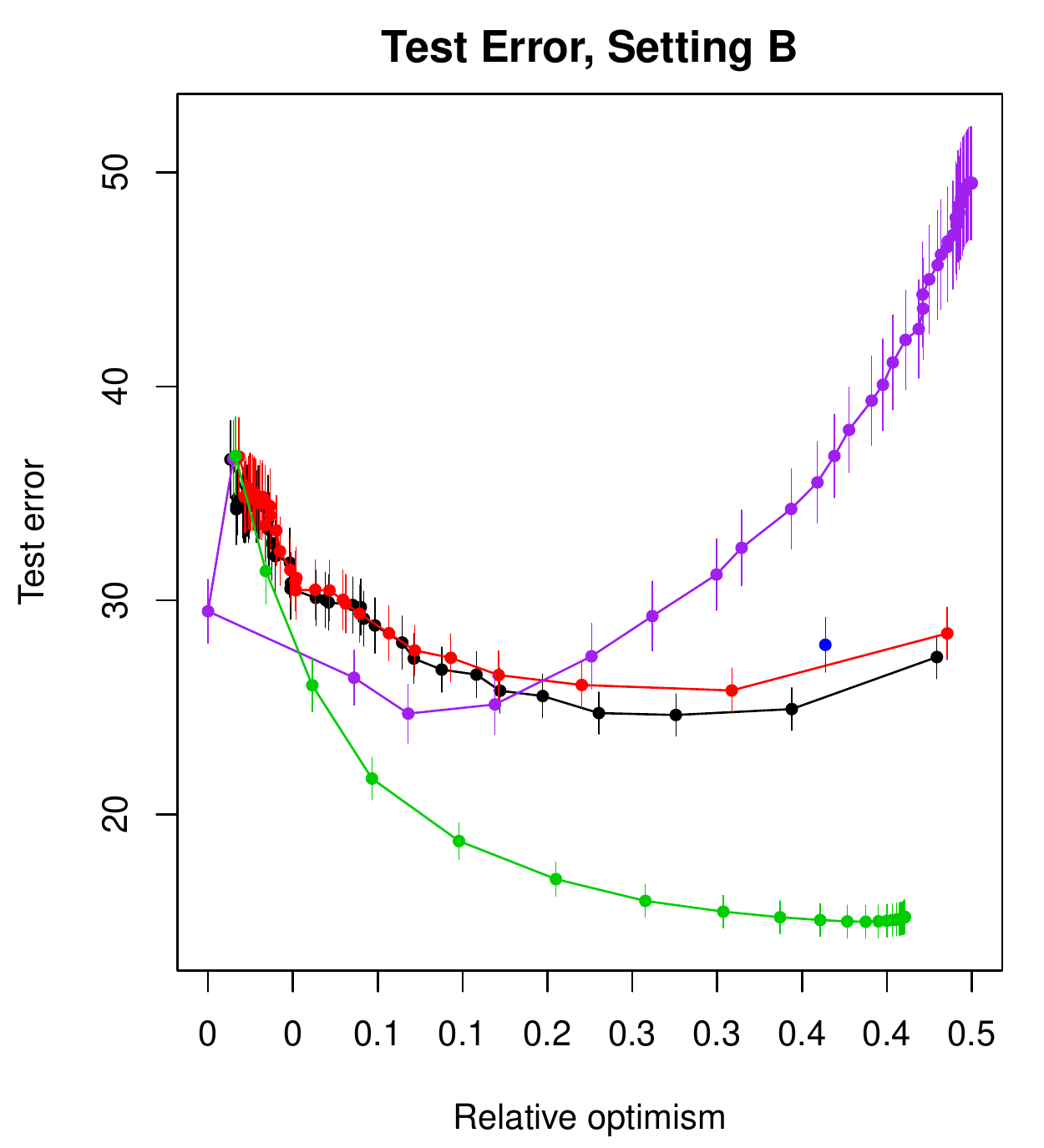} \\
\includegraphics[width=0.32\textwidth]{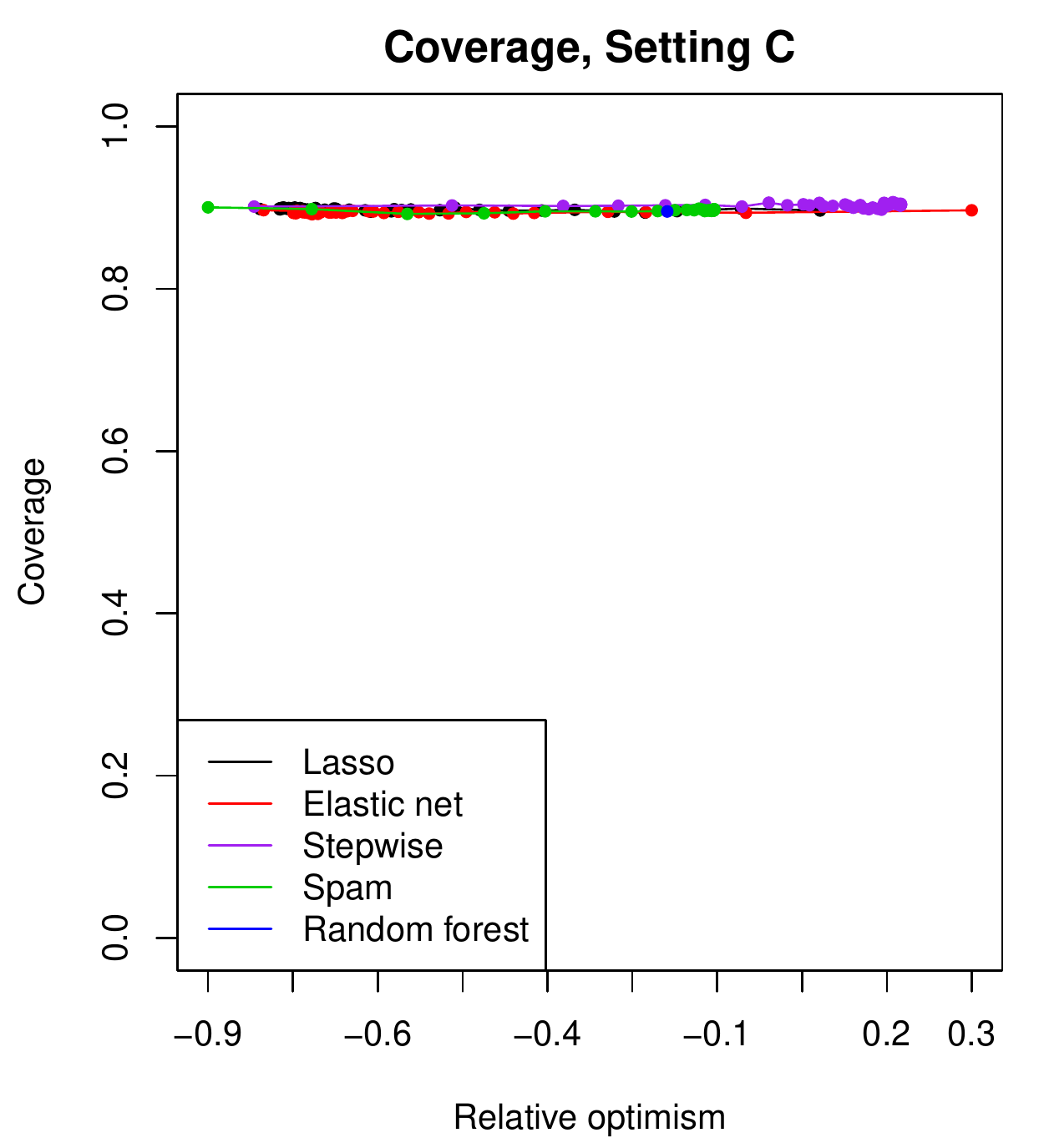} 
\includegraphics[width=0.32\textwidth]{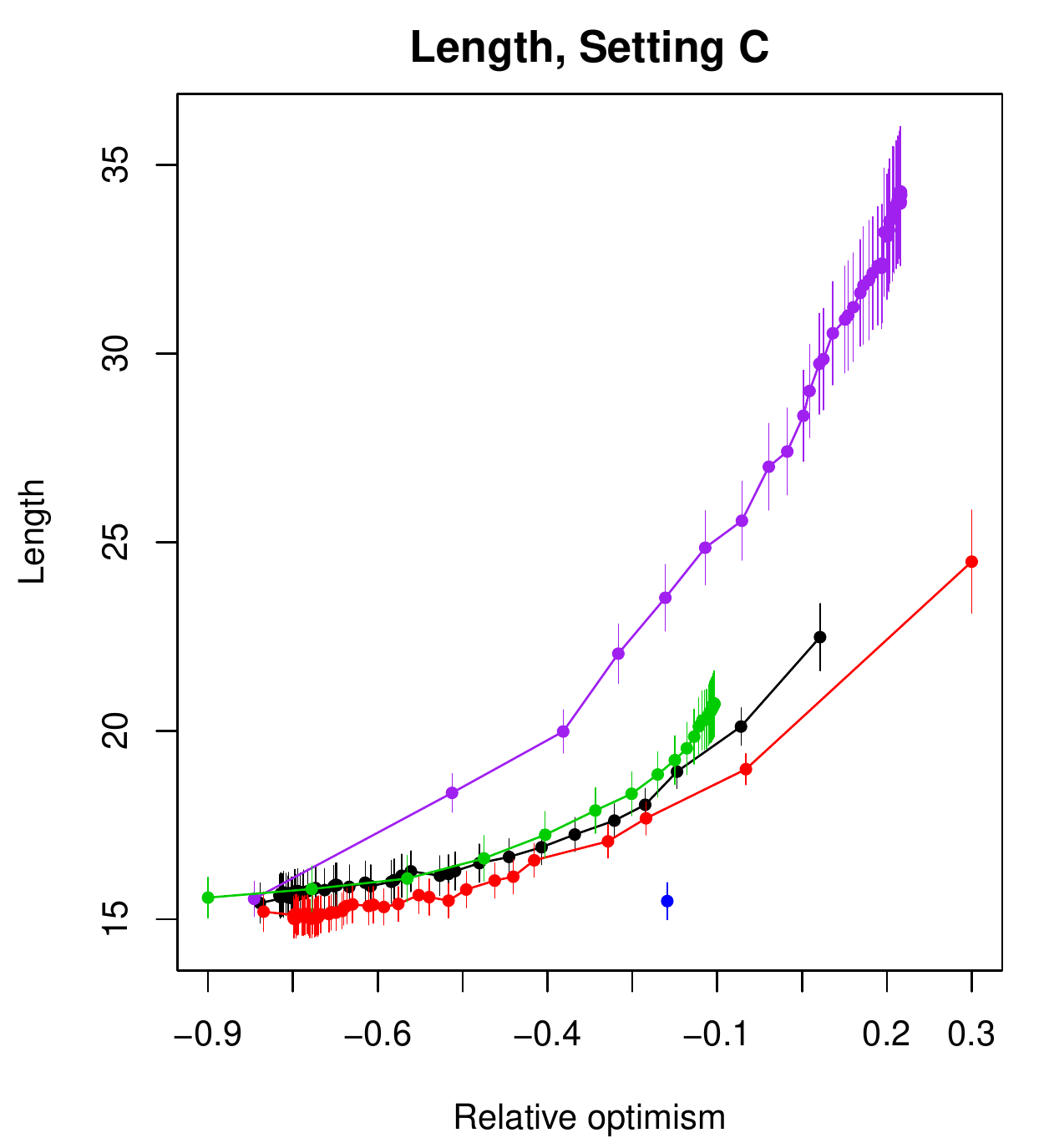} 
\includegraphics[width=0.32\textwidth]{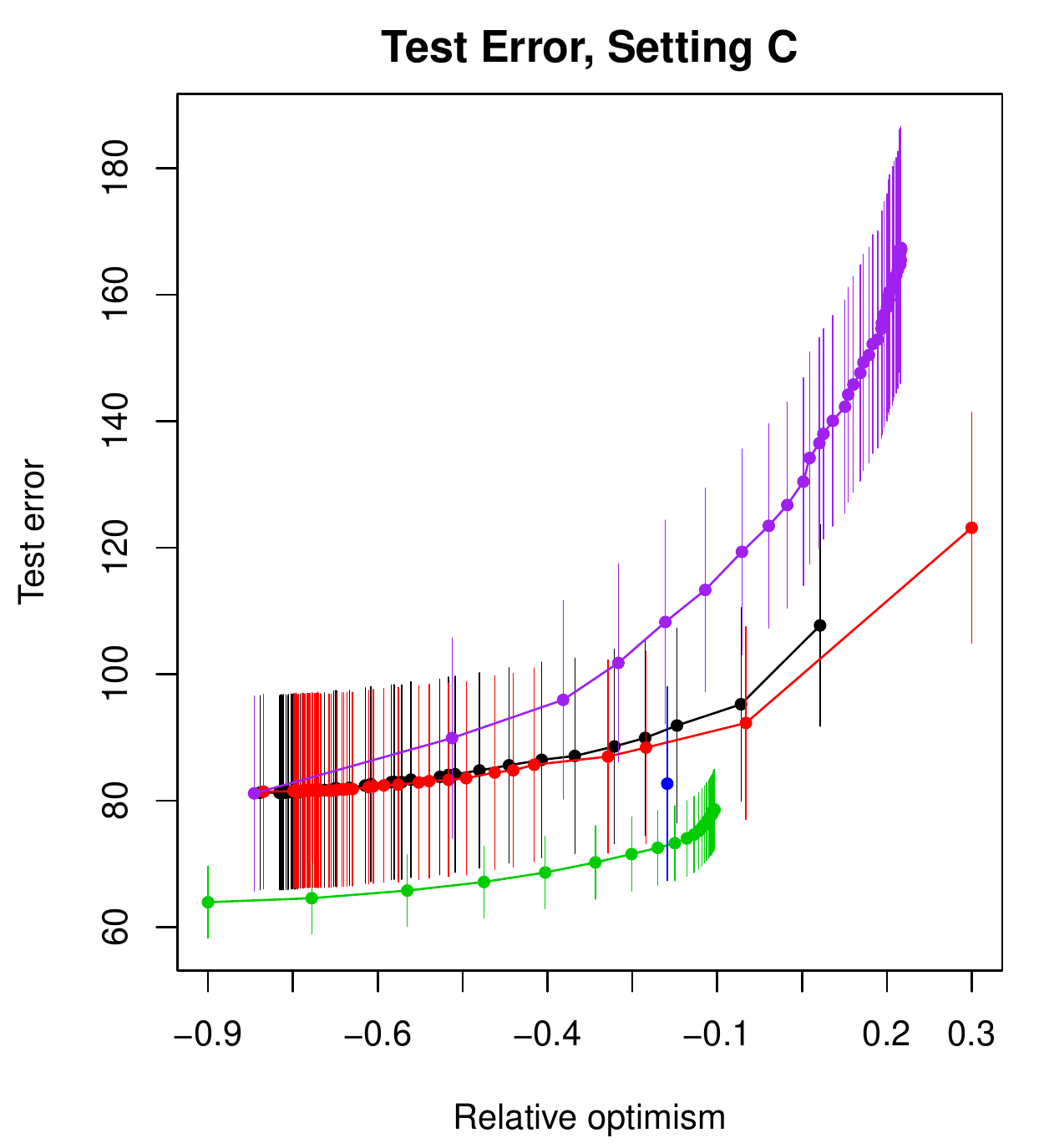} 
\caption{\it Comparison of conformal prediction intervals in
  high-dimensional problems with $n=200$, $d=2000$, across a variety of
  base estimators.}
\label{fig:many.hi}
\end{figure}

In Settings A and C (where the mean is linear), the mean function was
defined by choosing $s=5$ regression coefficients to be nonzero,
assigning them values $\pm 8$ with equal probability, and mulitplying
them against the standardized predictors. In Setting B (where the mean
is nonlinear), it is defined by multiplying these coefficients  
against B-splines transforms of the standardized predictors. To
demonstrate the effect of sparsity, we add a Setting D with
high-dimensionality that mimics Setting A except that the number of 
nonzero coefficients is $s=100$. Figures \ref{fig:many.lo}  
(low-dimensional), \ref{fig:many.hi} (high-dimensional) and
\ref{fig:many.hi.nonsparse} (high-dimensional, linear, nonsparse,
deferred to Appendix \ref{app:additional}) show the results of these
experiments.   

Each method is applied over a range of tuning parameter choices. For 
the sake of defining a common ground for comparisons, all values are
plotted against the relative optimism,\footnote{
In challenging settings (e.g., Setting C), the relative optimism can be
negative.  This is not an error, but occurs naturally for inflexible estimators
and very difficult settings.  This is unrelated to conformal inference and to
observations about the plot shapes.}
defined to be 
\begin{equation*}
\text{(relative optimism)}=\frac{(\text{test error}) - (\text{train
    error})}{(\text{test error})}.
\end{equation*}
The only exception is the random forest estimator, which gave stable
errors over a variety of tuning choices; hence it is represented by a
single point in each plot (corresponding to 500 trees in the
low-dimensional problems, and 1000 trees in the high-dimensional
problems).  All curves in the figures represent an average over 50
repetitions, and error bars indicating the standard errors.  In all
cases, we used the split conformal method for computational
efficiency. 

In the low-dimensional problems (\Cref{fig:many.lo}), the
best test errors are obtained by the linear methods (lasso, 
elastic net, stepwise) in the linear Setting A, and by SPAM in the
nonlinear (additive) Setting B.  In Setting C, all estimators perform
quite poorly.  We note that across all settings and estimators, no
matter the performance in test error, the coverage of the conformal  
intervals is almost exactly 90\%, the nominal level, and the
interval lengths seem to be highly correlated with test errors.  

In the high-dimensional problems (\Cref{fig:many.hi}), the
results are similar.  The regularized linear estimators perform best
in the linear Setting A, while SPAM dominates in the nonlinear
(additive) Setting B and performs slightly better in Setting C.  All
estimators do reasonably well in Setting A and quite terribly in 
Setting C, according to test error.  Nevertheless, across this range
of settings and difficulties, the coverage of the conformal
prediction intervals is again almost exactly 90\%, and the
lengths are highly correlated with test errors.    



\section{Extensions of Conformal Inference}
\label{sec::extensions}

The conformal and split conformal methods, combined with basically 
any fitting procedure in regression, provide finite-sample
distribution-free predictive inferences.
We describe some extensions of this framework to improve
the interpretability and applicability of conformal inference.

\subsection{In-Sample Split Conformal Inference}
\label{sec:in_sample}

Given samples $(X_i,Y_i)$, $i=1,\ldots,n$, and a method that outputs
a prediction band, we would often like to evaluate this band at some
or all of the observed points $X_i$, $i=1,\ldots,n$.  This is
perhaps the most natural way to visualize any prediction band.
However, the conformal prediction methods from
\Cref{sec::method} are designed to give a valid  prediction interval
at a future point $X_{n+1}$, from the same distribution as $\{X_i,
i=1,\ldots,n\}$, but not yet observed. 
If we apply the full or split conformal prediction methods at an
observed feature value, then it is not easy to establish finite-sample
validity of these methods. 

A simple way to obtain valid in-sample predictive inference from the 
conformal methods is to treat each $X_i$ as a new feature value and 
use the other $n-1$ points as the original features (running 
either the full or split conformal methods on these $n-1$ points).
This approach has two drawbacks.  First, it seriously degrades the
computational efficiency of the conformal methods---for full
conformal, it multiplies the cost of the (already expensive)
\Cref{alg:conf} by $n$, making it perhaps intractable for even
moderately large data sets; for split conformal, it multiplies the
cost of \Cref{alg:split} by $n$, making it as expensive as
the jackknife method in \Cref{alg:jack}. 
Second, if we denote by \smash{$C(X_i)$} the prediction
interval that results from this method at $X_i$, for $i=1,\ldots,n$,
then one might expect the empirical coverage
\smash{$\frac{1}{n} \sum_{i=1}^n \one\{Y_i \in C(X_i)\}$} to
be at least $1-\alpha$, but this is not easy to show due to the
complex dependence between the indicators. 

Our proposed technique overcomes both of these drawbacks, and is a
variant of the split conformal method that we call {\it rank-one-out}
or ROO split conformal inference.  The basic idea is quite similar to
split conformal, but the ranking is conducted in a leave-one-out
manner. The method is presented in \Cref{alg:roo}.  For
simplicity (as with our presentation of the split conformal method in
\Cref{alg:split}), we assume that $n$ is even, and only 
minor modifications are needed for $n$ odd. 
Computationally, ROO split conformal is very efficient.  First, the
fitting algorithm $\mathcal A$ (in the notation of
\Cref{alg:roo}) only needs to be run twice. Second, for each split, the
ranking of absolute residuals needs to be calculated just once; with
careful updating, it can be reused in order to calculate the prediction
interval for $X_i$ in $O(1)$ additional operations, for each
$i=1,\ldots,n$.   

\begin{algorithm}[tb]
\caption{Rank-One-Out Split Conformal}
\label{alg:roo}
\begin{algorithmic}
\State{\bf Input:} Data $(X_i,Y_i)$, $i=1,\ldots,n$, miscoverage level
$\alpha \in (0,1)$, regression algorithm $\mathcal A$
\State{\bf Output:} Prediction intervals at each $X_i$,
$i=1,\ldots,n$ 
\State Randomly split $\{1,\ldots,n\}$ into two equal-sized
subsets $\mathcal I_1$, $\mathcal I_2$
\For{$k \in \{1,2\}$}
\State \smash{$\hat\mu_k = \mathcal A\big(\{(X_i,Y_i):i\in \mathcal 
I_k\}\big)$} 
\For{$i \notin I_{k}$}
\State \smash{$R_i = |Y_i-\hat\mu_k(X_i)|$}
\EndFor
\For{$i \notin I_{k}$}
\State $d_i =$ the $m$th smallest value in $\{R_j : j \notin \mathcal
I_{k},~j\neq i\}$, where $m=\lceil n/2 (1-\alpha) \rceil$ 
\State \smash{$C_{\mathrm{roo}}(X_i) =
  [\hat\mu_k(X_i)-d_i, \hat\mu_k(X_i)+d_i]$} 
\EndFor
\EndFor
\State Return intervals
\smash{$C_{\mathrm{roo}}(X_i)$}, 
$i=1,\ldots,n$
\end{algorithmic}
\end{algorithm}

By symmetry in their construction, the ROO split conformal intervals
have the in-sample finite-sample coverage property
\begin{equation*}
\P\big(Y_i\in C_{\mathrm{roo}}(X_i)\big) \geq
1-\alpha, \quad \text{for all $i=1,\ldots,n$}.
\end{equation*}
A practically interesting performance measure is the empirical
in-sample average coverage \smash{$\frac{1}{n} \sum_{i=1}^n \one   
 \{ Y_i \in C_{\mathrm{roo}} (X_i)\}$}.  Our
construction in Algorithm \ref{alg:roo} indeed implies a weak
dependence among the random indicators in this average, which 
leads to a slightly worse coverage guarantee for the empirical in-sample average
coverage, with the difference from the nominal $1-\alpha$ level being of order
$\sqrt{\log n/n}$, with high probability.

\begin{theorem}
\label{thm:roo_asymptotic}
If $(X_i,Y_i)$, $i=1,\ldots,n$ are i.i.d., then for the ROO split
conformal band
\smash{$C_{\mathrm{roo}}$} constructed in 
\Cref{alg:roo}, there is an absolute constant $c>0$, such
that for all $\epsilon>0$, 
\begin{equation*}
\P\bigg(\frac{1}{n} \sum_{i=1}^n \one  
 \{ Y_i \in C_{\mathrm{roo}}(X_i)\} \geq
 1-\alpha-\epsilon \bigg) \geq 1-2\exp(-cn\epsilon^2).
\end{equation*}
Moreover, if we assume additionally that the residuals $R_i$, $i 
=1,\ldots,n$, have a continuous joint distribution, then for all
$\epsilon>0$,  
\begin{equation*}
\P\bigg(1-\alpha-\epsilon \leq \frac{1}{n} \sum_{i=1}^n \one   
 \{ Y_i \in C_{\mathrm{roo}}(X_i)\} \leq
 1-\alpha+\frac{2}{n}+\epsilon\bigg) \geq 
1-2\exp(-cn\epsilon^2).
\end{equation*}
\end{theorem}

The proof of \Cref{thm:roo_asymptotic} uses McDiarmid's
inequality. It is conceptually straightforward but requires a careful  
tracking of dependencies and is deferred until \Cref{sec:proof_2}.

\begin{remark}
An even simpler, and conservative approximation to
each in-sample prediction interval
\smash{$C_{\mathrm{roo}} (X_i)$} is  
\begin{equation}
\label{eq:roo_split_relax}
\tilde{C}_{\mathrm{roo}}(X_i) = [\hat\mu_k(X_i)-\tilde d_k,
\hat\mu_k(X_i)+\tilde d_k],
\end{equation}
where, using the notation of \Cref{alg:roo}, we define
\smash{$\tilde{d}_k$} to be the $m$th smallest element of the set
$\{R_i : i \in \notin I_k\}$, for $m=\lceil (1-\alpha)n/2
\rceil+1$. Therefore, now only a single sample quantile from the
fitted residuals is needed for each split.  As a price, each interval
in \eqref{eq:roo_split_relax} is wider
than its counterpart 
from \Cref{alg:roo} by at most one interquantile difference.  
Moreover, the results of \Cref{thm:roo_asymptotic} carry over to the 
prediction band \smash{$\tilde{C}_{\mathrm{roo}}$}: in the 
second probability statement (trapping the empirical in-sample average
coverage from below and above), we need only change the $2/n$ term
to $6/n$.
\end{remark}

In \Cref{sec:proof_1}, we prove \Cref{thm:split_asymptotic}
as a modification of \Cref{thm:roo_asymptotic}. 

\subsection{Locally-Weighted Conformal Inference}
\label{sec:local_weight}

The full conformal and split conformal methods both tend to produce 
prediction bands \smash{$C(x)$} whose width is roughly constant over
$x \in \R^d$.  In fact, for split conformal, the width is exactly
constant over $x$. For full conformal, the width can vary  
slightly as 
$x$ varies, but the difference is often negligible as long as the
fitting method is moderately stable.  This property---the width of  
\smash{$C(x)$} being roughly immune to $x$---is     
desirable if the spread of the residual $Y-\mu(X)$ does not vary
substantially as $X$ varies.  However, in some scenarios this will not
be true, i.e., the residual variance will vary nontrivially with
$X$, and in such a case we want the conformal band to adapt
correspondingly.  

We now introduce an extension to the conformal method that can account 
for nonconstant residual variance.  Recall that, in
order for the conformal inference method to have valid coverage, we
can actually use any conformity score function to generalize the
definition of (absolute) residuals as given in
\eqref{eq:conf_score_gen} of 
\Cref{rem:conformity_score}.    For the present extension, we
modify the definition of residuals in \Cref{alg:conf} by 
 scaling the fitted residuals inversely by
an estimated error spread.   Formally
\begin{equation}
\label{eq:locally_weighted_res}
R_{y,i} = \frac{|Y_i-\hat\mu_y(X_i)|}
{\hat\rho_y(X_i)}, \; i=1,\ldots,n, \quad\text{and}\quad
R_{y,n+1} = \frac{|y-\hat\mu_y(x)|}
{\hat\rho_y(x)},  
\end{equation}
where now \smash{$\hat\rho_y(x)$} denotes an estimate of the
conditional mean absolute deviation (MAD) of $(Y-\mu(X))|X=x$, as a
function of $x \in \R^d$. We choose to estimate the error spread by  
the mean absolute deviation of the
fitted residual rather than the standard deviation, since the former
exists in some cases in which the latter does not.
Here, the conditional mean \smash{$\hat\mu_y$} and conditional MAD
\smash{$\hat\rho_y$} can either be estimated jointly, or more simply,
the conditional mean \smash{$\hat\mu_y$} can be estimated first, and
then the conditional MAD \smash{$\hat\rho_y$} can be estimated using
the collection of fitted absolute residuals
\smash{$|Y_i-\hat\mu_y(X_i)|$}, $i=1,\ldots,n$ and
\smash{$|y-\hat\mu_y(X_{n+1})|$}.  With the locally-weighted
residuals in \eqref{eq:locally_weighted_res}, the validity and
accuracy properties of the full conformal inference method
carry over.

For the split conformal and the ROO split conformal methods, the
extension is similar. In \Cref{alg:split}, we instead use
locally-weighted residuals 
\begin{equation}
\label{eq:locally_weighted_res_split}
R_i = \frac{|Y_i-\hat\mu(X_i)|}
{\hat\rho(X_i)}, \; i \in \mathcal I_2,
\end{equation}
where the conditional mean \smash{$\hat\mu$} and conditional MAD
\smash{$\hat\rho$} are fit on the samples in $\mathcal I_1$, either 
jointly or in a two-step fashion, as explained above.  The output
prediction interval at a point $x$ must also be modified, now being 
\smash{$[\hat\mu(x)-\hat\rho(x)d, \;\hat\mu(x)+\hat\rho(x)d]$}.
In \Cref{alg:roo}, analogous modifications are performed.  
Using locally-weighted residuals, as in
\eqref{eq:locally_weighted_res_split}, the 
validity and accuracy properties of the split methods, both finite
sample and asymptotic, in
\Cref{thm:split_valid,thm:split_asymptotic,thm:roo_asymptotic}, again 
carry over.  The jackknife interval can also be extended in a similar
fashion. 

\begin{figure}[htbp]
\centering
\includegraphics[width=0.48\textwidth]{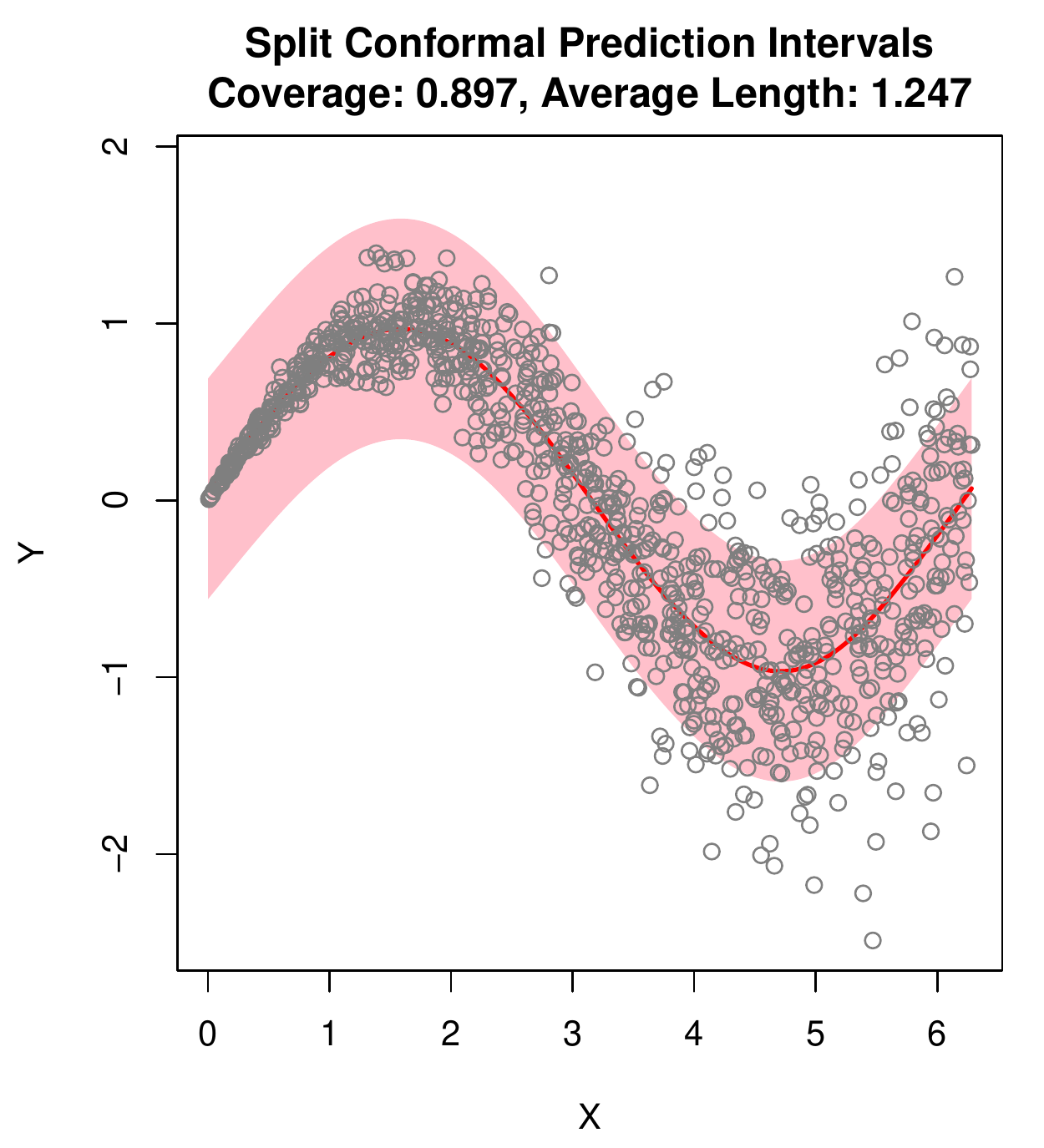} 
\includegraphics[width=0.48\textwidth]{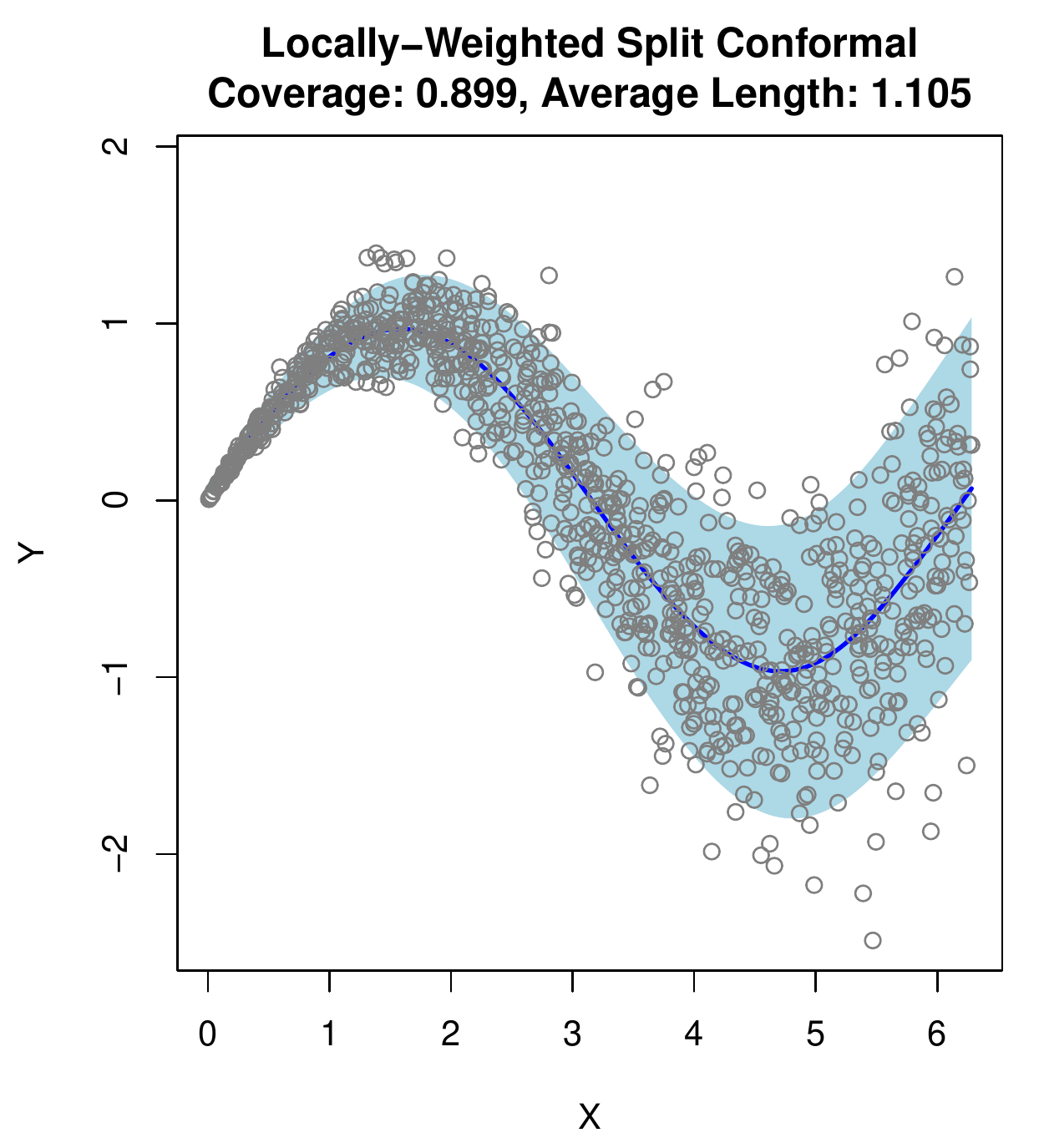} \\
\includegraphics[width=0.48\textwidth]{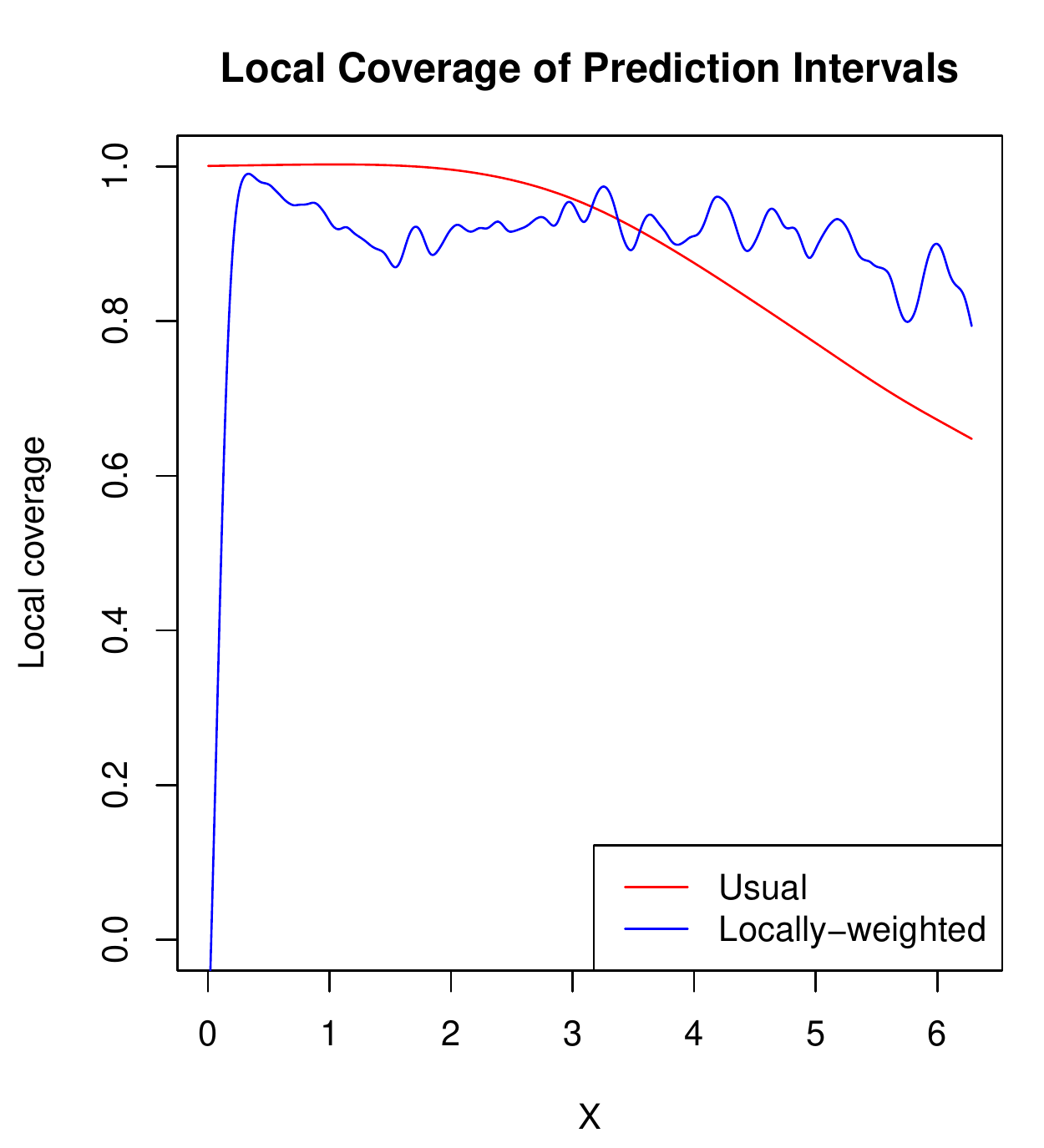} 
\includegraphics[width=0.48\textwidth]{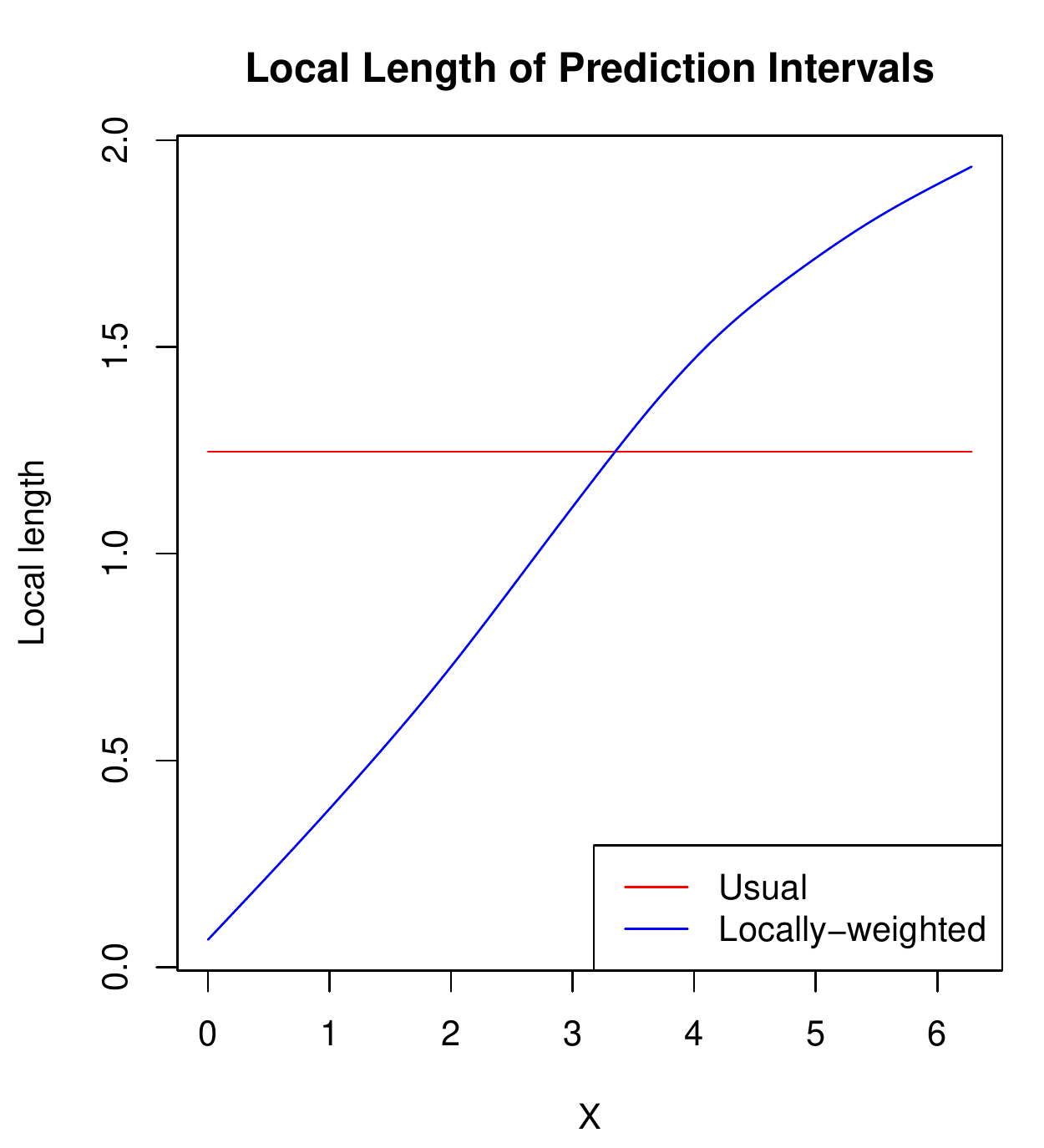} 
\caption{\it A simple univariate example of the usual (unweighted)
  split conformal and locally-weighted split conformal prediction
  bands. The top left 
  panel shows the split conformal band, and the top right shows the  
  locally-weighted split conformal band; we can see that the latter
  properly adapts to the heteroskedastic nature of the data, whereas
  the former has constant length over all $x$ (by construction).  The 
  bottom left and 
  right panels plot the empirical local coverage and local length
  measures (which have been mildly smoothed as functions of $x$ for
  visibility).  The locally-weighted split conformal method maintains
  a roughly constant level of local coverage across $x$, but its band
  has a varying local length; the usual split conformal method
  exhibits precisely the opposite trends.}   
\label{fig:local_weight}
\end{figure}

\Cref{fig:local_weight} displays a simple example of the
split conformal method using locally-weighted residuals.  We let
$n=1000$, drew i.i.d.\ copies $X_i \sim \mathrm{Unif}(0,2\pi)$,
$i=1,\ldots,n$, and let
\begin{equation*}
Y_i = \sin(X_i) + \frac{\pi |X_i|}{20} \epsilon_i, \; i=1,\ldots,n,
\end{equation*}
for i.i.d.\ copies $\epsilon_i \sim N(0,1)$, $i=1,\ldots,n$.  We 
divided the data set randomly into two halves $\mathcal I_1, \mathcal
I_2$, and fit the conditional mean estimator \smash{$\hat\mu$} on the 
samples from the first half $\mathcal I_1$ using a smoothing 
spline, whose tuning parameter was chosen by cross-validation.  This
was then used to produce a 90\% prediction band, according to the
usual (unweighted) split conformal strategy, that has
constant width by design; it is plotted, as a function of $x \in \R$,
in the top left panel of \Cref{fig:local_weight}.  For our
locally-weighted version, we then fit a conditional MAD estimator
\smash{$\hat\rho$} on \smash{$|Y_i-\hat\mu(X_i)|$}, $i \in \mathcal
I_1$, again using a smoothing spline, whose tuning parameter was
chosen by cross-validation.  Locally-weighted residuals were used
to produce a 90\% 
prediction band, with locally-varying width, plotted in the top right
panel of the figure. Visually, the locally-weighted
band adapts better to the heteroskedastic nature of the data. This
is confirmed by looking at the length of the locally-weighted band as
a function of $x$ in the bottom right panel.  It is also supported by
the improved empirical average length offered by the locally-weighted 
prediction band, computed over 5000 new draws from
$\mathrm{Unif}(0,2\pi)$, which is 1.105 versus 1.247 for the
unweighted band.  In terms of average coverage, again
computed empirically over the same 5000 new draws,
both methods are very close to the nominal 90\% level, with the
unweighted version at 89.7\% and the weighted version at 89.9\%.
Most importantly, the locally-weighted 
version does a better job here of maintaining a conditional
coverage level of around 90\% across all $x$, as shown in the bottom
left panel, as compared to the unweighted split conformal method,
which over-covers for smaller $x$ and under-covers for larger $x$. 

Lastly, it is worth remarking that if the noise is indeed
homoskedastic, then of course using such a locally-weighted conformal  
band will have generally a inflated (average) length compared to the
usual unweighted conformal band, due to the additional randomness in 
estimating the conditional MAD.  In Appendix \ref{app:additional}, we
mimic the setup in \Cref{fig:local_weight} but with
homoskedastic noise to demonstrate that, in 
this particular problem, there is not too much inflation in the 
length of the locally-weighted band compared to the usual band. 

\section{Model-Free Variable Importance: LOCO}
\label{sec::varimp}

In this section, we discuss the problem of
estimating the importance of each variable in a prediction model. 
A critical question is: how do we assess
variable importance when we are treating the working model as
incorrect? One possibility, if we are fitting a linear model 
with variable selection, is to interpret the coefficients as estimates
of the parameters in the best linear approximation to the mean 
function $\mu$.  This has been studied in, e.g.,
\citet{wass2014,Buja14Models,exactlar}. 
However, we take a different approach for two reasons.
First, our method is not limited to linear regression.
Second, the spirit of our approach is to focus on predictive
quantities and we want to measure variable importance 
directly in terms of prediction.  Our approach is similar in spirit to
the variable importance measure used in random forests
\citep{Breiman01}.  

Our proposal, {\it leave-one-covariate-out} or LOCO inference,
proceeds as follows.
Denote by \smash{$\hat\mu$} our estimate of the mean function, 
fit on data $(X_i,Y_i)$, $i \in \mathcal I_1$ for some $\mathcal I_1 
\subseteq \{1,\ldots,n\}$. To investigate the importance of the 
$j$th covariate, we refit our estimate of the mean  
function on the data set $(X_i(-j),Y_i)$, $i \in \mathcal I_1$, where  
in each $X_i(-j) = (X_i(1),\ldots,X_i(j-1),X_i(j+1),\ldots,X_i(d)) \in
\R^{d-1}$, we have removed the $j$th covariate.  Denote
by \smash{$\hat\mu_{(-j)}$} this refitted mean function, and denote 
the excess prediction error of covariate $j$, at a new 
i.i.d.\ draw $(X_{n+1},Y_{n+1})$, by
\begin{equation*}
\Delta_j(X_{n+1},Y_{n+1}) = |Y_{n+1}-\hat\mu_{(-j)}(X_{n+1})| -
|Y_{n+1}-\hat\mu(X_{n+1})|.
\end{equation*}
The random variable
$\Delta_j(X_{n+1},Y_{n+1})$ measures the increase in prediction error
due to not having access to covariate $j$ in our data set, and
will be the basis for inferential statements about variable
importance.  
There are two ways to look at $\Delta_j(X_{n+1},Y_{n+1})$, as
discussed below. 

\subsection{Local Measure of Variable Importance} 

Using conformal prediction bands, we can construct a valid
prediction interval for the random variable $\Delta_j(X_{n+1},Y_{n+1})$, as
follows.
Let $C$ denote a conformal prediction set for $Y_{n+1}$
given $X_{n+1}$, having coverage $1-\alpha$, constructed from either
the full or split methods---in the former, the index set used for the
fitting of \smash{$\hat\mu$} and \smash{$\hat\mu_{(-j)}$} is 
$\mathcal I_1 = \{1,\ldots,n\}$, and in the 
latter, it is $\mathcal I_1 \subsetneq \{1,\ldots,n\}$, a
proper subset (its complement $\mathcal I_2$ is used for computing the
appropriate sample quantile of residuals). Now define 
\begin{equation*}
W_j(x) = \left\{ |y-\hat\mu_{(-j)}(x)| - |y-\hat\mu(x)|\;:\; 
y\in C(x) \right\}. 
\end{equation*}
From the finite-sample validity of $C$, we immediately have
\begin{equation}
\label{eq:wj_valid}
\P\big( \Delta_j(X_{n+1},Y_{n+1}) \in W_j(X_{n+1}), \; \text{for all
  $j=1,\ldots,d$} \big) \geq 1-\alpha.
\end{equation}
It is important to emphasize that the prediction sets 
$W_1,\ldots, W_d$ are valid in finite-sample, without distributional 
assumptions. Furthermore, they are uniformly valid over $j$, and there is no
need to do any multiplicity adjustment.  One can decide to construct
$W_j(X_{n+1})$ 
at a single fixed $j$, at all $j=1,\ldots,d$, or at a randomly chosen
$j$ (say, the result of a variable selection procedure on the given
data $(X_i,Y_i)$, $i=1,\ldots,n$), and in each case the interval(s)
will have proper coverage.

As with the guarantees from conformal inference, the coverage
statement \eqref{eq:wj_valid} is marginal over $X_{n+1}$, and in
general, does not hold conditionally at $X_{n+1}=x$.  But, to
summarize the effect of covariate $j$, we can still plot the intervals 
$W_j(X_i)$ for $i=1,\ldots, n$, and loosely interpret these as making
local statements about variable importance. 

\begin{figure}[htb]
\centering
\includegraphics[width=0.32\textwidth]{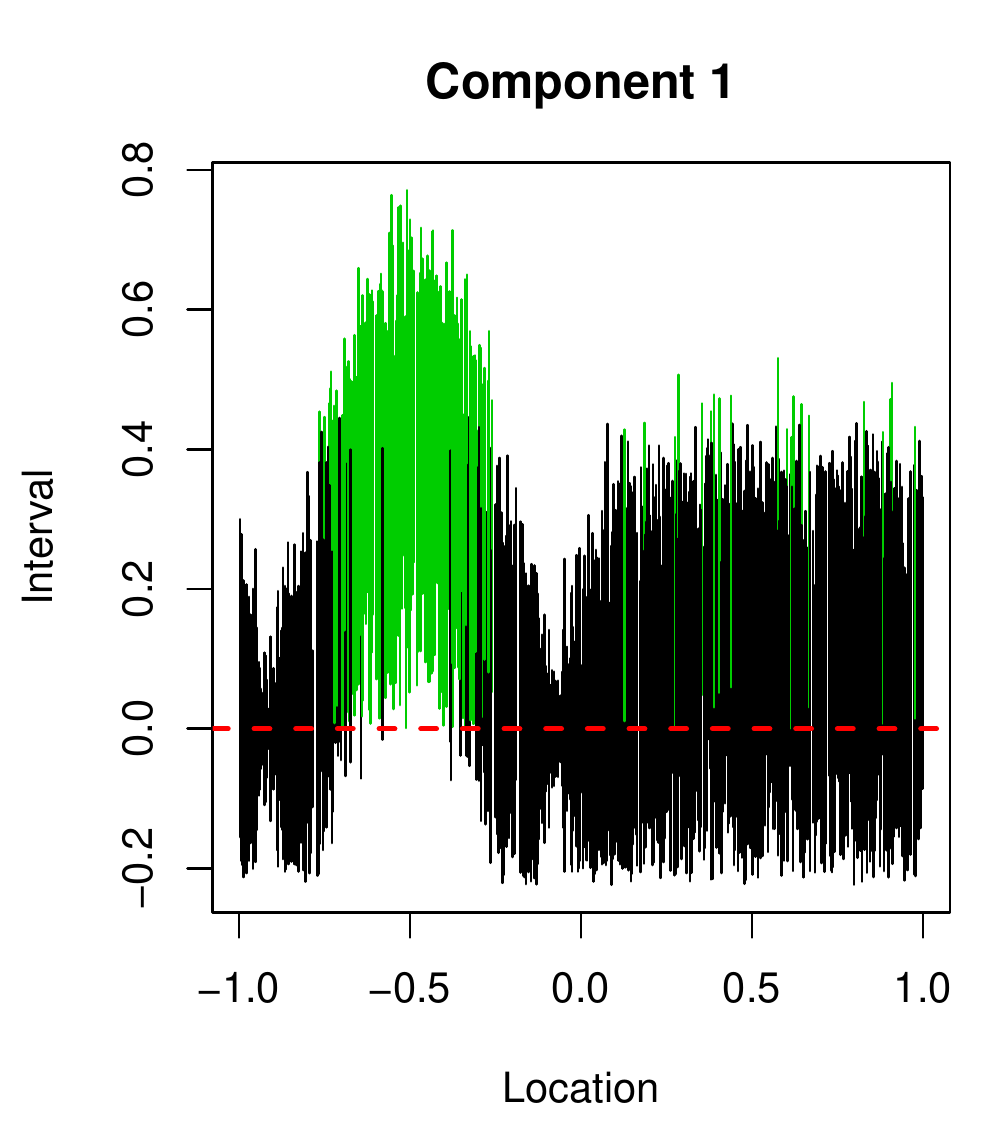}
\includegraphics[width=0.32\textwidth]{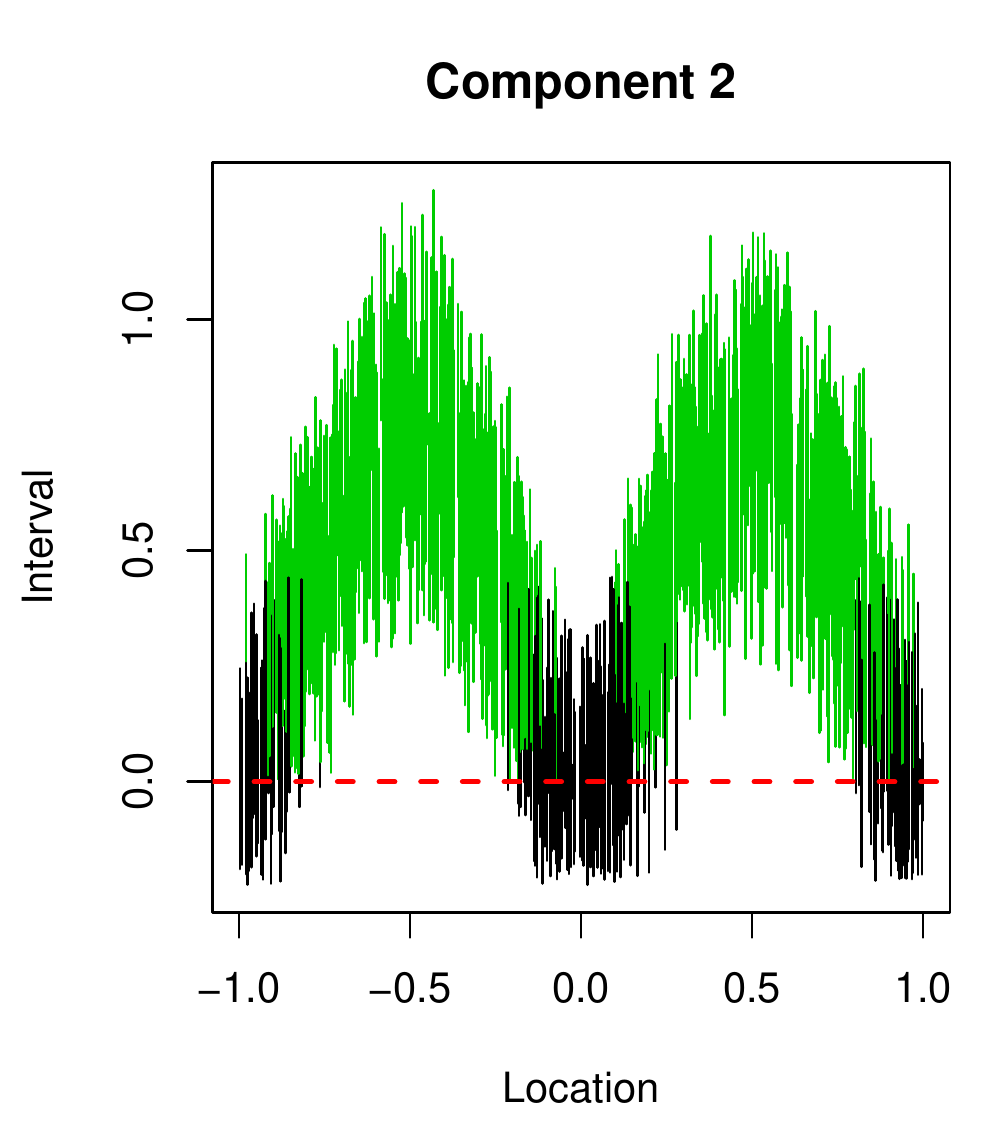}
\includegraphics[width=0.32\textwidth]{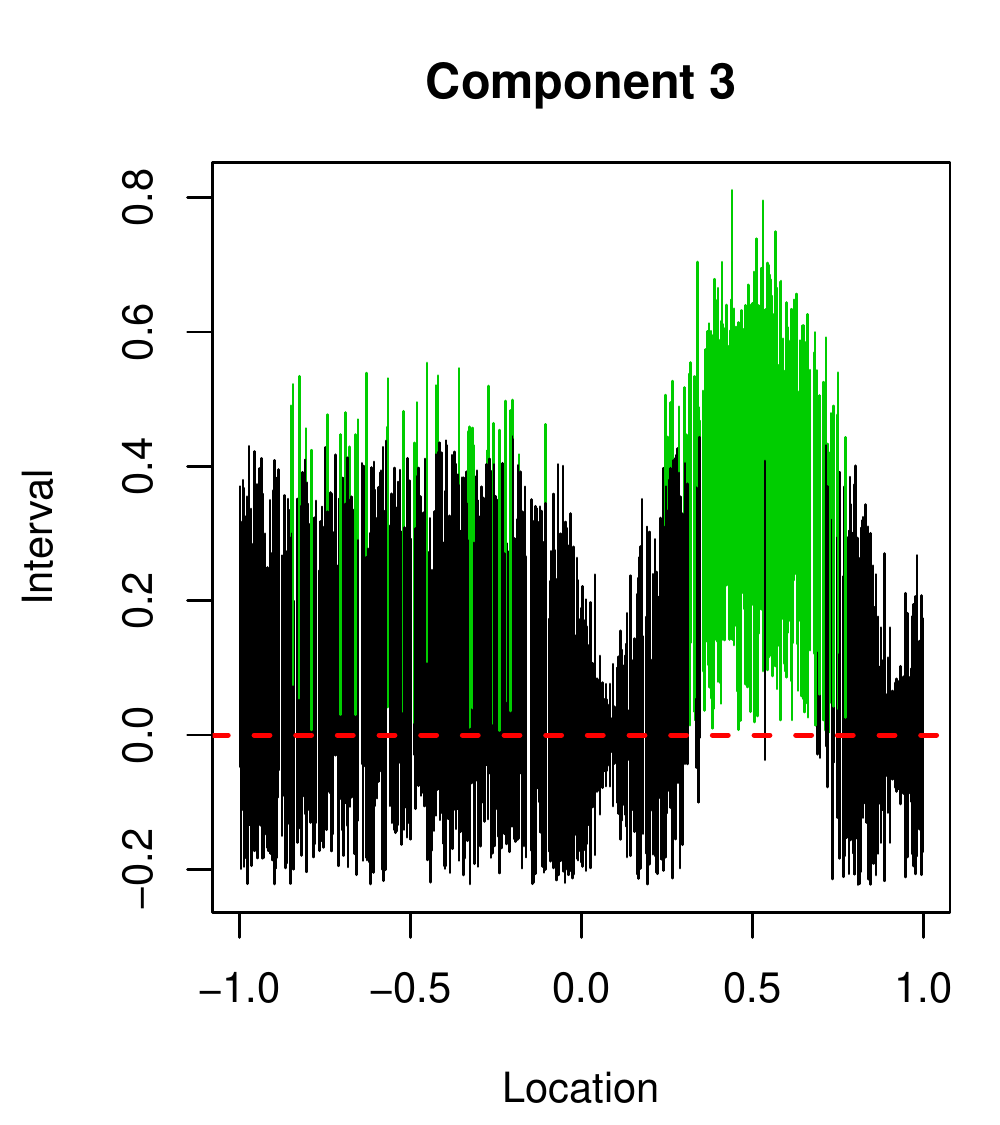} \\
\includegraphics[width=0.32\textwidth]{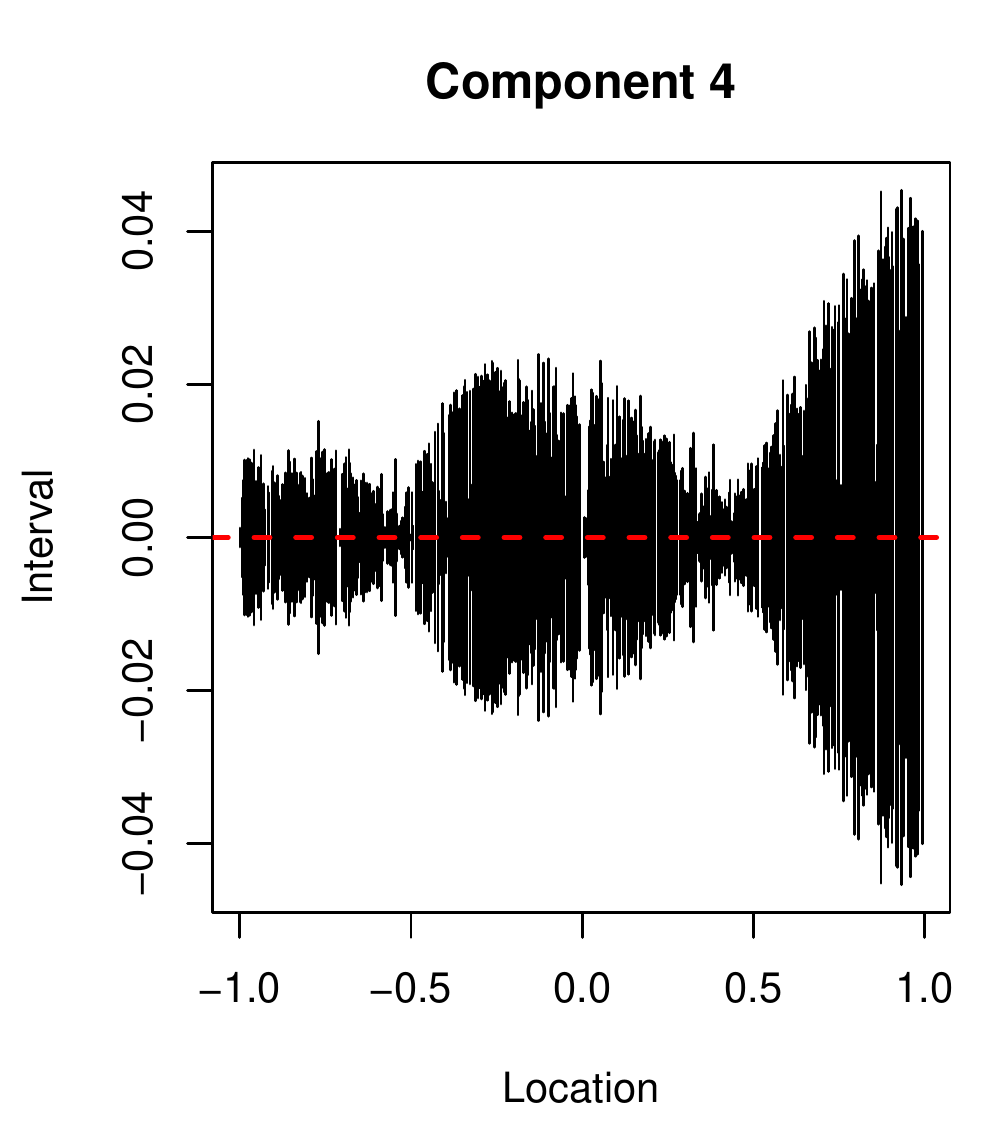}
\includegraphics[width=0.32\textwidth]{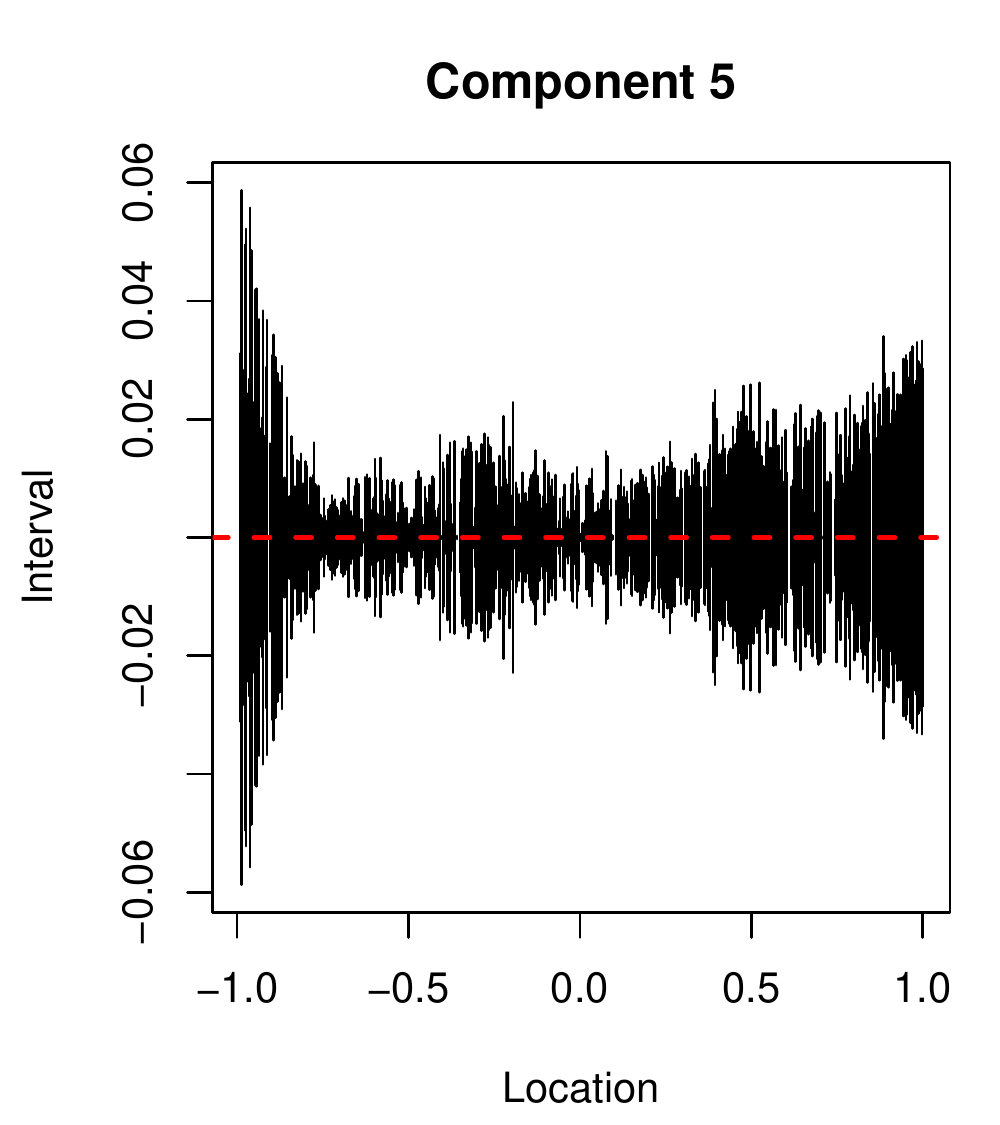}
\includegraphics[width=0.32\textwidth]{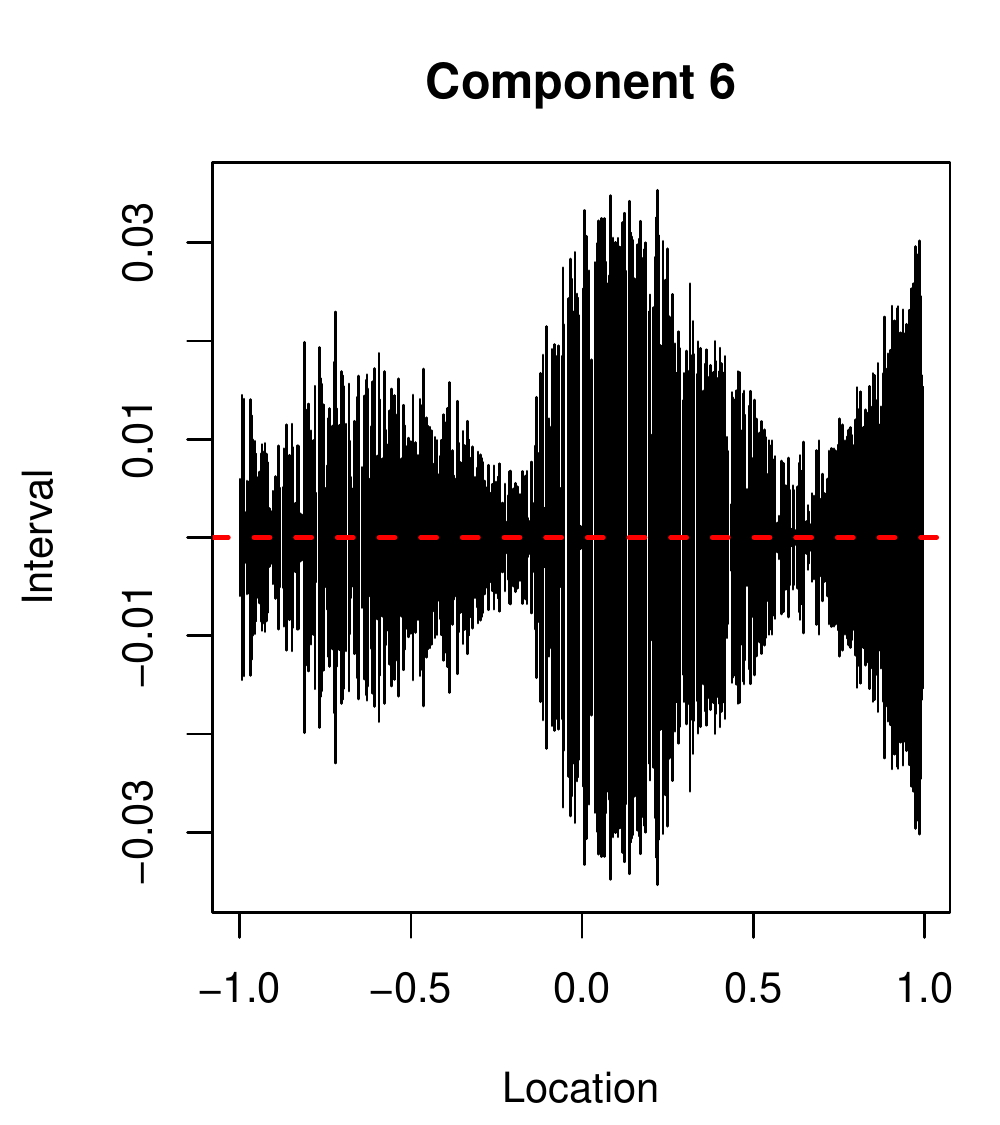}
\caption{\it In-sample prediction intervals for $\Delta_j(X_i)$ across
  all covariates $j=1,\ldots,6$ and samples $i=1,\ldots,1000$, in an 
  additive model setting described in the text. Each interval that lies
  strictly above zero is colored in green.}
\label{fig:loco_local}
\end{figure}

We illustrate this idea in a low-dimensional additive model, where
$d=6$ and the mean function is $\mu(x)=\sum_{j=1}^6 f_j(x(j))$, with
$f_1(t)=\sin(\pi(1+t))\one\{t<0\}$, $f_2(t)=\sin(\pi t)$,
$f_3(t)=\sin(\pi(1+t))\one\{t>0\}$, and $f_4=f_5=f_6=0$. We generated 
$n=1000$ i.i.d pairs $(X_i,Y_i)$, $i=1,\ldots,1000$, where each $X_i
\sim \mathrm{Unif}[-1,1]^d$ and $Y_i=\mu(X_i)+\epsilon_i$ for
$\epsilon_i \sim N(0,1)$.  We then computed each interval $W_j(X_i)$
using the ROO split conformal technique at the miscoverage level
$\alpha=0.1$, using an additive model as the base estimator (each
component modeled by a spline with 5 degrees of freedom). The
intervals are plotted in \Cref{fig:loco_local}.  We can see that many
intervals for components $j=1,2,3$ lie strictly above zero, indicating
that leaving out such covariates is damaging to the predictive accuracy
of the estimator.  Furthermore, the locations at which these intervals
lie above zero are precisely locations at which the underlying
components $f_1,f_2,f_3$ deviate significantly from zero.  On the
other hand, the intervals for components $j=4,5,6$ all contain zero,
as expected.

\subsection{Global Measures of Variable Importance}

For a more global measure of variable importance, we can focus 
on the distribution of $\Delta_j(X_{n+1},Y_{n+1})$, marginally over $(X_{n+1},Y_{n+1})$. We rely on a
splitting approach, where the index set used for the
training of \smash{$\hat\mu$} and \smash{$\hat\mu_{(-j)}$} is
$\mathcal I_1 \subsetneq \{1,\ldots,n\}$, a proper subset.  Denote by
$\mathcal I_2$ its complement, and by $\mathcal D_k = \{(X_i,Y_i):
i \in \mathcal I_k\}$, $k=1,2$ the data samples in each index
set. Define
\begin{equation*}
G_j(t) = \P\Big(\Delta_j(X_{n+1},Y_{n+1})\leq t \;\big|\;
\mathcal D_1 \Big), \; t \in \R,
\end{equation*}
the distribution function of $\Delta_j(X_{n+1},Y_{n+1})$ conditional
on the data $\mathcal D_1$ in the first half of the data-split.  We
will now infer parameters of $G_j$ such as its mean $\theta_j$ or
median $m_j$. For the former parameter, 
\begin{equation*}
\theta_j = \E\Big[\Delta_j(X_{n+1},Y_{n+1}) \;\big|\; \mathcal D_1\Big],  
\end{equation*}
we can obtain the asymptotic $1-\alpha$ confidence interval
\begin{equation*}
\bigg[\hat\theta_j - \frac{z_{\alpha/2}s_j}{\sqrt{n/2}}, \; 
\hat\theta_j + \frac{z_{\alpha/2}s_j}{\sqrt{n/2}}\bigg],
\end{equation*}
where \smash{$\hat\theta_j = (n/2)^{-1}
\sum_{i\in\mathcal I_2} \Delta_j(X_i,Y_i)$} is the sample mean,
$s_j^2$ is the analogous sample variance, measured on $\mathcal D_2$,
and $z_{\alpha/2}$ is the $1-\alpha/2$ quantile of the standard normal
distribution. Similarly, we can perform a one-sided hypothesis test of 
\begin{equation*}
H_0: \theta_j \leq 0 \quad\text{versus}\quad H_1: \theta_j > 0 
\end{equation*}
by rejecting when \smash{$\sqrt{n/2}\cdot\hat\theta_j/s_j> z_\alpha$}.  
Although these inferences are asymptotic, the convergence to its
asymptotic limit is uniform (say, as governed by the Berry-Esseen
Theorem) and independent of the feature dimension
$d$ (since $\Delta_j(X_{n+1},Y_{n+1})$ is always univariate).
To control for multiplicity, we suggest replacing $\alpha$ in the
above with $\alpha /|S|$ where $S$ is the set of variables whose 
importance is to be tested.  

Inference for the parameter $\theta_j$ requires existence of the first
and second moments for the error term. In practice it may be more
stable to consider the median parameter 
$$
m_j = \mathrm{median}\Big[\Delta_j(X_{n+1},Y_{n+1}) \;\big|\; 
\mathcal D_1\Big].
$$
We can conduct nonasymptotic inferences about $m_j$ using standard,
nonparametric tests such as the sign test or the Wilcoxon signed-rank
test, applied to $\Delta_j(X_i,Y_i)$, $i \in \mathcal I_2$. This
allows us to test   
\begin{equation*}
H_0: m_j \leq 0 \quad\text{versus}\quad H_1: m_j > 0 
\end{equation*}
with finite-sample validity under essentially no assumptions on the
distribution $G_j$ (the sign test only requires continuity, and the
Wilcoxon test requires continuity and symmetry). Confidence intervals for
$m_j$ can be obtained by inverting the (two-sided) versions of the
sign and Wilcoxon tests, as well.
Again, we suggest replacing $\alpha$ with $\alpha/|S|$ to adjust for
multiplicity, where $S$ is the set of variables to be tested.

\begin{figure}[htb]
\centering
\includegraphics[width=0.65\textwidth]{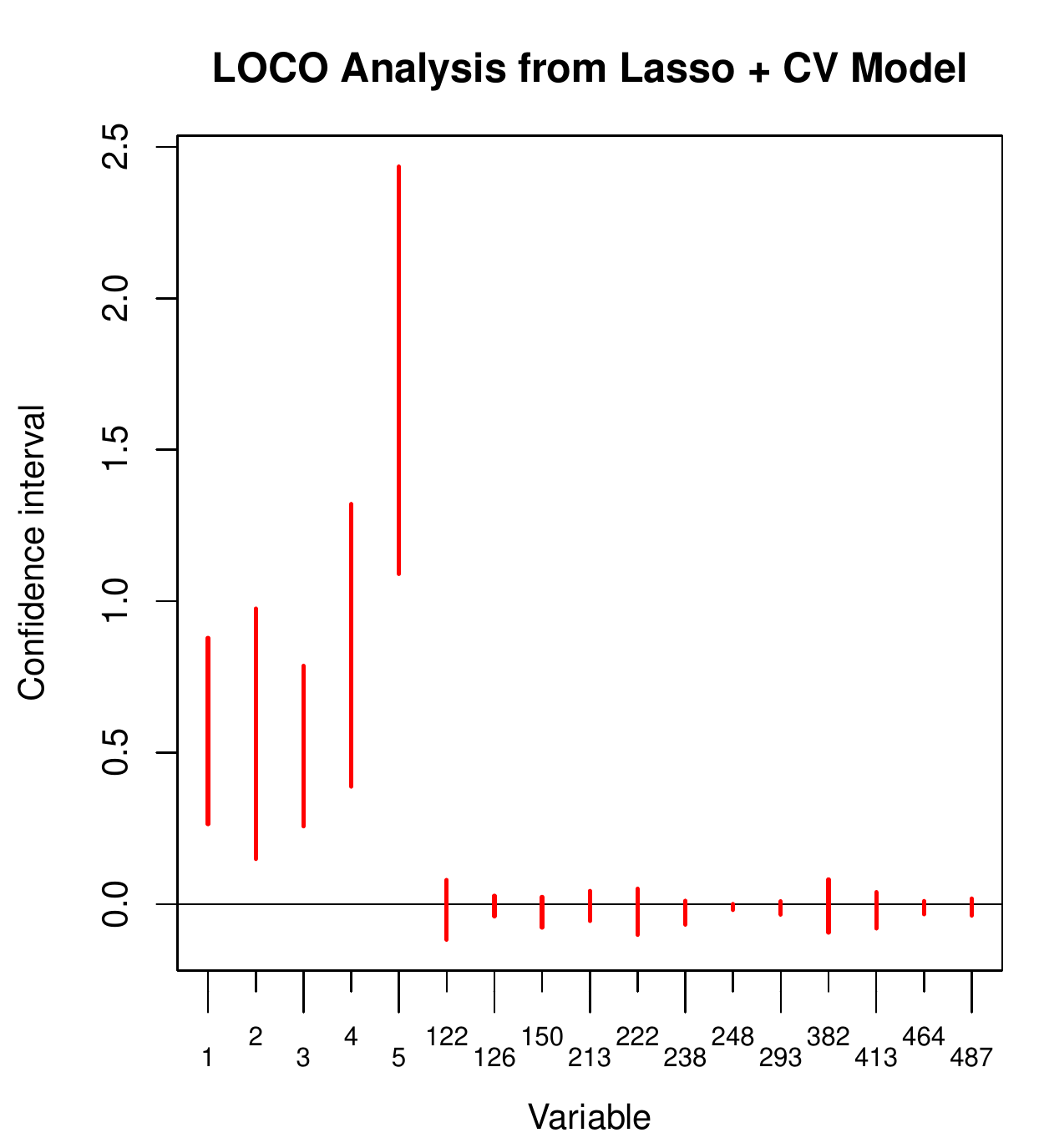}
\caption{\it Wilcoxon-based confidence intervals for the median excess
  test error due to variable dropping, for all selected variables in a 
  high-dimensional linear regression example with $n=200$ and $d=500$
  described in the text.}
\label{fig:loco_global}
\end{figure}

We finish with an example of a high-dimensional linear regression
problem with $n=200$ observations and $d=500$ variables.  The mean
function $\mu(x)$ was defined to be a linear function of
$x(1),\ldots,x(5)$ only,
with coefficients drawn i.i.d.\ from $N(0,4)$.  We drew $X_i(j) \sim
N(0,1)$ independently across all $i=1,\ldots,200$ and
$j=1,\ldots,500$, and then defined the responses by
$Y_i=\mu(X_i)+\epsilon_i$, for $\epsilon_i \sim N(0,1)$,
$i=1,\ldots,200$.  A single data-split was applied, and the on the
first half we fit the lasso estimator \smash{$\hat\mu$} with the
tuning parameter $\lambda$ chosen by 10-fold cross-validation.  The
set of active predictors $S$ was collected, which had size $|S|=17$;
the set $S$ included the 5 truly relevant variables, but also 12
irrelevant ones.  We then refit the lasso estimator
\smash{$\hat\mu_{(-j)}$} using the same cross-validation, with covariate $j$
excluded, for each $j \in S$.   On the second half of the data, we
applied the Wilcoxon rank sum test to compute confidence intervals for
the median excess test error due to variable dropping, $m_j$, for each 
$j \in S$.  These intervals were properly corrected for multiplicity:
each was computed at the level $1-0.1/17$ in order to obtain a
simultaneous level $1-0.1=0.9$ of coverage.  Figure
\ref{fig:loco_global} shows the results.  We can see that the
intervals for the first 5 variables are well above zero, and those for
the next 12 all hover around zero, as desired.

The problem of inference after model selection is an important but
also subtle topic and we are only dealing with the issue briefly
here. In a future paper we will thoroughly compare several approaches
including LOCO.  

%

\section{Conclusion}
\label{section::conclusion}

Current high-dimensional inference methods make strong
assumptions while little is known about their
robustness against model misspecification.  We have shown that if
we focus on prediction bands, almost all existing point estimators can be
used to build valid prediction bands, even when the model is grossly
misspecified, as long as the data are i.i.d.  Conformal inference
is similar to the jackknife, bootstrap, and cross-validation in the use of
symmetry of data.  A remarkable difference in conformal inference is
its ``out-of-sample fitting''.  That is, unlike most existing
prediction methods which fit a model using the training sample and
then apply the fitted model to any new data points for prediction,
the full conformal method refits the model each time when a new
prediction is requested at a new value $X_{n+1}$.  An important and
distinct consequence of such an ``out-of-sample fitting'' is the
guaranteed finite-sample coverage property.

The distribution-free coverage offered by conformal intervals is 
marginal.  The conditional coverage may be larger than
$1-\alpha$ at some values of $X_{n+1}=x$ and smaller than $1-\alpha$
at other values. This should not be viewed as a disadvantage of  
conformal inference, as the statistical accuracy of the
conformal prediction band is strongly tied to the base estimator.  In  
a sense, conformal inference broadens the scope and value of any
regression estimator at nearly no cost: if the estimator is
accurate (which usually requires an approximately correctly specified
model, and a proper choice of tuning parameter), then the conformal 
prediction band is 
near-optimal; if the estimator is bad, then we still have valid marginal
coverage.  As a result, it makes sense to use a conformal prediction
band as a diagnostic and comparison tool for regression function
estimators.

There are many directions in conformal inference that are worth
exploring.  Here we give a short list.  First, it would be interesting
to better understand the trade-off between the full and split 
conformal methods.
The split conformal method is fast, but at the cost of less accurate
inference.  Also, in practice it would be desirable to reduce the
additional randomness caused by splitting the data.  In this paper we
showed that aggregating results from multiple splits (using a 
Bonferonni-type correction) leads to wider bands. It would be practically
appealing to develop novel methods that more efficiently combine
results from multiple splits.  Second, it would be interesting to see
how conformal inference can help with model-free variable selection.
Our leave-one-covariate-out (LOCO) method is a first step in this
direction.  However, the current version of LOCO based on excess
prediction error can only be implemented with the split conformal
method due to computational reasons.  When split conformal is used,
the inference is then conditional on the model fitted in the first
half of the data.  The effect of random splitting inevitably raises an
issue of selective inference, which needs to be appropriately
addressed. In a future paper, we will report on
detailed comparisons of LOCO with other approaches to high-dimensional inference.



\bibliographystyle{apa-good}
\bibliography{paper}

\begin{thebibliography}{34}
\expandafter\ifx\csname natexlab\endcsname\relax\def\natexlab#1{#1}\fi
\expandafter\ifx\csname url\endcsname\relax
  \def\url#1{{\tt #1}}\fi
\expandafter\ifx\csname urlprefix\endcsname\relax\def\urlprefix{URL }\fi

\bibitem[{Belloni et~al.(2012)Belloni, Chen, Chernozhukov, \&
  Hansen}]{BelloniCCH12}
Belloni, A., Chen, D., Chernozhukov, V., \& Hansen, C. (2012).
\newblock Sparse models and methods for optimal instruments with an application
  to eminent domain.
\newblock {\em Econometrica\/}, {\em 80\/}(6), 2369--2429.

\bibitem[{Berk et~al.(2013)Berk, Brown, Buja, Zhang, \& Zhao}]{posi}
Berk, R., Brown, L., Buja, A., Zhang, K., \& Zhao, L. (2013).
\newblock Valid post-selection inference.
\newblock {\em Annals of Statistics\/}, {\em 41\/}(2), 802--837.

\bibitem[{Bickel et~al.(2009)Bickel, Ritov, \& Tsybakov}]{BickelRT09}
Bickel, P.~J., Ritov, Y., \& Tsybakov, A.~B. (2009).
\newblock Simultaneous analysis of lasso and dantzig selector.
\newblock {\em The Annals of Statistics\/}, (pp. 1705--1732).

\bibitem[{Breiman(2001)}]{Breiman01}
Breiman, L. (2001).
\newblock Random forests.
\newblock {\em Machine Learning\/}, {\em 45\/}(1), 5--32.

\bibitem[{Buhlmann(2013)}]{buhlsignif}
Buhlmann, P. (2013).
\newblock Statistical significance in high-dimensional linear models.
\newblock {\em Bernoulli\/}, {\em 19\/}(4), 1212--1242.

\bibitem[{Buja et~al.(2014)Buja, Berk, Brown, George, Pitkin, Traskin, Zhang,
  \& Zhao}]{Buja14Models}
Buja, A., Berk, R., Brown, L., George, E., Pitkin, E., Traskin, M., Zhang, K.,
  \& Zhao, L. (2014).
\newblock Models as approximations: How random predictors and model violations
  invalidate classical inference in regression.
\newblock ArXiv: 1404.1578.

\bibitem[{Bunea et~al.(2007)Bunea, Tsybakov, \& Wegkamp}]{BuneaTW07}
Bunea, F., Tsybakov, A., \& Wegkamp, M. (2007).
\newblock Sparsity oracle inequalities for the lasso.
\newblock {\em Electronic Journal of Statistics\/}, {\em 1\/}, 169--194.

\bibitem[{Burnaev \& Vovk(2014)}]{burnaev2014efficiency}
Burnaev, E., \& Vovk, V. (2014).
\newblock Efficiency of conformalized ridge regression.
\newblock {\em Proceedings of the Annual Conference on Learning Theory\/}, {\em
  25\/}, 605--622.

\bibitem[{Butler \& Rothman(1980)}]{Rothman}
Butler, R., \& Rothman, E. (1980).
\newblock Predictive intervals based on reuse of the sample.
\newblock {\em Journal of the American Statistical Association\/}, {\em
  75\/}(372), 881--889.

\bibitem[{Efroymson(1960)}]{stepwise}
Efroymson, M.~A. (1960).
\newblock Multiple regression analysis.
\newblock In {\em Mathematical Methods for Digital Computers\/}, vol.~1, (pp.
  191--203). Wiley.

\bibitem[{Fithian et~al.(2014)Fithian, Sun, \& Taylor}]{optimalinf}
Fithian, W., Sun, D., \& Taylor, J. (2014).
\newblock Optimal inference after model selection.
\newblock ArXv: 1410.2597.

\bibitem[{Hebiri(2010)}]{Hebiri10}
Hebiri, M. (2010).
\newblock Sparse conformal predictors.
\newblock {\em Statistics and Computing\/}, {\em 20\/}(2), 253--266.

\bibitem[{Javanmard \& Montanari(2014)}]{montahypo2}
Javanmard, A., \& Montanari, A. (2014).
\newblock Confidence intervals and hypothesis testing for high-dimensional
  regression.
\newblock {\em Journal of Machine Learning Research\/}, {\em 15\/}, 2869--2909.

\bibitem[{Lee et~al.(2016)Lee, Sun, Sun, \& Taylor}]{exactlasso}
Lee, J., Sun, D., Sun, Y., \& Taylor, J. (2016).
\newblock Exact post-selection inference, with application to the lasso.
\newblock {\em Annals of Statistics\/}, {\em 44\/}(3), 907--927.

\bibitem[{Lei(2014)}]{Lei14}
Lei, J. (2014).
\newblock Classification with confidence.
\newblock {\em Biometrika\/}, {\em 101\/}(4), 755--769.

\bibitem[{Lei et~al.(2015)Lei, Rinaldo, \& Wasserman}]{LeiRW14}
Lei, J., Rinaldo, A., \& Wasserman, L. (2015).
\newblock A conformal prediction approach to explore functional data.
\newblock {\em Annals of Mathematics and Artificial Intelligence\/}, {\em
  74\/}(1), 29--43.

\bibitem[{Lei et~al.(2013)Lei, Robins, \& Wasserman}]{LeiRW13}
Lei, J., Robins, J., \& Wasserman, L. (2013).
\newblock Distribution free prediction sets.
\newblock {\em Journal of the American Statistical Association\/}, {\em 108\/},
  278--287.

\bibitem[{Lei \& Wasserman(2014)}]{LeiW14}
Lei, J., \& Wasserman, L. (2014).
\newblock Distribution-free prediction bands for non-parametric regression.
\newblock {\em Journal of the Royal Statistical Society: Series B\/}, {\em
  76\/}(1), 71--96.

\bibitem[{Meinshausen \& Buhlmann(2010)}]{stabselect}
Meinshausen, N., \& Buhlmann, P. (2010).
\newblock Stability selection.
\newblock {\em Journal of the Royal Statistical Society: Series B\/}, {\em
  72\/}(4), 417--473.

\bibitem[{Papadopoulos et~al.(2002)Papadopoulos, Proedrou, Vovk, \&
  Gammerman}]{papadopoulos2002inductive}
Papadopoulos, H., Proedrou, K., Vovk, V., \& Gammerman, A. (2002).
\newblock Inductive confidence machines for regression.
\newblock In {\em Machine Learning: ECML 2002\/}, (pp. 345--356). Springer.

\bibitem[{Ravikumar et~al.(2009)Ravikumar, Lafferty, Liu, \&
  Wasserman}]{Ravikumar09}
Ravikumar, P., Lafferty, J., Liu, H., \& Wasserman, L. (2009).
\newblock Sparse additive models.
\newblock {\em Journal of the Royal Statistical Society: Series B\/}, {\em
  71\/}(5), 1009--1030.

\bibitem[{Steinberger \& Leeb(2016)}]{steinberger2016}
Steinberger, L., \& Leeb, H. (2016).
\newblock Leave-one-out prediction intervals in linear regression models with
  many variables.
\newblock ArXiv: 1602.05801.

\bibitem[{Thakurta \& Smith(2013)}]{ThakurtaS13}
Thakurta, A.~G., \& Smith, A. (2013).
\newblock Differentially private feature selection via stability arguments, and
  the robustness of the lasso.
\newblock In {\em Conference on Learning Theory\/}, (pp. 819--850).

\bibitem[{Tian \& Taylor(2015{\natexlab{a}})}]{tian2015asymptotics}
Tian, X., \& Taylor, J. (2015{\natexlab{a}}).
\newblock Asymptotics of selective inference.
\newblock ArXiv: 1501.03588.

\bibitem[{Tian \& Taylor(2015{\natexlab{b}})}]{tian2015selective}
Tian, X., \& Taylor, J. (2015{\natexlab{b}}).
\newblock Selective inference with a randomized response.
\newblock ArXiv: 1507.06739.

\bibitem[{Tibshirani(1996)}]{lasso}
Tibshirani, R. (1996).
\newblock Regression shrinkage and selection via the lasso.
\newblock {\em Journal of the Royal Statistical Society: Series B\/}, {\em
  58\/}(1), 267--288.

\bibitem[{Tibshirani et~al.(2016)Tibshirani, Taylor, Lockhart, , \&
  Tibshirani}]{exactlar}
Tibshirani, R.~J., Taylor, J., Lockhart, R., , \& Tibshirani, R. (2016).
\newblock Exact post-selection inference for sequential regression procedures.
\newblock {\em Journal of the American Statistical Association\/}, {\em
  111\/}(514), 600--620.

\bibitem[{van~de Geer et~al.(2014)van~de Geer, Buhlmann, Ritov, \&
  Dezeure}]{vdgsignif}
van~de Geer, S., Buhlmann, P., Ritov, Y., \& Dezeure, R. (2014).
\newblock On asymptotically optimal confidence regions and tests for
  high-dimensional models.
\newblock {\em Annals of Statistics\/}, {\em 42\/}(3), 1166--1201.

\bibitem[{Vovk(2013)}]{VVV}
Vovk, V. (2013).
\newblock Conditional validity of inductive conformal predictors.
\newblock {\em Machine Learning\/}, {\em 92\/}, 349--376.

\bibitem[{Vovk et~al.(2005)Vovk, Gammerman, \& Shafer}]{VovkGS05}
Vovk, V., Gammerman, A., \& Shafer, G. (2005).
\newblock {\em Algorithmic Learning in a Random World\/}.
\newblock Springer.

\bibitem[{Vovk et~al.(2009)Vovk, Nouretdinov, \& Gammerman}]{vovk2009line}
Vovk, V., Nouretdinov, I., \& Gammerman, A. (2009).
\newblock On-line predictive linear regression.
\newblock {\em The Annals of Statistics\/}, {\em 37\/}(3), 1566--1590.

\bibitem[{Wasserman(2014)}]{wass2014}
Wasserman, L. (2014).
\newblock Discussion: {A} significance test for the lasso.
\newblock {\em Annals of Statistics\/}, {\em 42\/}(2), 501--508.

\bibitem[{Zhang \& Zhang(2014)}]{zhangconf}
Zhang, C.-H., \& Zhang, S. (2014).
\newblock Confidence intervals for low dimensional parameters in high
  dimensional linear models.
\newblock {\em Journal of the Royal Statistical Society: Series B\/}, {\em
  76\/}(1), 217--242.

\bibitem[{Zou \& Hastie(2005)}]{enet}
Zou, H., \& Hastie, T. (2005).
\newblock Regularization and variable selection via the elastic net.
\newblock {\em Journal of the Royal Statistical Society: Series B\/}, {\em
  67\/}(2), 301--320.

\end{thebibliography}

\newpage

\appendix


\section{Technical Proofs}\label{sec:proof}

\subsection{Proofs for \Cref{sec::method}}
\label{sec:proof_1}

\begin{proof}[Proof of \Cref{thm:conf_valid}]
The first part (the lower bound) comes directly from the
definition of the conformal interval in \eqref{eq:conf_int}, and the 
(discrete) p-value property in \eqref{eq:pi_pvalue}.  We focus on the
second part (upper bound).  Define $\alpha'=\alpha-1/(n+1)$. 
By assuming a continuous joint distribution of the fitted 
residuals, we know that the values \smash{$R_{y,1},\ldots,R_{y,n+1}$}
are all distinct with probability one. The set
$C_{\mathrm{conf}}$ in \eqref{eq:conf_int}
is equivalent to the set of all points $y$ such that $R_{y,n+1}$ ranks  
among the $\lceil(n+1)(1-\alpha)\rceil$ smallest of all
$R_{y,1},\ldots,R_{y,n+1}$.  Consider now the set 
$D(X_{n+1})$ consisting of points $y$ such
that $R_{y,n+1}$ is among the $\lceil(n+1)\alpha'\rceil$ largest.   
Then by construction 
\begin{equation*}
\P\big(Y_{n+1} \in D(X_{n+1})\big) \geq \alpha',
\end{equation*}
and yet $C_{\mathrm{conf}}(X_{n+1})\cap D(X_{n+1}) = \emptyset$, which
implies the result.  
\end{proof}

\begin{proof}[Proof of \Cref{thm:split_valid}]
The first part (lower bound) follows directly by symmetry between the
residual at $(X_{n+1},Y_{n+1})$ and those at $(X_i,Y_i)$, $i \in
\mathcal I_2$. We prove the upper bound in the second part.  Assuming
a continuous joint distribution of residual, and hence no ties, the 
set \smash{$C_{\mathrm{split}}(X_{n+1})$} excludes
the values of $y$ such that \smash{$|y-\hat\mu(X_{n+1})|$} is among
the $(n/2)-\lceil(n/2 +1)(1-\alpha)\rceil$ largest in 
$\{R_i : i\in\mathcal I_2\}$. Denote the set of these excluded points
as $D(X_{n+1})$. Then again by symmetry,
\begin{equation*}
\P\big(Y_{n+1} \in D(X_{n+1})\big) \geq
\frac{(n/2)-\lceil(n/2 +1)(1-\alpha)\rceil}{n/2+1} \geq
\alpha-2/(n+2),
\end{equation*}
which completes the proof.
\end{proof}

\begin{proof}[Proof of \Cref{thm:mult_split}]
Without loss of generality, we assume that the sample size is $2n$.
The individual split conformal interval has length infinity if
$\alpha/N < 1/n$. Therefore, we only need to consider $2\le N\le
\alpha n\le n$.  Also, in this proof we will ignore all rounding
issues by directly working with the empirical quantiles.  The
differences caused by rounding are negligible.
  
For $j=1,\ldots,N$, the $j$th
split conformal prediction band at $X$, \smash{$C_{{\rm
      split},j}(X)$}, is an interval with half-width \smash{$\hat
  F^{-1}_{n,j}(1-\alpha/N)$}, where \smash{$\hat F_{n,j}$} is the
empirical CDF of fitted absolute residuals in the ranking subsample in 
the $j$th split.
  
We focus on the event 
$$
\left\{\max_{j=1,\ldots,N}\|\hat
\mu_j-\tilde\mu\|_\infty< \eta_n\right\}, 
$$
which has probability 
at least $1-N\rho_n \ge 1-n\rho_n\rightarrow 1$.  On this event, 
the length of \smash{$C^{(N)}_{\rm split}(X)$} is at least 
$$
2 \min_{j=1,\ldots,N} \tilde F_{n,j}^{-1}(1-\alpha/N) - 2\eta_n, 
$$
where \smash{$\tilde F_{n,j}$} is the empirical CDF of the absolute
residuals about \smash{$\tilde\mu$} in the ranking subsample in the
$j$th split.   
  
Note that the split conformal band \smash{$C_{{\rm split},1}(X)$} from 
a single split has length no more than 
\smash{$2 \tilde 
F_{n,1}^{-1}(1-\alpha)+2\eta_n$} on the event we focus on.  Therefore,  
it suffices to show that
\begin{equation}\label{eq:mult_vs_single}
\P\left(\tilde F_{n,1}^{-1}(1-\alpha)< 
\tilde F^{-1}_{n,j}(1-\alpha/N)-2\eta_n, \; j=1,\ldots,N \right)
\rightarrow 1. 
\end{equation}

Let \smash{$\tilde F$} be the CDF of $|Y-\tilde\mu(X)|$.  
Note that it is \smash{$\tilde F_{n,j}$}, instead of \smash{$\hat
  F_{n,j}$}, that corresponds to \smash{$\tilde F$}.
By the Dvoretzky-Kiefer-Wolfowitz inequality, we have
\begin{align*}
\P\left(\tilde F^{-1}_{n,j}(1-\alpha/N)\le \tilde F^{-1}(1-\alpha/1.6)\right) & \le
\P\left(\|\tilde F_{n,j}-\tilde F\|_\infty \ge \alpha(1/1.6-1/N)\right)\\
  &\le  \P\left(\|\tilde F_{n,j}-\tilde F\|_\infty \ge \alpha/8\right)\\
  &\le  2\exp(-n \alpha^2/32).
\end{align*}
Using a union bound,
\begin{align*}
\P\left( \min_{j=1,\ldots,N} \tilde F_{n,j}^{-1}(1-\alpha/N)\le 
\tilde F^{-1}(1-\alpha/1.6)\right)\le 2N \exp(-n\alpha^2/32).
\end{align*}
On the other hand,
\begin{align*}
\P\left(\tilde F_{n,1}^{-1}(1-\alpha)\ge\tilde F^{-1}(1-\alpha/1.4)\right)
&\le \P\left(\|\tilde F_{n,1}-\tilde F\|_\infty \ge \alpha(1-1/1.4)\right)\le
2\exp(-n\alpha^2/8).
\end{align*}
So with probability at least
$1-2\exp(-n\alpha^2/8)-2N\exp(-n\alpha^2/32)$ we have 
$$
\min_{j=1,\ldots,N}\tilde F_{n,j}^{-1}(1-\alpha/N) - \tilde F_{n,1}^{-1}(1-\alpha)
\ge \tilde F^{-1}(1-\alpha/1.6) - \tilde F^{-1}(1-\alpha/1.4)>0.
$$
Therefore we conclude \eqref{eq:mult_vs_single}, as $\eta_n=o(1)$. 
\end{proof}

\begin{proof}[Proof of \Cref{thm:split_asymptotic}] 
Comparing the close similarity of \smash{$\tilde d_1$} in
\eqref{eq:roo_split_relax} and $d$ in Algorithm \ref{alg:split},   
we see that \smash{$\tilde d_1= d$} if we choose the target coverage
levels to be $1-\alpha$ for the regular split conformal band 
\smash{$C_{\mathrm{split}}$}, and
$1-(\alpha+2\alpha/n)$ for the modified ROO split conformal band
\smash{$\tilde{C}_{\mathrm{roo}}$}.  The desired result
follows immediately by replacing $\alpha$ by $\alpha+2\alpha/n$ in
Theorem \ref{thm:roo_asymptotic}, as it applies to  
\smash{$\tilde{C}_{\mathrm{roo}}$} (explained in the above
remark). 
\end{proof}

\subsection{Proofs for \Cref{sec::theory}}
\label{sec:proof_3}

\begin{proof}[Proof of \Cref{thm:compare_oracle}] 
For any $t>0$, by Fubini's theorem and independence between $\epsilon$
and $(\Delta_n,X)$, 
\begin{align}
    F_n(t) &= \mathbb{P}(|Y-\hat\mu_n(X)| \leq t)\nonumber\\
    &=\mathbb{P}(-t +\Delta_n(X) \leq \epsilon \leq t + \Delta_n(X))\nonumber\\
&= \mathbb{E}_{\hat\mu_n,X}[ F_0(t+\Delta_n(X)) - F_0(-t+\Delta_n(X))],\label{eq:F_n(t)}
\end{align}
where $F_0$ is the CDF of $\epsilon$.

Let $f_0$ be the density function of $F_0$. We can approximate $F_0$ at any $t$ using first order Taylor expansion
\begin{equation*}
  F_0(t+\delta) = F_0(t)+\delta f_0(t) +\delta^2 R(t,\delta) ,
\end{equation*}
where
$R(t,\delta)=0.5\int_{0}^1(1-u)f'_0(t+u\delta)du$ satisfies $\sup_{t,\delta}|R(t,\delta)|\le M/4$.

Next, using symmetry of $F_0$ we have $f_0(t) = f_0(-t)$ for all $t$, the RHS of \eqref{eq:F_n(t)} becomes
\begin{align*}
\mathbb{E}_{\hat\mu_n,X}[ &F_0(t+\Delta_n(X)) - F_0(-t+\Delta_n(X))]
  \\ &= \mathbb{E}_{\hat\mu_n,X} [ F_0(t) + 
\Delta_n(X) f_0(t) + \Delta_n^2(X) R(t,\Delta_n(X)) \\
&\qquad\qquad - F_0(-t)-\Delta_n(X) f_0(-t)
  -\Delta_n^2(X)R(-t,\Delta_n(X))]\\  
&= F_0(t) - F_0(-t) + \mathbb{E}_{\hat\mu_n,X}[ \Delta_n^{2}(X)W]\\
&= F(t)+ \mathbb{E}_{\hat\mu_n,X}[ \Delta_n^{2}(X)W],
\end{align*}
where
$W = R(t,\Delta_n(X)) - R(-t,\Delta_n(X))$. Equation \eqref{eq:Linfty} follows
immediately since $|W| \leq M/2$, almost surely.

Next we show equation (\ref{eq::next}). Because $F$ has density at least $r>0$
in an open neighborhood of $q_\alpha$,
if $t<q_\alpha-\delta$ for some $\delta >(M/2r)\mathbb
E(\Delta_n^2(X))$ then 
\begin{align*}
  F_n(t)\leq & F(q_\alpha-\delta)+(M/2)\mathbb E(\Delta_n^2(X))\\ 
  &\leq F(q_\alpha)-\delta r +(M/2)\mathbb E(\Delta_n^2(X))\\ 
  &<  1-\alpha. 
\end{align*}
Thus $q_{n,\alpha}\ge q_\alpha - (M/2r)\mathbb E\Delta_n^2(X)$.
Similarly we can show that $q_{n,\alpha}\le q_\alpha + (M/2r)\mathbb
E\Delta_n^2(X)$, and hence establish the claimed result.
\end{proof}

\begin{proof}[Proof of \Cref{lemma::split}]
  Without a loss of generality, we assume that the split conformal
  band is obtained using $2n$ samples.
  Let $\tilde q_\alpha$ be the $\alpha$ upper quantile of
  $|Y-\tilde\mu(X)|$.  We first show that 
  \begin{equation}\label{eq:tilde_q_q_n}
    |\tilde q_\alpha-q_{n,\alpha}|\le \rho_n/r + \eta_n,
  \end{equation}
  where $r>0$ is the assumed lower bound on \smash{$\tilde{f}$} in an
  open neighborhood of its $\alpha$ upper quantile. To see this, note
  that  
  \begin{align*}
   \mathbb{P}(|Y-\hat\mu_n(X)|\le \tilde q_{\alpha+\rho_n}-\eta_n)
 &\leq
     \mathbb{P}\Big(|Y-\hat\mu_n(X)|\le\tilde
   q_{\alpha+\rho_n}-\eta_n, \;\|\tilde\mu-\hat\mu_n\|_\infty\le
   \eta_n\Big) 
    +\rho_n\\
    &\leq \mathbb P(|Y-\tilde\mu(X)|\le \tilde q_{\alpha+\rho_n}) 
    +\rho_n\\
   &=1-\alpha-\rho_n+\rho_n=1-\alpha.
  \end{align*}
  Thus \smash{$q_{n,\alpha}\ge \tilde q_{\alpha+\rho_n}-\eta_n\ge
    \tilde q_\alpha-\rho_n/r-\eta_n$}. 
Similarly, \smash{$q_{n,\alpha}\le \tilde
  q_{\alpha-\rho_n}+\eta_n\le \tilde q_\alpha+\rho_n/r+\eta_n$}. 

The width of split conformal band is \smash{$2\hat
  F_n^{-1}(1-\alpha)$}, where \smash{$\hat F_n$} denotes
the empirical CDF of \smash{$|Y_{i}-\hat\mu_n(X_{i})|$},
$i=1,\ldots,n$, and \smash{$\hat\mu_n=\mathcal
  A_n(\{(X_i,Y_i):i=n+1,\ldots,2n\})$}.  
On the event \smash{$\{\|\hat\mu_n-\tilde\mu\|_\infty\le \eta_n\}$}, 
we have \smash{$ |Y_{i}-\hat\mu_n(X_{i})|-|Y_{i}-\tilde\mu(X_{i})|\le
  \eta_n$} for $1=1,\ldots,n$. Therefore, denoting by \smash{$\tilde
  F_n$} the empirical CDF of \smash{$|Y_i-\tilde\mu(X_i)|$},
$i=1,\ldots,n$, we have
\begin{equation}\label{eq:hat_F_n_tilde_F_n}
\mathbb P\left(|\hat F_n^{-1}(1-\alpha)-\tilde F_n^{-1}(1-\alpha)|\le \eta_n\right)\ge 1-\rho_n.
\end{equation}
Using standard empirical quantile theory for i.i.d.\ data and using
the assumption that $\tilde{f}$ is bounded from below by $r>0$ in a
neighborhood of its $\alpha$ upper quantile, we have
\begin{equation}\label{eq:e_quantile_tilde_F}
  \tilde F_n^{-1}(1-\alpha)=\tilde q_{\alpha}+O_\P(n^{-1/2}).
\end{equation}
Combining \eqref{eq:tilde_q_q_n}, \eqref{eq:hat_F_n_tilde_F_n}, and
\eqref{eq:e_quantile_tilde_F}, we conclude that
$$
  |\hat F_n^{-1}(1-\alpha) -
  q_{n,\alpha}|=O_\P(\eta_n+\rho_n+n^{-1/2}),
$$
which gives the result.
\end{proof}

\begin{proof}[Proof of \Cref{thm:full_practical}]
 We focus on the event 
 $$\{\|\hat\mu_n-\tilde\mu\|_\infty\le \eta_n\}\cap
 \left\{\sup_{y\in\mathcal Y}\|\hat\mu_n-\hat\mu_{n,(X,y)}\|_\infty\le
   \eta_n\right\},$$ 
 which, by assumption, has probability at least $1-2\rho_n\rightarrow
 1$.  On this event, we have
\begin{align}
&\Big|  |Y_i-\hat\mu_{n,(X,y)}(X_i)| - |Y_i-\tilde\mu(X_i)|\Big|\le
  2\eta_n, \quad i=1,\ldots,n,\label{eq:residual_approx1}\\
&\Big| |y-\hat\mu_{n,(X,y)}(X)| - |y - \tilde\mu(X)| \Big| \le
  2\eta_n.\label{eq:residual_approx2} 
\end{align}

With \eqref{eq:residual_approx1} and \eqref{eq:residual_approx2}, by
the construction of full conformal prediction interval we can directly
verify the following two facts. 
\begin{enumerate}
  \item  $y\in C_{n,{\rm conf}}(X)$ if $|y-\tilde\mu(X)|\le \tilde
    F_n^{-1}(1-\alpha)-4\eta_n$, and 
  \item $y\notin C_{n,{\rm conf}}(X)$ if $|y-\tilde\mu(X)|\ge \tilde
    F_n^{-1}(1-(\alpha-3/n))+4\eta_n$, 
\end{enumerate}
where $\tilde F_n$ is the empirical CDF of
\smash{$|Y_i-\tilde\mu(X_i)|$}, $i=1,\ldots,n$. 

Therefore, the length of $C_{n,{\rm conf}}(X)$ satisfies
$$
\nu_{n,{\rm conf}}(X) = 2 \tilde q_\alpha + O_\P(\eta_n+n^{-1/2}).
$$
The claimed result follows by further combining the above equation
with \eqref{eq:tilde_q_q_n}. 
\end{proof}

\begin{proof}[Proof of \Cref{thm:split_super}]
Without loss of generality, we assume that the split conformal band is
obtained using $2n$ data points. The proof consists of two
steps. First we establish that 
\smash{$\hat\mu_n(X)-\mu(X)=o_\P(1)$}.  Second  
we establish that \smash{$\hat F_n^{-1}(1-\alpha)-q_\alpha=o_\P(1)$},
where \smash{$\hat F_n$} is the empirical CDF of
\smash{$|Y_i-\hat\mu_n(X_i)|$}, $i=1,\ldots,n$, 
and \smash{$\hat\mu_n=\mathcal A_n(\{(X_i,Y_i):i=n+1,\ldots,2n\})$}. 


We now show the first part.  We will focus on the
event that 
\smash{$\{\mathbb E_X(\hat\mu_n(X)-\mu(X))^2\le \eta_n\}$}, which has
probability at least $1-\rho_n$ by Assumption A4. On this event,
applying 
Markov's inequality, we have that \smash{$\mathbb P(X\in B_n^c \,|\,
\hat\mu_n)\ge 1-\eta_n^{1/3}$}, where
\smash{$B_n=\{x:|\hat\mu_n(x)-\mu(x)|\ge \eta_n^{1/3}\}$}. 
Hence we conclude that \smash{$\mathbb
P_{X, \widehat{\mu}_n }(|\hat\mu_n(X)-\mu(X)|\ge \eta_n^{1/3})\leq 
\eta_n^{1/3} + \rho_n \rightarrow 0$} as $n \rightarrow \infty$, and
the first part of the proof is complete. 

For the second part, define 
$\mathcal I_1=\{i \in \{1,\ldots,n\} : X_i\in B_n^c\}$ and $\mathcal
I_2=\{1,\ldots,n\}\backslash \mathcal I_1$. Note that $B_n$ is
independent of $(X_i,Y_i)$, $i=1,\ldots,n$. Using Hoeffding's
inequality 
conditionally on \smash{$\hat\mu_n$}, we have 
\smash{$|\mathcal I_2| \le 
n\eta_n^{1/3}+c\sqrt{n \log n}=o(n)$} with probability tending to 1,  
for some absolute constant $c>0$. This also holds unconditionally on
\smash{$\hat\mu_n$}.   

Let \smash{$\hat G_{n,1}$} be the empirical CDF of
\smash{$|Y_i-\hat\mu_n(X_i)|$}, $i\in\mathcal I_1$, 
and \smash{$\tilde G_{n,1}$} be the empirical CDF of
\smash{$|Y_i-\mu(X_i)|$}, $i\in\mathcal I_1$.  By definition of
$\mathcal I_1$ we know that 
$$
\Big| |Y_i-\hat\mu_n(X_i)| -
|Y_i-\mu(X_i)| \Big| \le \eta_n^{1/3}, 
$$ 
for all $i\in\mathcal I_1$. All
empirical quantiles of \smash{$\hat G_{n,1}$} and 
\smash{$\tilde G_{n,1}$} are at most
\smash{$O_\P(\sqrt{n})$} apart, because 
\smash{$|\mathcal I_1|=n(1+o_\P(1))$}.    

The half-width of \smash{$C_{n,{\rm split}}(X)$} is 
\smash{$\hat F_n^{-1}(1-\alpha)$}.  According to the definition of
$\mathcal I_1$, we have 
$$
\hat G_{n,1}^{-1}\left(1-\frac{n\alpha}{|\mathcal I_1|}\right)
 \le  \hat F_n^{-1}(1-\alpha) \le \hat
  G_{n,1}\left(1-\frac{n\alpha-|\mathcal I_2|}{|\mathcal
  I_1|}\right). 
$$
 Both $n\alpha/|\mathcal I_1|$ and $(n\alpha-|\mathcal I_2|)/|\mathcal
 I_1|$ are $\alpha+o_\P(1)$.  As a result we conclude that 
$$
\hat F_n^{-1}(1-\alpha)-q_\alpha =o_\P(1),
$$
and the second part of the proof is complete.
\end{proof}

\begin{proof}[Proof of \Cref{thm::consistent}]
  Using the same arguments as in the proof of \Cref{thm:split_super},
  we can define the set $B_n$ and index sets $\mathcal I_1$, $\mathcal
  I_2$. 
  Now we consider the event $\{X\in B_n^c\}$, which has probability  
  tending to 1. Then on this event, by definition of $B_n$ and the
  fact that \smash{$\eta_n\le \eta_n^{1/3}$}, we have  
  \begin{align}
  &\Big|  |Y_i-\hat\mu_{n,(X,y)}(X_i)| - |Y_i-\tilde\mu(X_i)|\Big|\le
    2\eta_n^{1/3}, \quad i\in\mathcal
    I_1,\label{eq:residual_approx3}\\  
  &\Big| |y-\hat\mu_{n,(X,y)}(X)| - |y - \tilde\mu(X)| \Big| \le
    2\eta_n^{1/3}.\label{eq:residual_approx4} 
  \end{align}
  
  By definition of $C_{n,{\rm conf}}(X)$ and following the same reasoning as in the proof of 
\Cref{thm:full_practical}, we can verify the following facts:
\begin{enumerate}
  \item  $y\in C_{n,{\rm conf}}(X)$ if $|y-\tilde\mu(X)|\le \tilde
    G_{n,1}^{-1}\left(1-\frac{n\alpha}{|\mathcal
        I_1|}\right)-4\eta_n^{1/3}$, and
  \item $y\notin C_{n,{\rm conf}}(X)$ if $|y-\tilde\mu(X)|\ge \tilde
    G_{n,1}^{-1}\left(1-\frac{n\alpha-|\mathcal I_2|-3}{|\mathcal
        I_1|}\right)+4\eta_n^{1/3}$, 
\end{enumerate}
where \smash{$\tilde G_{n,1}$} is the empirical CDF of 
\smash{$|Y_i-\tilde\mu(X_i)|$}, $i\in\mathcal I_1$.  

 Both $n\alpha/|\mathcal I_1|$ and $(n\alpha-|\mathcal
 I_2|-3)/|\mathcal I_1|$ are $\alpha+o_\P(1)$, and hence 
$$
\tilde G_{n,1}^{-1}\left(1-\frac{n\alpha}{|\mathcal
    I_1|}\right)=q_\alpha+o_\P(1), \quad 
\tilde G_{n,1}^{-1}\left(1-\frac{n\alpha-|\mathcal I_2|-3}{|\mathcal
    I_1|}\right)=q_\alpha+o_\P(1). 
$$
Thus the lower (upper) end point of \smash{$C_{n,{\rm conf}}(X)$} is
\smash{$q_\alpha+o_\P(1)$} below (above) $\mu(X)$, and the proof is 
complete. 
\end{proof}

\subsection{Proofs for \Cref{sec::extensions}}
\label{sec:proof_2}

\begin{proof}[Proof of \Cref{thm:roo_asymptotic}]
  For notational simplicity, we assume that $\mathcal
  I_1=\{1,\ldots,n/2\}$, and  $R_i$, $i=1,\ldots,n/2$ are in
  increasing order. 
    Let $m=\lceil (1-\alpha)n/2\rceil$.  Then $\one \{Y_i\in C_{\rm
      roo}(X_i)\}=\one\{R_i\le  d_i\}$ where $d_i$ is the $m$th  
    smallest value in $R_1,\ldots,R_{i-1},R_{i+1},\ldots,R_{n/2}$.  
  Now we consider changing a sample point, say, $(X_j,Y_j)$, in
  $\mathcal I_1$ and denote the 
  resulting possibly unordered residuals by
  \smash{$R_1',\ldots,R_{n/2}'$}, and define $d_i'$ correspondingly.  
  Consider the question: for which values of \smash{$i\in\mathcal 
  I_1\backslash \{j\}$} can we have \smash{$\one\{R_i\le d_i\}\neq
  \one\{R_i' \le d_i'\}$}?  

  Recall that by assumption $R_1\le R_2\le \ldots\le R_{n/2}$.
  If $i\le m-1$ and $i\neq j$, then $d_i\ge R_{m}$, $d_i'\ge R_{m-1}$,
  $R_i=R'_i$, and hence $\one\{R_i\le d_i\}=\one\{R_i' \le d_i'\}=1$.   
  If $i\ge m+2$ and $i\neq j$, then using similar reasoning we have
  $\one\{R_i\le d_i\}=\one\{R_i' \le d_i'\}=0$.  Therefore, changing a
  single data point can change $\one\{Y_i\in C_{\rm roo}(X_i)\}$ for
  at most three values of $i$ (i.e., $i=m,m+1,j$).  As the
  input sample points are independent, we can use McDiarmid's
  inequality, which gives
  \begin{align*}
    \P\left( \frac{2}{n}\sum_{i\in\mathcal I_1}\one\{Y_i\in C_{\rm
    roo}(X_i)\}\le 1-\alpha-\epsilon\right)\leq \exp(-cn\epsilon^2).  
  \end{align*}
  The claim follows by switching $\mathcal I_1$ and $\mathcal I_2$ and
  adding the two inequalities up. 

  Now we consider the other direction. We must only show that $\mathbb
  P(Y_j\notin C_{\rm roo}(X_j)) \geq \alpha-2/n$. Under the continuity 
  assumption, with probability one the residuals are all distinct. Let
  \smash{$j\in\mathcal I_{k}$} for $k=1$ or $2$. By construction,
  $C_{\rm roo}(X_j)$ does not contain the $y$ values such that
  \smash{$|y-\hat\mu_{3-k}(X_j)|$} 
  is among the $n/2 - \lceil(n/2)(1-\alpha)\rceil$ largest of
  $\{R_i:i\in\mathcal I_k\backslash \{j\}\}$. 
  Denote this set by $D_{\rm roo}(X_j)$.  Then the standard conformal
  argument implies that 
  $$
  \mathbb P(Y_i\in D_{\rm roo}(X_i)) \geq \frac{n/2 -
    \lceil(n/2)(1-\alpha)\rceil}{n/2}\ge \alpha-\frac{2}{n}, 
  $$
and we can establish the corresponding exponential deviation
inequality using the same reasoning as above. 

  For \smash{$\tilde C_{\rm roo}(X_j)$}, the lower bound follows from
  that of $C_{\rm roo}(X_j)$ because  
  \smash{$\tilde C_{\rm roo}(X_j)\supseteq C_{\rm roo}(X_j)$}.  To
  prove the upper bound, note that \smash{$\tilde C_{\rm roo}(X_j)$}
  does not contain $y$ such that \smash{$|y-\hat\mu_{3-k}(X_j)|$} is
  among the $(n/2)-\lceil(n/2)(1-\alpha)\rceil-1$ largest of
  $\{R_i:i\in \mathcal I_k \backslash \{j\}\}$. 
  Hence it does not contain points $y$ such that
  \smash{$|y-\hat\mu_{3-k}(X_j)|$} is among the
  $(n/2)-\lceil(n/2)(1-\alpha)\rceil-2$ largest of $\{R_i:i\in
  \mathcal I_k\backslash\{j\}\}$. Comparing this with the argument for 
  $C_{\rm roo}$, the extra $-2$ in the ranking changes $2/n$ to $6/n$ 
  in the second probability statement in the theorem. 
\end{proof}

\section{Additional Experiments}
\label{app:additional}

We present two additional experiments.

\begin{figure}[htbp]
\centering
\includegraphics[width=0.32\textwidth]{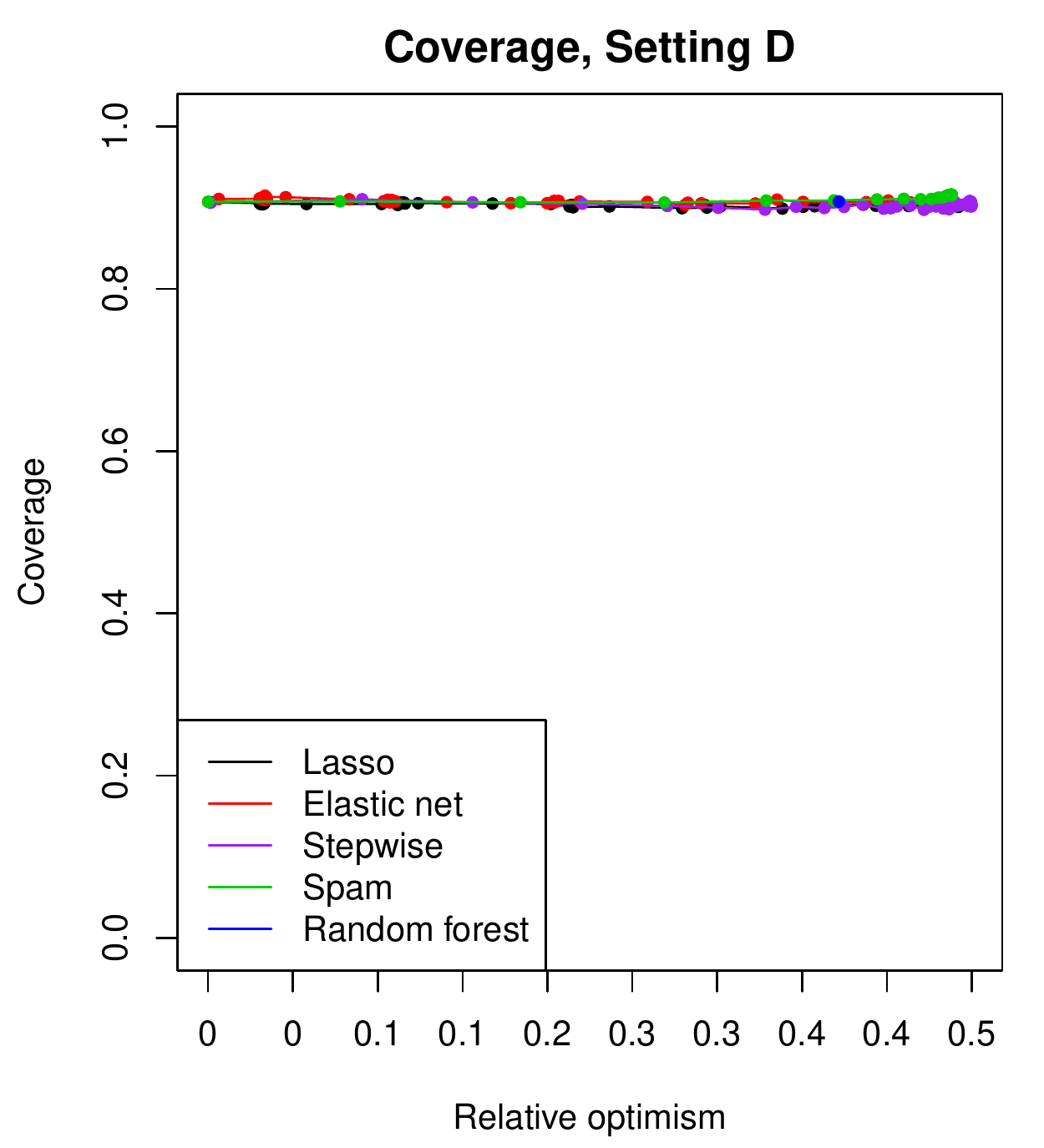} 
\includegraphics[width=0.32\textwidth]{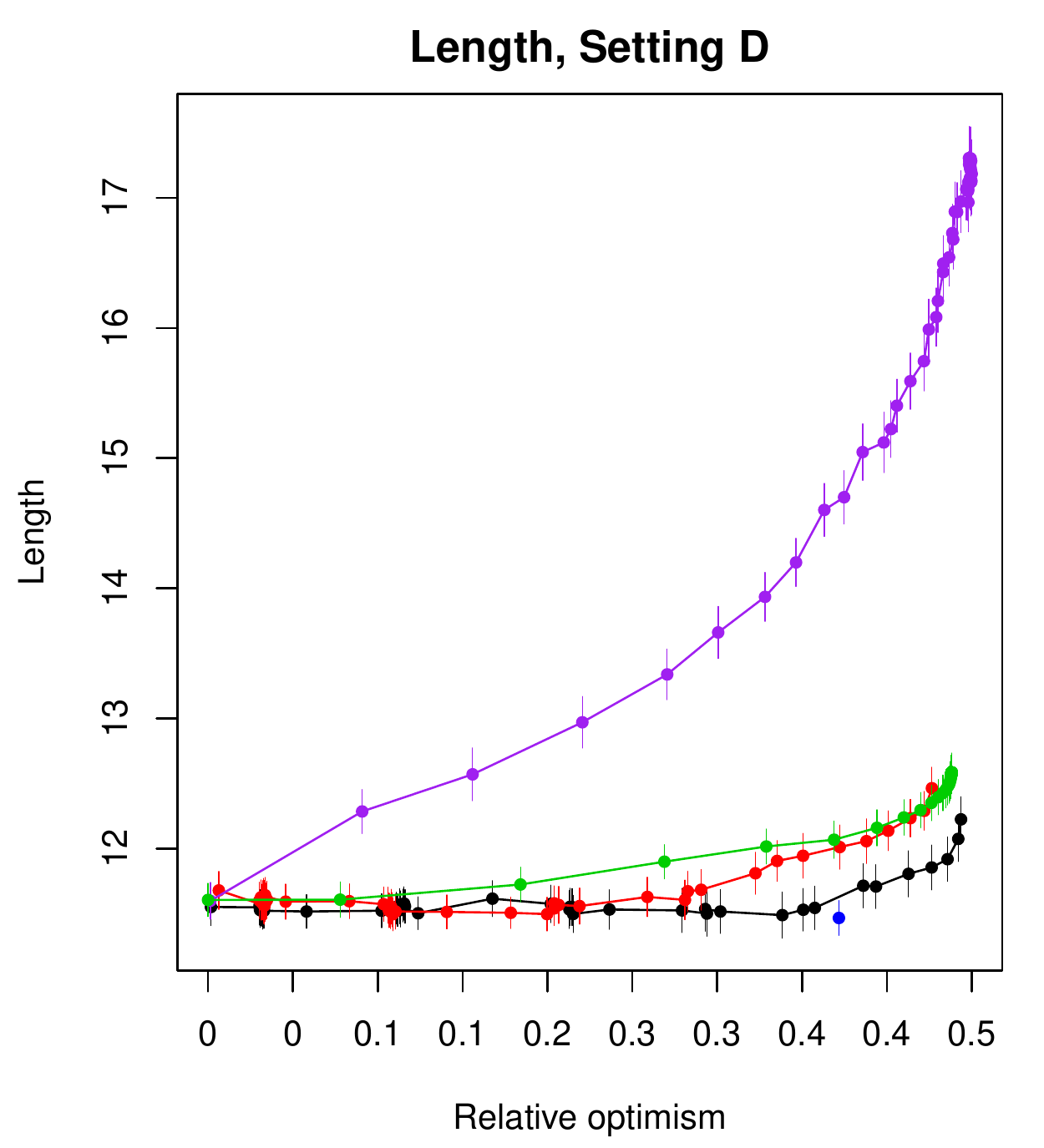} 
\includegraphics[width=0.32\textwidth]{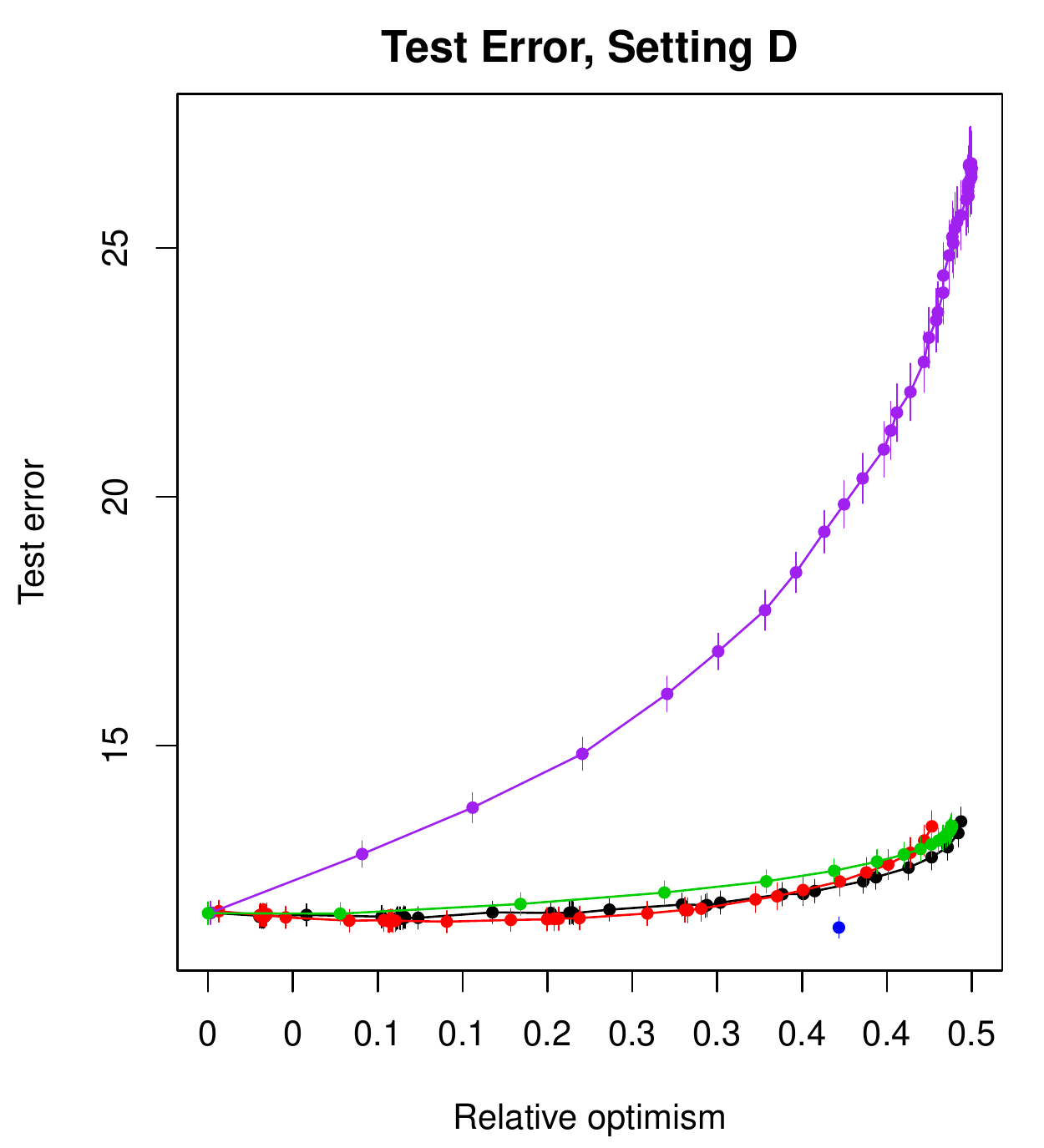}
\caption{\it More comparisons of conformal prediction intervals in 
  high-dimensional problems with $n=200$, $d=2000$; here the setup
  mimics Setting A of \Cref{fig:many.hi}, but without sparsity:
  the number of active variables in the linear model for the mean is  
  $s=100$.}   
\label{fig:many.hi.nonsparse}
\end{figure}

\Cref{fig:many.hi.nonsparse} shows the results for the same
simulation 
setup as in the first panel of \Cref{fig:many.hi}, except with a
nonsparse mean function: the mean is a linear combination of 
$s=100$ of the underlying features.  The message is that, while no
methods perform well in terms of test error, all conformal bands still  
achieve exactly (more or less) 90\% average coverage, as prescribed. 

\begin{figure}[htbp]
\centering
\includegraphics[width=0.48\textwidth]{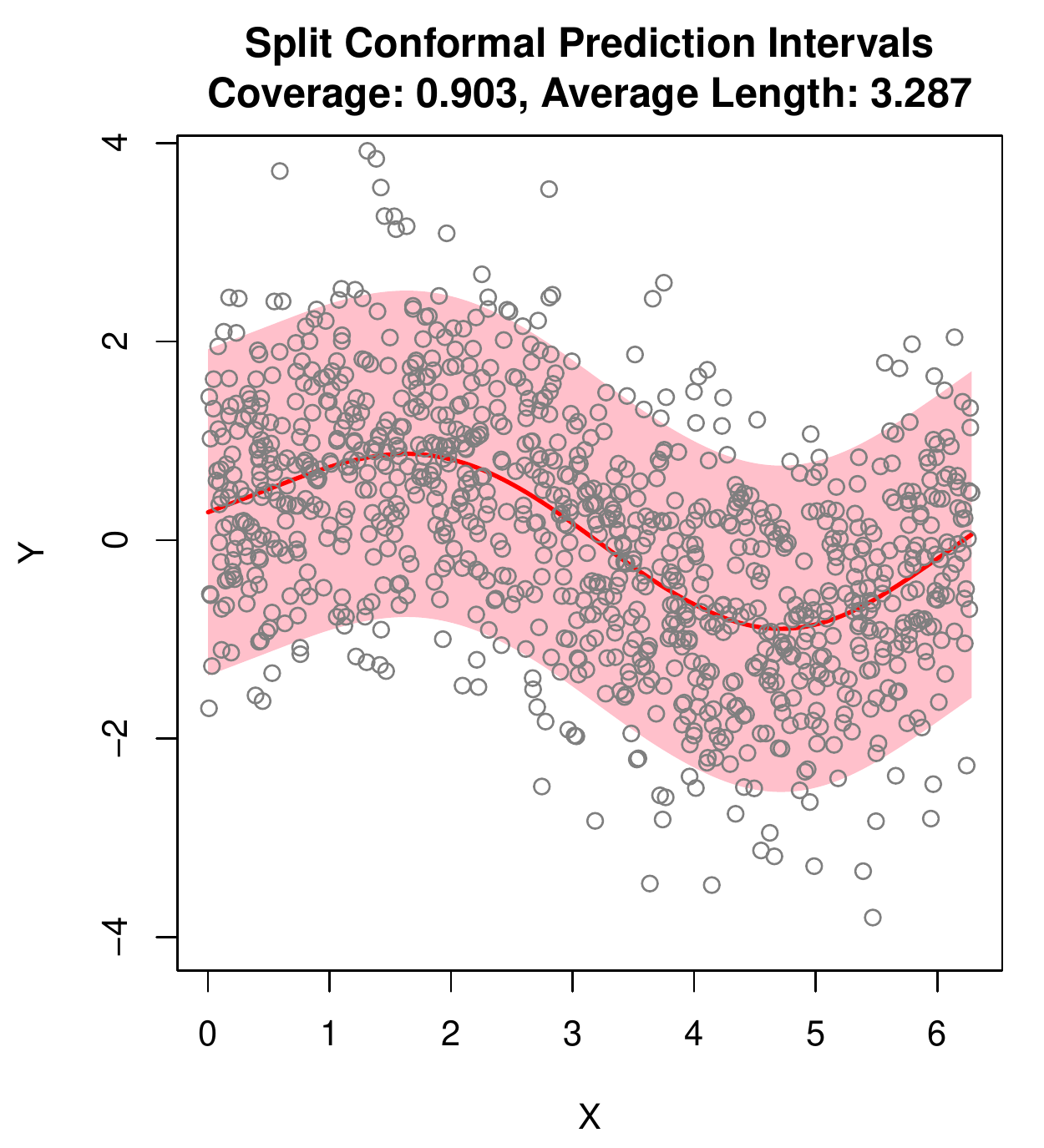} 
\includegraphics[width=0.48\textwidth]{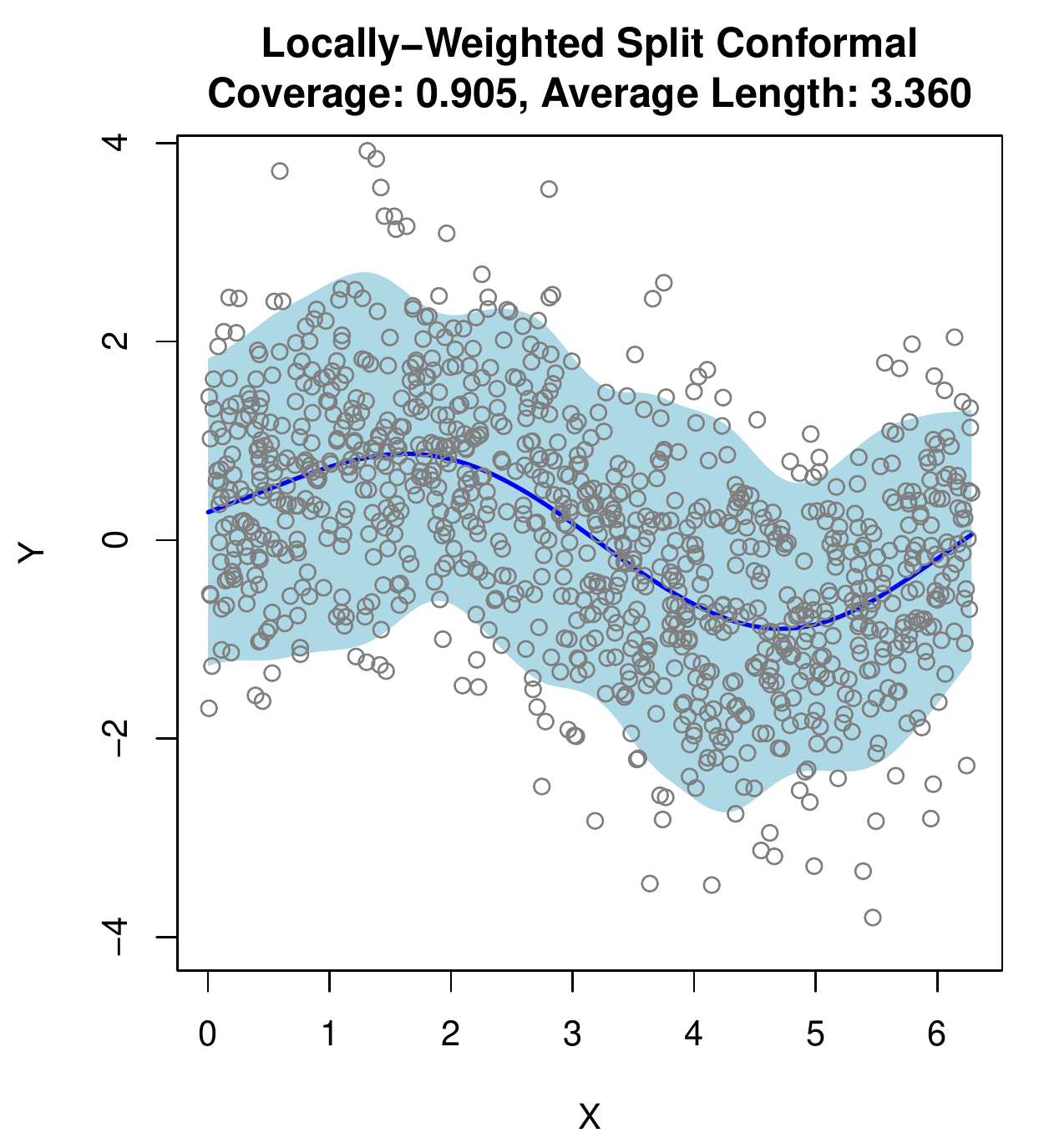} \\
\includegraphics[width=0.48\textwidth]{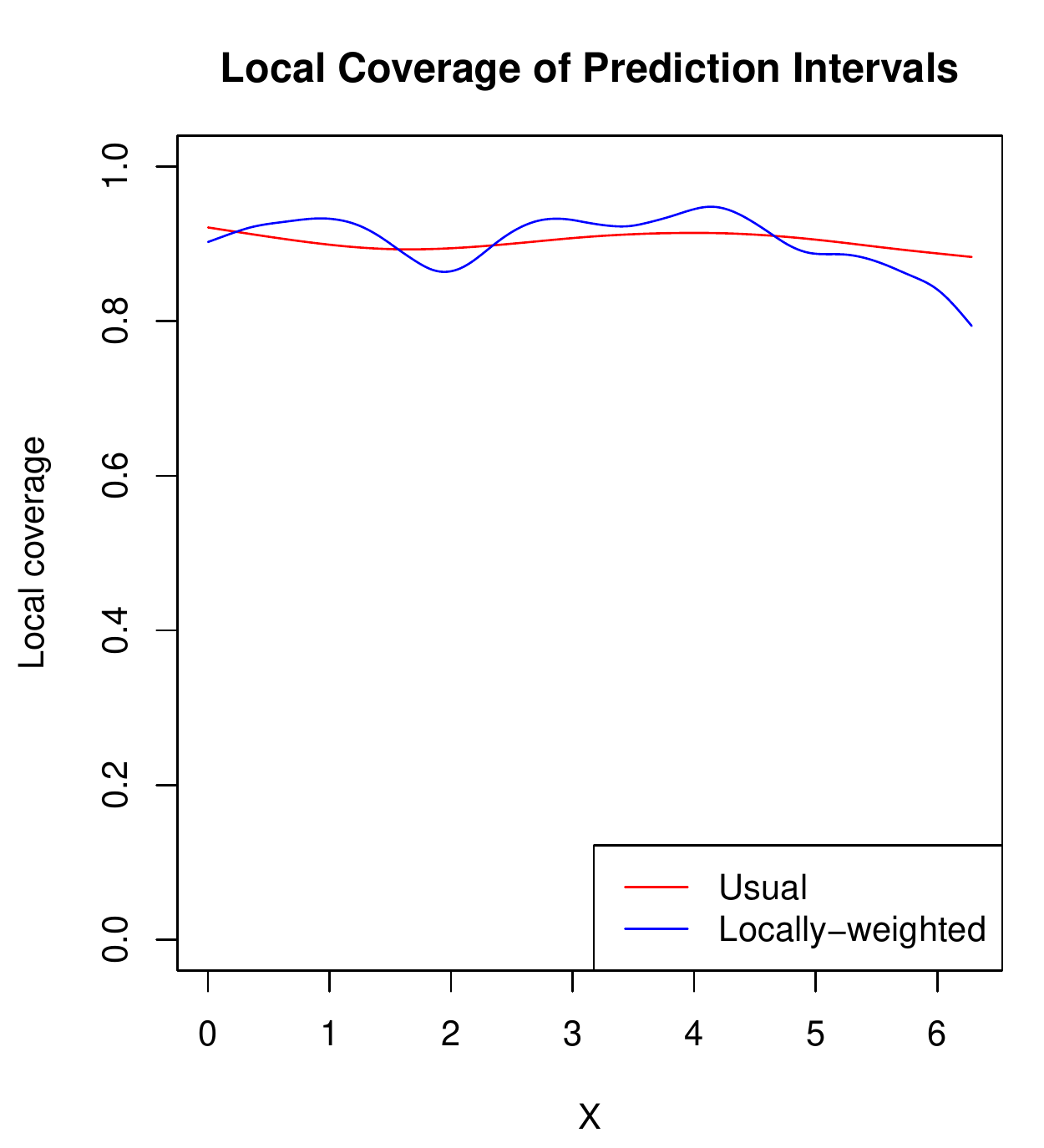} 
\includegraphics[width=0.48\textwidth]{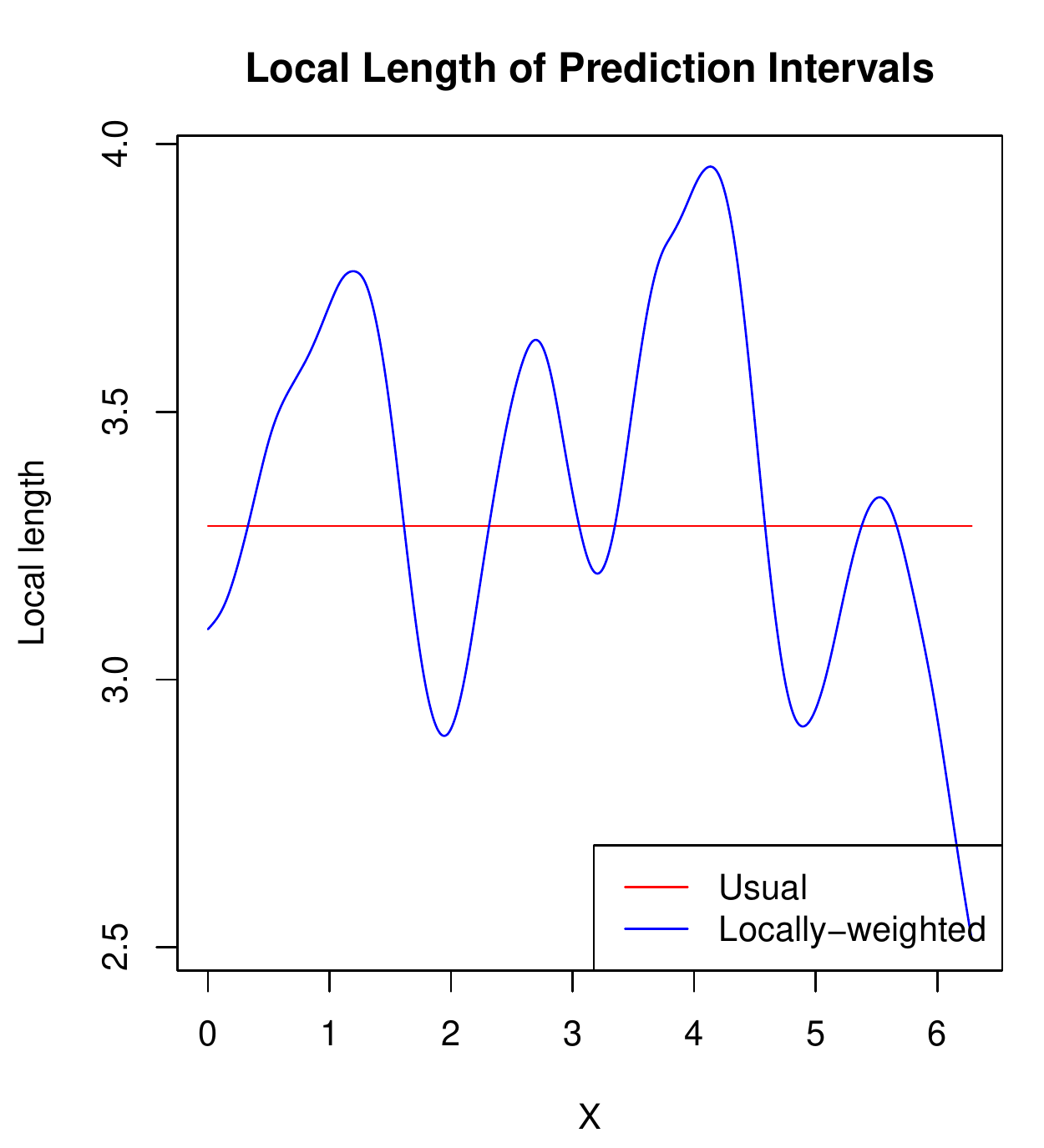} 
\caption{\it Another example of the usual (unweighted)
  split conformal and locally-weighted split conformal prediction
  bands, in the same setup as \Cref{fig:local_weight}, except without
  heteroskedasticity.}
\label{fig:local_weight_homo}
\end{figure}

\Cref{fig:local_weight_homo} displays the results for the same
simulation setup as in \Cref{fig:local_weight}, but without
heteroskedasticity in the noise distribution. We can see that the
locally-weighted method produces a band with only mildly varying local
length, and with essentially constant local coverage.  Overall, the
average length of the locally-weighted method is not much worse than
the usual unweight conformal method.

\end{document}